\documentclass[11pt]{article}
\pdfoutput=1
  
\usepackage{a4wide}
\usepackage[fleqn]{amsmath}
\usepackage{hyperref}
\usepackage{color}

\RequirePackage{amsmath,amssymb,amsxtra,amsthm}

\RequirePackage[T1]{fontenc}
\RequirePackage[utf8]{inputenc}

\RequirePackage{mathrsfs}

\RequirePackage{setspace}
\RequirePackage{array}
\RequirePackage{booktabs}

\RequirePackage{braket}

\RequirePackage[pdftex,final]{graphicx}
\graphicspath{{Plots/}}

\usepackage{cite}

\newcommand{\be}{\begin{equation}}
\newcommand{\ee}{\end{equation}}
\newcommand{\bea}{\begin{eqnarray}}
\newcommand{\eea}{\end{eqnarray}}
\newcommand{\IR}{\mathbb{R}}
\newcommand{\IC}{\mathbb{C}}
\newcommand{\IZ}{\mathbb{Z}}

\newcommand{\cF}{\mathcal{F}}
\newcommand{\cV}{\mathcal{V}}

\newcommand{\cH}{\mathcal{H}}
\newcommand{\cL}{\mathcal{L}}
\newcommand{\cP}{\mathcal{P}}

\newcommand{\cM}{\mathcal{M}}

\newcommand{\cN}{\mathcal{N}}
\newcommand{\cE}{\mathcal{E}}

\newcommand{\cO}{\mathcal{O}}
\newcommand{\cR}{\mathcal{R}}
\newcommand{\cT}{\mathcal{T}}

\newcommand{\de}{\mathrm{d}}

\newcommand{\I}{\mathrm{i}}
\newcommand{\zetastar}{\zeta^\star}

\newcommand{\nn}{\nonumber}

\newcommand{\RN}{{\rm R.N.}}
\newcommand{\Res}{{\rm Res}}
\renewcommand{\Re}{{\rm Re}}
\renewcommand{\Im}{{\rm Im}}

\def\pa{\partial}

\numberwithin{equation}{section}
\numberwithin{table}{section}
\author{
  \begin{minipage}{0.97\linewidth}
    \vspace{1cm}
    \begin{center}
        {\Large \textbf{\textbf{Ioannis Florakis}$^{1}$ and  \textbf{Boris Pioline}$^{1,2}$}}
    \end{center}
    \vspace{.5cm} \hspace{1.3cm}\begin{minipage}{.75\linewidth}
      {\it \begin{small}
          \begin{itemize}
         \item[${}^1$] CERN Theory Division, 1211 Geneva 23, Switzerland
         \item[${}^2$] Laboratoire de Physique Th\'eorique et Hautes Energies, CNRS UMR 7589,
         \\
         Universit\'e Pierre et Marie Curie - Paris 6, 4 place Jussieu,
         75252 Paris cedex 05, France
          \end{itemize}
        \end{small}
        }
    \end{minipage}
    \vspace{1cm}
  \end{minipage}
}

\date{}

\title{\vspace{3cm}
  \begin{huge}
    \textbf{On the Rankin-Selberg method \\[2mm] for higher genus 
string amplitudes}
  \end{huge}
}

\begin{document}

\begin{titlepage}
  \maketitle
  \thispagestyle{empty}

  \vspace{-14cm}
  \begin{flushright}
    CERN-TH-2016-022 \\ arXiv:1602.00308v2
   \end{flushright}

  \vspace{11cm}

\begin{center}
\textsc{Abstract}
\\
\end{center}
Closed string amplitudes at genus $h\leq 3$ are given by integrals of Siegel modular functions on a fundamental domain of the Siegel upper half-plane. When the integrand is of rapid decay near the cusps, the integral can be computed by the Rankin-Selberg method, which consists of inserting an Eisenstein series $\cE_h(s)$ in the integrand, computing the integral by the orbit method, and  finally extracting the residue at a suitable value of $s$. String amplitudes, however, typically involve integrands with polynomial or even exponential growth at the cusps, and a renormalization scheme is required to treat infrared divergences. 
Generalizing Zagier's extension of the Rankin-Selberg method  at genus one, we develop the Rankin-Selberg method for Siegel modular functions of degree 2 and 3 with polynomial growth near the cusps. In particular, we show that the renormalized modular integral of the Siegel-Narain partition function of an even self-dual lattice of signature $(d,d)$ is proportional to a residue of the Langlands-Eisenstein series  attached to the $h$-th antisymmetric tensor representation of the T-duality group $O(d,d,\IZ)$.
\vfill
{\small
\begin{itemize}
\item[E-mail:] {\tt ioannis.florakis@cern.ch}\\
{\tt boris.pioline@cern.ch}
\end{itemize}
}
\vfill

\end{titlepage}

\setstretch{1.1}

\tableofcontents


\section{Introduction}

According to the current rules of perturbative superstring theory, scattering amplitudes at $h$-loops are expressed as integrals of a suitable superconformal correlation function on the moduli space $\mathfrak{M}_h$ of super-Riemann surfaces of genus $h$ \cite{D'Hoker:2005ia,Witten:2012bh}. Although there is in general no global holomorphic projection onto the moduli space $\cM_h$ of ordinary Riemann surfaces \cite{Donagi:2013dua}, for most practical purposes the integral over $\mathfrak{M}_h$ can be reduced
to an integral over $\cM_h$, possibly supplemented with boundary contributions from nodal curves, which incorporate the infrared singularities due to massless degrees of freedom. Even after this reduction has been performed, there are 
very few cases where the integral over $\cM_h$ can be evaluated explicity, due to the complexity of the integrand and of the integration domain, a quotient of the Teichm\"uller space $\cT_h$ by the mapping class group $\Gamma_h$. For $h\leq 3$, $\cT_h$ is isomorphic (via the period map $\Sigma\to\Omega$) to the Siegel upper half plane 
$\cH_h$ of degree $h$ (away from the separating degeneration locus), while $\Gamma_h$ is isomorphic to the Siegel modular group $PSp(2h,\IZ)=Sp(2h,\IZ)/\{\pm 1\}$ (or a congruence subgroup thereof, if one keeps track of spin structures). 

At genus one, several efficient methods for integrating modular functions on the fundamental domain $\cF_1$ of the Poincar\'e upper-half plane have been developped, starting with the integration-by-parts method \cite{Lerche:1987qk} and the orbit method of \cite{McClain:1986id,O'Brien:1987pn,Dixon:1990pc,Harvey:1996gc} in the physics literature, and the Rankin-Selberg-Zagier method for automorphic functions of polynomial growth in the math literature \cite{MR656029}. With the recent advances of \cite{Angelantonj:2011br,Angelantonj:2012gw,Angelantonj:2013eja,Angelantonj:2015rxa}\footnote{Other attempts to use the Rankin-Selberg method
to compute string amplitudes or probe the asymptotic density of string states  include \cite{Cardella:2008nz,Cacciatori:2010js,Cardella:2010bq,Angelantonj:2010ic,Cacciatori:2011fc,Cacciatori:2011qd}.}, it is now possible to evaluate analytically the integral over $\cF_1$ of any (modular invariant) product of an almost weakly holomorphic modular form times a lattice partition function. This covers most of the cases relevant for  BPS-saturated amplitudes at one-loop.

At genus two or three, the only examples computed so far are those where the integrand is constant \cite{D'Hoker:2005ht,Gomez:2013sla}, or proportional to the so-called Kawazumi-Zhang 
invariant $\varphi_{\rm KZ}(\Omega)$ \cite{D'Hoker:2013eea,D'Hoker:2014gfa}, or proportional to the Narain partition function $\Gamma_{d,d,h}$ of the even self-dual lattice  of signature $(d,d)$ \cite{Obers:1999um,Green:2010wi}. In the first case, the volume of the fundamental domain $\cF_h$ of the Siegel modular group of degree $h$ is well-known since \cite{0013.24901}, while  in the second case the integral of $\varphi_{\rm KZ}$ could be computed by integration by 
parts \cite{D'Hoker:2014gfa}. In the last case, the integral $\int_{\cF_h} \de\mu_h \Gamma_{d,d,h}$ is divergent for $d\geq h+1$ and must be regularized. It was conjectured in \cite{Obers:1999um} that the renormalized integral should be proportional to the sum of the Langlands-Eisenstein series $\cE^{\star,SO(d,d)}_{S,C}(s=h)$ attached to the spinorial representations of the T-duality group $SO(d,d,\IZ)$. 

In \cite{Pioline:2014bra}, some preliminary steps were taken towards evaluating the integral $\int_{\cF_h} \de\mu_h \Gamma_{d,d,h}$ by the Rankin-Selberg method. Recall that for a modular function $F$ of rapid decay, the integral
$\int_{\cF_h} \de\mu_h \, F(\Omega)$ can be deduced from the `Rankin-Selberg transform'
\be
\label{DefRS}
\cR_h^\star(F,s) = \int_{\cF_h} \de\mu_h \, \cE_h^\star(s,\Omega)\, F(\Omega)\ ,
\ee
where $\cE_h^\star(s,\Omega)$ is the (completed) non-holomorphic Siegel-Eisenstein series,
by using the known
fact \cite{0397.10021} that $\cE^\star(s,\Omega)$ is a meromorphic function of $s\in\IC$ with a pole at $s=\tfrac{h+1}{2}$, and with constant residue $r_h= \frac12 \, \prod_{j=1}^{\lfloor h/2 \rfloor}\, \zetastar(2j+1)$ :
\be
\label{RSresidue}
\int_{\cF_h} \de\mu_h \, F(\Omega) = \frac{1}{r_h} \Res_{s=\tfrac{h+1}{2}}\, \cR_h^\star(F,s)\ .
\ee
The Rankin-Selberg transform \eqref{DefRS} can in turn be computed by representing the Eisenstein series $\cE^\star_h(s,\Omega) $ for $\Re(s)>\tfrac{h+1}{2}$ as
a sum over images under $\Gamma_h$,
\be
\label{defEstarh}
\cE^\star_h(s,\Omega) = \zetastar(2s) \, \prod_{j=1}^{\lfloor h/2 \rfloor} \zetastar(4s-2j)\, 
\sum_{\gamma\in\Gamma_{h,\infty}\backslash \Gamma_h} |\Omega_2|^{s}\vert \gamma\ ,\qquad 
\Gamma_{h,\infty}= \Gamma_h \cap \{ \begin{pmatrix} A & B \\ 0 & D \end{pmatrix}\}\,,
\ee
where $\Omega=\Omega_1+\I\Omega_2$, $|\Omega_2|\equiv \det\Omega_2$,
and unfolding the integration domain against the sum over images, leading to 
\be
\label{Rstarhunfold}
\cR_h^\star(F,s) =\zetastar(2s) \, \prod_{j=1}^{\lfloor h/2 \rfloor} \zetastar(4s-2j)\, 
\int_{\Gamma_{\infty,h}\backslash \cH_h} \de\mu_h \, |\Omega_2|^{s}\, F(\Omega)\ .
\ee
The integral over $\Omega_1$ projects $F(\Omega)$ onto the zero-th Fourier coefficient 
$F^{(0)}(\Omega_2)$ with respect to $\Gamma_{h,\infty}$, while the remaining integral
over $\Omega_2$ runs over a fundamental domain of the action of $PGL(h,\IZ)$ on the space
$\cP_h$ of positive definite symmetric matrices. The latter can be viewed as a generalized Mellin transform of $F^{(0)}(\Omega_2)$.  Notably it inherits the analyticity property and 
invariance under $s\mapsto \tfrac{h+1}{2}-s$ of the Eisenstein series $\cE^\star_h(s,\Omega)$.
In the case where $F(\Omega)=|\Omega_2|^w |\Psi(\Omega)|^2$, where $\Psi(\Omega)$ is a holomorphic cusp form of weight $w$, the integral
over $\cP_h/PGL(h,\IZ)$ can be computed using again the unfolding method and identity (\ref{GenEuler}) of appendix \S\ref{sec_lath}. This
expresses $\cR_h^\star(F,s)$ as an L-series generalizing the symmetric square  L-series 
for $h=1$ \cite{0021.39201,zbMATH03355125,zbMATH03475535,0715.11025}, establishing its analytic properties and functional equation. 

If however $F(\Omega)$ does not  decay sufficiently rapidly at the cusps,  as is typically the case in string amplitudes, the integrals  \eqref{DefRS} and \eqref{Rstarhunfold} must be regularized. For $h=1$, it was shown in \cite{MR656029} that for modular functions with polynomial growth at the cusp, a suitably  renormalized version of the integral  \eqref{DefRS} is given by the Mellin transform of $F^{(0)}-\varphi$ (still denoted by $\cR_h^\star(F,s)$), where $\varphi(\Omega_2)$ is the non-decaying part of $F(\Omega)$; the latter has a meromorphic continuation in $s$ with additional poles over and above those of the Eisenstein series $\cE^\star_h(s,\Omega)$. Moreover, the renormalized integral of $F$ differs from (one over $r_h$ times) the residue of $\cR_h^\star(F,s)$ at $s=\tfrac{h+1}{2}$ by a computable term whenever the order of the pole is greater than one. In the context of string theory, 
the regularization of a physical amplitude $\int_{\cF_h} \de\mu_h \, F(\Omega)$ by inserting 
an Eisenstein series in the  integrand and then extracting the residue at $s=\tfrac{h+1}{2}$,
can be viewed as an analogue of dimensional regularization in quantum field theory, where the number of non-compact space-time dimensions is analytically continued from $D$ to $D-(2s-h-1)$.
 
 Assuming that a similar procedure could be carried out for $h>1$, it was found
in \cite{Pioline:2014bra} that for the particular case $F=\Gamma_{d,d,h}$, the renormalized Rankin-Selberg transform
\eqref{Rstarhunfold} (after subtracting by hand all non-decaying terms from $F$) is proportional to the Langlands-Eisenstein series $\cE^{\star,SO(d,d)}_{\Lambda^h V}(s+\tfrac{d-1-h}{2})$ attached to the $h$-th antisymmetric power of the fundamental representation; and that
the renormalized integral of $F$ is equal to the residue of the same\footnote{This is compatible with the fact that the same integral is proportional to the residue of $\cE^{\star,SO(d,d)}_{S,C}(s)$ at $s=h$,  thanks to identities between Langlands-Eisenstein series.} at $s=\tfrac{h+1}{2}$, up to an undetermined correction term
$\delta$ when the order of the pole is greater than one. The arguments in \cite{Pioline:2014bra} were heuristic, however, and one motivation for the present work is to put the claim of \cite{Pioline:2014bra} on solid footing.

More generally, the goal of this work is to extend the Rankin-Selberg-Zagier method \cite{MR656029} to Siegel modular functions of degree $h\geq 2$ which are regular inside $\cH_h$ and have at most polynomial growth at the cusps. 
The main challenge is to find a convenient cut-off which removes all divergences, while allowing to unfold the integration domain against the sum in the Siegel-Eisenstein series. In general, divergences originate from regions of $\cF_h$ where a diagonal block of size $h_2\times h_2$ in the lower-right corner of $\Omega_2$ is scaled to infinity, while keeping the remaining entries in $\Omega$ finite.
The stabilizer of this cusp inside $PSp(2h,\IZ)$ is a parabolic subgroup of the Siegel modular group with Levi component equal to $[Sp(2h_1,\IZ)\times GL(h_2,\IZ)]/\mathbb Z_2$, where $h_1+h_2=h$.  In the language of string theory (so for $h=2,3$ only), this divergence is interpreted as a
 $h_2$-loop infrared subdivergence for $1\leq h_2<h$, or as a primitive infrared divergence for $h_2=h$.  To subtract these divergences, we shall apply the following strategy (already suggested in \cite{Angelantonj:2011br}):  
\begin{enumerate}
\item[i)] construct an increasing family of compact domains $\cF_h^\Lambda\subset \cF_h$, $\Lambda\in\IR^+$ such that $\lim_{\Lambda\to\infty} \cF_h^{\Lambda} = \cF_h$; the regularized integral
$\cR^{\star,\Lambda}_h(F,s)=\int_{\cF_h^\Lambda} \de\mu_h \,\cE_h^\star(s)\, F$ on
the `truncated fundamental domain' $\cF_h^\Lambda$ is then manifestly finite and has the same analytic structure as $\cE_h^\star(s)$ as a function of $s\in\IC$;
\item[ii)] find an invariant differential operator $\Diamond$ which annihilates the non-decaying part of the integrand $F$; 
\item[iii)] relate the regularized integrals $\cR^{\star,\Lambda}_h(F,s)$ and $\cR^{\star,\Lambda}_h(\Diamond F,s)$  using integration by parts;
\item[iv)] apply the standard Rankin-Selberg method to the decaying function $\Diamond F$, {\it i.e.} compute $\lim_{\Lambda\to\infty} \cR_h^{\star,\Lambda}(\Diamond F,s)$ in terms of the 
(generalized) Mellin transform of the  constant term $\Diamond F^{(0)}$;
\item [v)] relate the Mellin transform of $\Diamond F^{(0)}$ to the regularized Mellin transform of $F^{(0)}$.
\end{enumerate}
In the body of this paper, we develop this strategy in detail in degree one (revisiting \cite{MR656029,Angelantonj:2011br}), degree two, and degree three. 
In appendix \S\ref{app_gendegree}, we collect relevant facts about Siegel-Eisenstein series 
and invariant differential operators valid in any degree, which could be used to extend
our procedure beyond degree three.  In appendix \S\ref{sec_lath}, we compute the renormalized Rankin-Selberg transform for the Siegel-Narain lattice partition function in arbitrary degree.
Aside from this example, it would be interesting to use our techniques to study the analytic properties of L-series associated to non-cuspidal Siegel modular forms.

\begin{figure}
\begin{center}
\includegraphics[height=14cm]{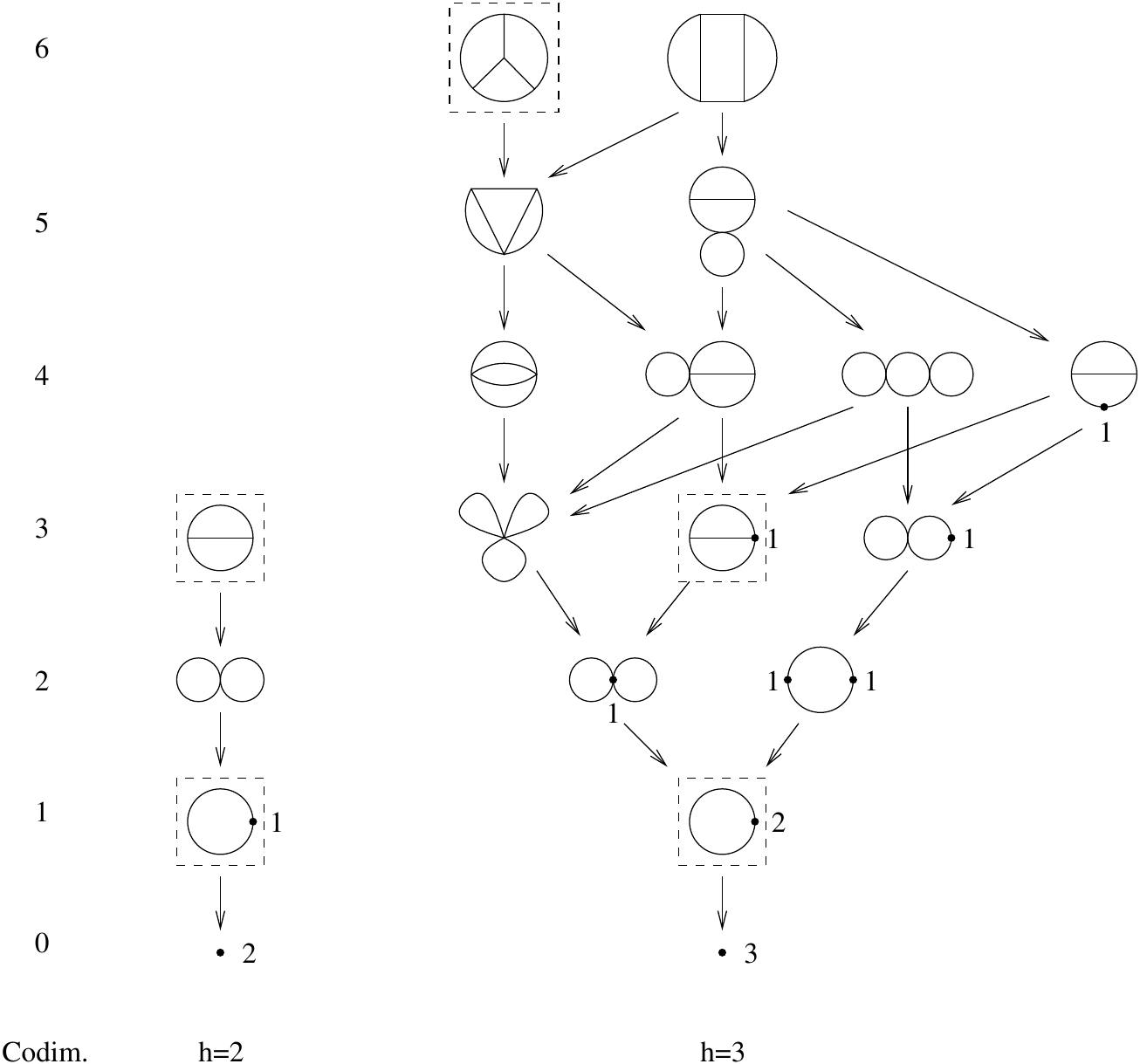}%
\end{center}
\caption{Non-separating degenerations of Riemann surfaces of genus two and three (see  e.g. Figures 1 and 4 in \cite{zbMATH06126971} for the full set including separating degenerations).
The boxed ones correspond to those where the period matrix reaches a boundary of the fundamental domain $\cF_h$ in the Siegel upper half-plane.\label{figdeg}}
\end{figure}

Since we only consider integrals of Siegel modular functions which are regular inside $\cH_h$, our procedure only applies to string amplitudes of genus $h\leq 3$, whose integrand is regular in all
separating degeneration limits, as well as in non-separating degenerations which do not correspond to cusps of the Siegel upper half-plane (see Figure \ref{figdeg}). Still, it is applicable to a number of 
amplitudes of physical interest, such as the two-loop $D^4 \cR^4$ \cite{D'Hoker:2005ht} and three-loop $D^6 \cR^4$ \cite{Gomez:2013sla} couplings in type II string theory compactified on $T^d$,
which are proportional to the renormalized modular integral of  the lattice partition function  $\Gamma_{d,d,h}$ for $h=2$ and $h=3$,  respectively. Using the techniques developed in the present paper, we establish that the
two-loop $D^4 \cR^4$ and the three-loop $D^6 \cR^4$ amplitudes are given by residues of the 
Langlands-Eisenstein series $\cE^{\star,SO(d,d)}_{\Lambda^2 V}(s')$ and 
$\cE^{\star,SO(d,d)}_{\Lambda^3 V}(s')$ at $s'=d/2$, respectively. Moreover,  in appendix  \S\ref{lap_gen3}, using similar techniques as in \cite{Pioline:2015nfa},  we show that the three-loop $D^6 \cR^4$ amplitude  satisfies a Laplace-type differential equation as a function of the moduli of the torus $T^d$, with anomalous source terms originating from boundaries of the moduli space; the coefficients of these anomalous terms agree perfectly with those predicted 
from S-duality \cite{Pioline:2015yea}. 

An important challenge is to extend our techniques to Siegel modular forms with singularities inside the Siegel upper half-plane. The  special case of the modular integral of the genus-two Kawazumi-Zhang invariant $\varphi_{\rm KZ}(\Omega)$, relevant for two-loop $D^6 \cR^4$ amplitudes in type II string theory on tori, was considered in \cite{Pioline:2015nfa}, leading to a novel construction
of $\varphi_{\rm KZ}(\Omega)$ \cite{Pioline:2015qha}. It would be interesting
to consider other examples with more severe singularities on  the separating degeneration locus,
such as $D^2 H^4$ amplitudes in type II string theory compactified on $K_3$ \cite{Lin:2015dsa}.

\medskip

\noindent {\bf Acknowledgements}.
B.P. is grateful to Rodolfo Russo for collaboration on the related work \cite{Pioline:2015nfa}  which paved the way for this project. I.F. would like to thank the ICTP Trieste for its kind hospitality during the final stages of this work.

\section{Degree one\label{sec_one}}

As a warm-up for the higher degree cases discussed in this work, let us recover the main results of \cite{MR656029}
using the method outlined above. Let $F(\tau)$ be an automorphic function of $SL(2,\IZ)$ with polynomial growth\footnote{A more general growth condition $\varphi(\tau_2)=\sum_{i=1}^{\ell} c_i\, \tau_2^{\sigma_i} (\log \tau_2)^{n_i}/n_i!$ was considered in \cite{MR656029}. For simplicity we shall assume $n_i=0$. The case $n_i>0$ can be dealt with by raising the $i$-th factor in \eqref{Diamond1}
to the power $n_i$. We could also allow the coefficients $c_i$ to be arbitrary periodic functions of $\tau_1$. The subsequent analysis is unchanged provided $c_i$ is understood to represent $\int_{-\frac12}^{\frac12} c_i(\tau_1) \de\tau_1$.} 
at the cusp $\tau=\I\infty$,
\be
\label{growthone}
F(\tau) = \varphi(\tau_2) + \cO(\tau_2^{-N})\quad \forall N>0\ , \qquad \varphi(\tau_2)=\sum_{i=1}^{\ell} c_i\, \tau_2^{\sigma_i} \ ,
\ee
where $c_i\in \IC, \sigma_i\in \IC$. We define the regularized Rankin-Selberg transform of $F$ by
\be
\cR_1^{\star,\Lambda}(F,s) \equiv  \int_{\cF_1^{\Lambda}} \de\mu_1\, \cE_1^\star(s,\tau)\,  F(\tau) \,,
\ee
where 
\be
\cE^{\star}_1(s,\tau)  \equiv \frac12\zetastar(2s)\,\sum_{(c,d)\in \IZ^2,\atop (c,d)=1}
\frac{\tau_2^s}{|c\tau+d|^{2s}} \,,
\ee
is the completed non-holomorphic Eisenstein series of weight 0, 
$\cF_1^{\Lambda}=\{ \tau\in \cH_1, |\tau|>1, -\frac12 \leq \tau_1 < \frac12, \tau_2<\Lambda\}$ is the `truncated fundamental domain', and $\de\mu_1=\de\tau_1\de\tau_2/\tau_2^2$ (normalized such that the volume of the fundamental domain is $\cV_1=\lim_{\Lambda\to\infty} \int_{\cF_1^{\Lambda}} \de\mu_1=2\zetastar(2)=\frac{\pi}{3}$). 

The non-decaying part $\varphi$ is annihilated by the operator
\be
\label{Diamond1}
\Diamond= \prod_{i=1}^{\ell} \left( \Delta - \sigma_i(\sigma_i-1) \right)\ .
\ee
Using the Chowla-Selberg formula 
\be
\label{kron1}
\cE^{\star}_1(s,\tau) 
=\zetastar(2s) \,\tau_2^s + \zetastar(2s-1)\, \tau_2^{1-s} + \dots\ ,
\ee
where the dots denote exponentially suppressed terms as $\tau_2\to\infty$, 
and integrating by parts $\ell$ times,
one finds that the regularized Rankin-Selberg transform of $\Diamond F$  is given by 
\be
\label{R1parts}
\begin{split}
\cR_1^{\star,\Lambda}(\Diamond F,s)
=& 
\prod_{i=1}^{\ell} \left[ s(s-1)-\sigma_i(\sigma_i-1)\right]\,  
\cR_1^{\star,\Lambda}(F,s)\\
&+\sum_{j=1}^{\ell} c_j \left[ \zetastar(2s)\, \prod_{i=1\dots \ell} (s-\sigma_i)\, 
\prod_{i=1\dots \ell \atop i\neq j} (s-1+\sigma_i)\, \Lambda^{\sigma_j+s-1} \right.\\
& \left. 
- \zetastar(2s-1)\, \prod_{i=1\dots \ell \atop i\neq j} (s-\sigma_i)\, 
\prod_{i=1\dots \ell } (s-1+\sigma_i)\, \Lambda^{\sigma_j-s} 
 \right]  + \dots
\end{split}
\ee
where the dots denote exponentially suppressed terms as $\Lambda\to\infty$. 
Thus, reorganizing terms and assuming that $s$ does not coincide with any of the $\sigma_i$ or $1-\sigma_i$,
\be
\label{R1parts2}
\begin{split}
\cR_1^{\star,\Lambda}( F,s) =& \frac{\cR_1^{\star,\Lambda}(\Diamond F,s)}{D_1(s)}\,  
&+\sum_{j=1}^{\ell} c_j \left( \frac{\zetastar(2s)\, \Lambda^{\sigma_j+s-1}}{\sigma_j+s-1}
 + \frac{\zetastar(2s-1)\,\Lambda^{\sigma_j-s}}{\sigma_j-s} 
 \right)  + \dots
\end{split}
\ee
where
\be
D_1(s) = \prod_{i=1}^{\ell} \left[ s(s-1)-\sigma_i(\sigma_i-1)\right]\ .
\ee
On the other hand, since $\Diamond F$ is decaying faster than any power of $\tau_2$, the standard Rankin-Selberg method shows that
$\cR_1^{\star,\Lambda,s}(\Diamond F)$ has a finite limit at $\Lambda\to\infty$, given by 
\be
\label{R1parts3}
\cR_1^{\star}(\Diamond F,s) =\zetastar(2s)\, \int_0^{\infty} \de\tau_2\, \tau_2^{s-2} \, \int_{-1/2}^{1/2}\de\tau_1\,
\Diamond F\ ,
\ee
and that $\cR_1^{\star,\Lambda}(\Diamond F)$ has a meromorphic continuation in the $s$ plane, invariant under
$s\mapsto 1-s$, with only simple poles at $s=0$ and $s=1$, due to the poles of $\cE^\star_1(s)$. 
Using the fact that $\Diamond\varphi=0$ and integrating by parts, we have
\be
\cR_1^{\star}(\Diamond F,s) =\zetastar(2s)\, D_1(s)\, 
\int_0^{\infty} \de\tau_2\, \tau_2^{s-2} \, (F^{(0)}-\varphi) \,,
\ee
where $F^{(0)}(\tau)= \int_{-1/2}^{1/2}\de\tau_1 F$. Thus, 
if we define the renormalized integral of $\cE_1^\star(s)$ times $F$ by subtracting the divergent
terms in \eqref{R1parts2},
\be
\label{RNE1loop}
\begin{split}
\cR_1^{\star}( F,s) \equiv & \RN \int_{\cF_1} \de \mu_1\, \cE_1^\star(s)\, F \\
& \equiv 
\lim_{\Lambda\to\infty} \left[ \cR_1^{\star,\Lambda}( F) -\sum_{j=1}^{\ell} c_j \left( \frac{\zetastar(2s)\, \Lambda^{\sigma_j+s-1}}{\sigma_j+s-1}
 + \frac{\zetastar(2s-1)\,\Lambda^{\sigma_j-s}}{\sigma_j-s} \right)\right]\ ,
 \end{split}
\ee
then it follows from \eqref{R1parts2} and \eqref{R1parts3} that
\be
\label{eqboxed1}
\boxed{\cR_1^{\star}( F,s) =   
\zetastar(2s)\, \int_0^{\infty} \de\tau_2\, \tau_2^{s-2} \, (F^{(0)}-\varphi) =
\frac{\cR_1^{\star}(\Diamond F,s)}{D_1(s)}\ .}
\ee
Thus the renormalized integral $\cR_1^{\star}( F,s)$ is equal to the Mellin transform of the subtracted 
constant term $F^{(0)}-\varphi$, and has a meromorphic continuation to the $s$ plane, invariant under
$s\mapsto 1-s$, with  only simple poles at $s\in \{0,1,\sigma_i, 1-\sigma_i, i=1\dots \ell\}$ 
(assuming for now that no $\sigma_i$ is equal to 1). For $\sigma\notin\{0,1\}$, the residue 
at $s=\sigma$ originates from the subtraction in \eqref{RNE1loop},
\be
\Res_{s=\sigma}\, \cR_1^{\star}( F,s) =
 \sum_{i=1 \atop \sigma=\sigma_i}^{\ell} \zetastar(2\sigma_i-1)\, c_i\
-  \sum_{i=1\atop  \sigma=1-\sigma_i}^{\ell} \zetastar(2\sigma_i-1)\, c_i\ .
\ee  
Since the
residue of $\cE^\star_1(s)$ at $s=1$ is a constant $r_1=\Res_{s=1}\zetastar(2s-1)=\tfrac12$, it is natural to define the renormalized integral
of $F$ as twice the residue of $\cR_1^{\star}( F,s)$ at that point,
\be
\label{RN1}
\RN \int_{\cF_1} \de \mu_1\,F = \frac{1}{r_1} \Res_{s=1}\cR_1^{\star}( F,s)
=  \lim_{\Lambda\to\infty} \left[ \int_{\cF_1^{\Lambda}} \de \mu_1 \, F- \sum_{j=1}^{\ell} \frac{c_j \,\Lambda^{\sigma_j-1}}{\sigma_j-1} \right]\ .
\ee
If, however, one of the $\sigma_i$'s coincides with $1$, then $\cR_1^{\star}( F,s)$ has a double pole
at $s=0$ and $s=1$, with coefficient $\tfrac12\sum_{\sigma_j=1} c_j $. We may then define the renormalized integral as
\be
\RN \int_{\cF_1} \de \mu_1\,F =  \lim_{\Lambda\to\infty} \left[ \int_{\cF_1^{\Lambda}} \de \mu_1 \, F
- \sum_{j=1\dots \ell\atop \sigma_j\neq 1} \frac{c_j \,\Lambda^{\sigma_j-1}}{\sigma_j-1}
- \sum_{j=1\dots \ell\atop \sigma_j=1} c_j \log\Lambda \right]\ ,
\ee
in which case it differs from twice the residue of $\cR_1^{\star}( F,s)$ at $s=1$ by an additive constant,
\be
\RN \int_{\cF_1} \de \mu_1\,F = 2\,\Res_{s=1} \cR_1^{\star}( F,s)
+ \sum_{j\atop \sigma_j=1} c_j\, \log(4\pi e^{-\gamma}) + \frac{\pi}{3}
\sum_{j\atop \sigma_j=0} c_j\ .
\ee

As a prime example of a one-loop modular integral relevant for string theory, let us consider the case
$F = \Gamma_{d,d,1}$, where $\Gamma_{d,d,h}$ is the  Siegel-Narain partition function for the even self-dual lattice of signature $(d,d)$, which we define here for arbitrary genus $h$ ($\Omega=\tau$ for $h=1$):
\be
\label{defGammaddh}
\Gamma_{d,d,h}(g,B;\Omega)= |\Omega_2|^{d/2}\, \sum_{(m_i^I, n^{i,I})\in\IZ^{2dh}}\, 
e^{-\pi\cL^{IJ}\Omega_{2,IJ}+2\pi \I m_i^I n^{i,J}\Omega_{1,IJ}}\ ,
\ee
where 
\be
\label{defcL}
\cL^{IJ}=(m_i^I+B_{ik} n^{I,k}) g^{ij} (m_j^J+B_{jl} n^{J,l}) + n^{i,I} g_{ij} n^{j,J}\ .
\ee
At genus one,
$\cL^{11}\equiv\cM^2$ 
is the squared mass of a closed string with momentum $m_i^1$ and winding number $n^{i,1}$ along a $d$-dimensional torus with metric $g_{ij}$ and Kalb-Ramond field $B_{ij}$. The positive definite matrix $g_{ij}$ and antisymmetric matrix $B_{ij}$ can be viewed as coordinates on the Grassmannian $G_{d,d}=O(d,d)/[O(d)\times O(d)]$, which parametrizes even self-dual lattices of signature $(d,d)$. In this case, $\varphi(\tau_2)=\tau_2^{d/2}$, and the renormalized integral  \eqref{eqboxed1} is given by
\be
\label{RN1Gdd}
\cR_1^{\star}( \Gamma_{d,d,1},s) 
= \frac{\zetastar(2s)\, \Gamma(s+\tfrac{d}{2}-1)}{\pi^{s+\tfrac{d}{2}-1}}\,
\sum_{(m_i, n^i)\in\IZ^{2d} \atop m_i n^i=0, (m_i,n^i)\neq 0} \cM^{-2s-d+2} \, .
\ee
This is recognized as the  Langlands-Eisenstein series of $SO(d,d,\IZ)$ attached to
the vector representation,
\be
\cR_1^{\star}( F,s) = 2\,\cE^{\star,SO(d,d)}_V(s+\tfrac{d}{2}-1)\ .
\ee
In particular, it follows from the above that  $\cE^{\star,SO(d,d)}_{V}(s)$ has a meromorphic continuation 
to the $s$ plane, invariant under $s\mapsto d-1-s$, with simple poles (for $d\neq 2)$ at 
$s=0, \frac{d}{2}-1, \frac{d}{2}, d-1$
and residues 
\bea
\label{resE1}
\Res_{s=d-1}  \cE^{\star,SO(d,d)}_{V}(s) &=&  \frac12\zetastar(d-1) \,,\\
\Res_{s=\tfrac{d}{2}}  \cE^{\star,SO(d,d)}_{V}(s) &=&  
\frac14 \RN \int_{\cF_1}\de\mu_1\, 
\Gamma_{d,d,1}\,,
\eea
where we define the renormalized integral $\RN \int_{\cF_1}\de\mu_1\, 
\Gamma_{d,d,1}$, for all values of $d$, as 
\be
\label{defRN1Gdd}
 \RN \int_{\cF_1}\de\mu_1\, 
\Gamma_{d,d,1} = \lim_{\Lambda\to\infty} \left[
\int_{\cF_1^{\Lambda} }\de\mu_1\, 
\Gamma_{d,d,1} -
\left( \frac{\Lambda^{\tfrac{d}{2}-1}}{\tfrac{d}{2}-1} \Theta(d-2) + \delta_{d,2}\, \log\Lambda\right)\right] \,.
\ee
Here, $\Theta(x)=1$ if $x>0$ and $0$ otherwise. This analytic structure is consistent with the one derived from the Langlands constant term formula (see e.g. Appendix A in \cite{Pioline:2015yea}).

For $d=2$, $\cE^{\star,SO(2,2)}_{V}(s)$ has a double
pole at $s=1$, 
\be
\cE^{\star,SO(2,2)}_{V}(s) =
\frac{1}{4(s-1)^2} +  \frac1{4(s-1)} \left[ 2\gamma  -\log\left( 16\pi^2 T_2 U_2 
|\eta(T)\eta(U)|^4 \right)\right]\ ,
\ee
where we used $\cE^{\star,SO(2,2)}_{V}(s) = \cE^\star_1(s,T) \, \cE^\star_1(s,U)$, \cite{Angelantonj:2011br} and 
the Kronecker limit formula
\be
\label{Estar11}
\cE^\star_1(s,\tau) = \frac{1}{2(s-1)}
-\frac12 \log\left[ 4\pi\tau_2 |\eta(\tau)|^4 e^{-\gamma} \right] + \cO(s-1) \ .
\ee
The residue at the pole differs by an  additive constant 
from (1/4 times) the renormalized integral defined by \eqref{defRN1Gdd},
\be
\RN \int_{\cF_1}\de\mu_1\, \Gamma_{2,2,1} =  -\log\left( 4\pi e^{-\gamma} T_2 U_2 
|\eta(T)\eta(U)|^4 \right)\ .
\ee
The latter also differs by an additive constant from the integral computed  in \cite{Dixon:1990pc} using a different renormalization prescription. 

For $d=1$, using $\cM^2=\frac{m^2}{R^2}+n^2 R^2$, where $R$ is the radius of the circle $T^1$, one finds
\be
\cE^{\star,SO(1,1)}_{V}(s) = \zetastar(2s)\, \zetastar(2s+1)\, (R^{2s}+R^{-2s})\ .
\ee

For $d=0$, the Rankin-Selberg transform \eqref{RN1Gdd} vanishes, but the renormalized integral 
$\RN \int_{\cF_1}\de\mu_1$ is still non-zero, and equal to $\cV_1=2\zetastar(2)$.

\section{Degree two\label{sec_two}}

In this section we develop the Rankin-Selberg method for Siegel modular functions of degree two with at most polynomial growth at the cusp, following the strategy outlined in the introduction. Our aim is to define the 
renormalized integral
\be
\label{DefRS2}
\cR_2^\star(F,s) = \RN \int_{\cF_2} \de\mu_2 \, \cE_2^\star(s,\Omega)\, F(\Omega)\ ,
\ee
and relate it to the generalized Mellin transform of the zero-th Fourier coefficient $F^{(0)}(\Omega_2)$, defined with a suitable subtraction. 

\subsection{Regularizing the divergences}

\begin{figure}
\begin{center}
\begin{picture}(0,0)%
\includegraphics{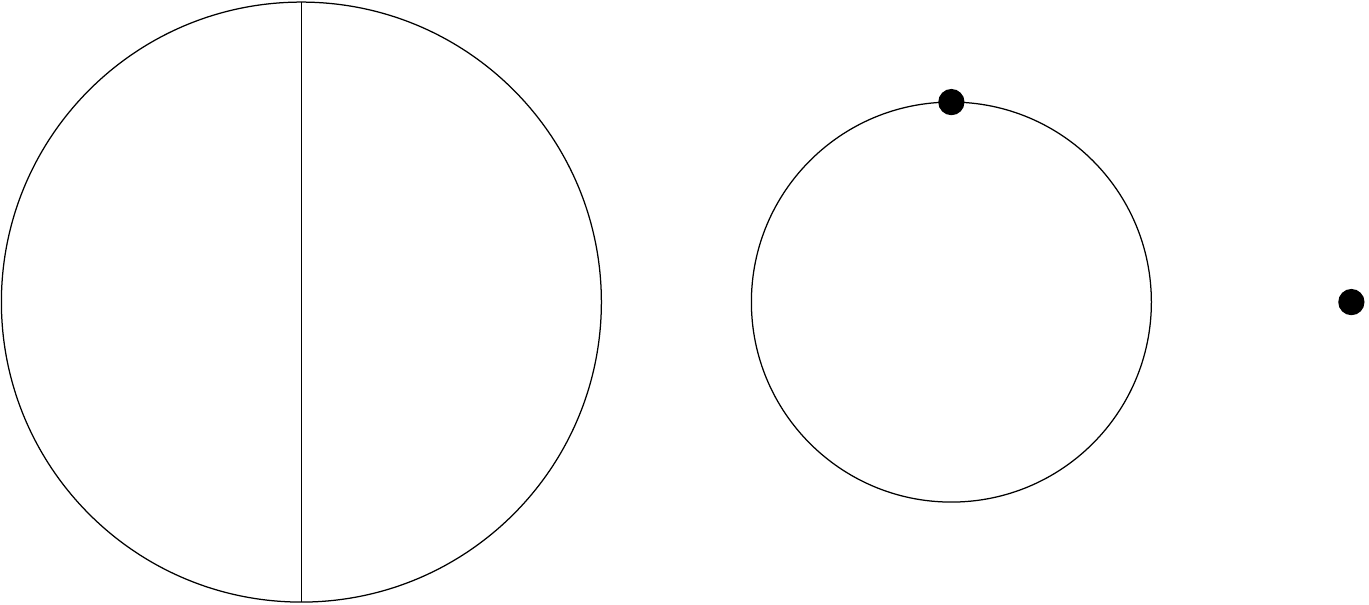}%
\end{picture}%
\setlength{\unitlength}{3158sp}%
\begingroup\makeatletter\ifx\SetFigFont\undefined%
\gdef\SetFigFont#1#2#3#4#5{%
  \reset@font\fontsize{#1}{#2pt}%
  \fontfamily{#3}\fontseries{#4}\fontshape{#5}%
  \selectfont}%
\fi\endgroup%
\begin{picture}(8201,3624)(893,-4573)
\put(8956,-2461){\makebox(0,0)[lb]{\smash{{\SetFigFont{10}{12.0}{\familydefault}{\mddefault}{\updefault}{\color[rgb]{0,0,0}2}%
}}}}
\put(1051,-2761){\makebox(0,0)[lb]{\smash{{\SetFigFont{10}{12.0}{\familydefault}{\mddefault}{\updefault}{\color[rgb]{0,0,0}$L_1$}%
}}}}
\put(2326,-2761){\makebox(0,0)[lb]{\smash{{\SetFigFont{10}{12.0}{\familydefault}{\mddefault}{\updefault}{\color[rgb]{0,0,0}$L_2$}%
}}}}
\put(4126,-2761){\makebox(0,0)[lb]{\smash{{\SetFigFont{10}{12.0}{\familydefault}{\mddefault}{\updefault}{\color[rgb]{0,0,0}$L_3$}%
}}}}
\put(7426,-2461){\makebox(0,0)[lb]{\smash{{\SetFigFont{10}{12.0}{\familydefault}{\mddefault}{\updefault}{\color[rgb]{0,0,0}$t$}%
}}}}
\put(8926,-3211){\makebox(0,0)[lb]{\smash{{\SetFigFont{10}{12.0}{\familydefault}{\mddefault}{\updefault}{\color[rgb]{0,0,0}$\Omega$}%
}}}}
\put(6541,-1261){\makebox(0,0)[lb]{\smash{{\SetFigFont{10}{12.0}{\familydefault}{\mddefault}{\updefault}{\color[rgb]{0,0,0}1}%
}}}}
\put(6556,-1936){\makebox(0,0)[lb]{\smash{{\SetFigFont{10}{12.0}{\familydefault}{\mddefault}{\updefault}{\color[rgb]{0,0,0}$\rho$}%
}}}}
\end{picture}
\end{center}
\caption{Two-loop sunset diagram, and its one-loop and two-loop counterterms, corresponding to regions II, I, 0 defined in \eqref{defF2split}. \label{fig_gen2}}
\end{figure}
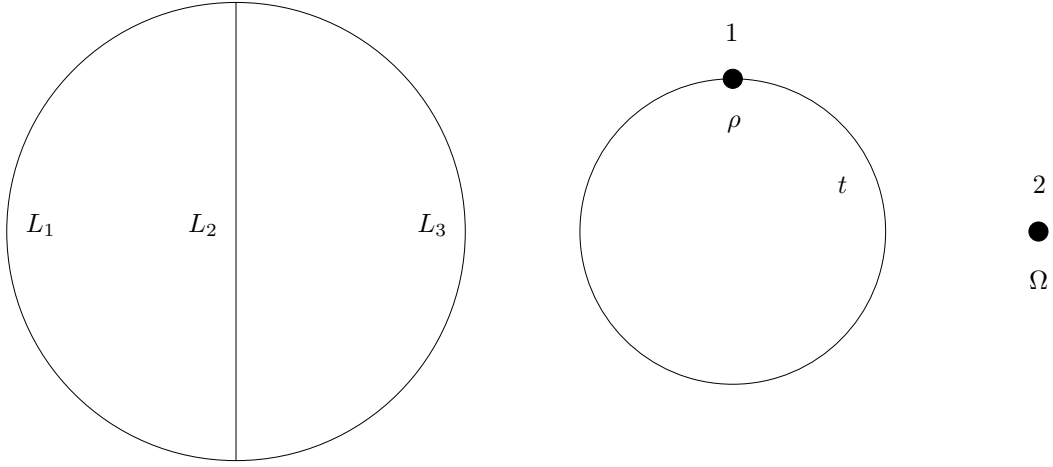

Our first task is to understand the possible sources of divergence in the integral \eqref{DefRS2}. For this we choose
the standard  fundamental domain $\cF_2$ from \cite{zbMATH03144647},
\be
\label{defF2}
\begin{split}
(1) & \qquad -\frac12 < \Re(\Omega_{11}), \Re(\Omega_{12}), \Re(\Omega_{22}) \leq \frac12 \\
(2) & \qquad 0< 2 \Im(\Omega_{12}) \leq \Im(\Omega_{11}) \leq \Im(\Omega_{22})     \\
(3) & \qquad  |C\Omega+D|>1 \ \mbox{for all}\  \begin{pmatrix} A & B \\ C& D \end{pmatrix} \in Sp(4,\IZ)
\end{split}
\ee
Note that (3) states that $|\Omega_2|$ attains the maximal possible value in the orbit of $Sp(4,\IZ)$, while (2) amounts to the requirement that $\Omega_2$ lies in the Minkowski fundamental domain for the action of $GL(2,\IZ)$ on positive quadratic forms. This domain is conveniently parametrized by three ordered positive real numbers $0<L_2<L_1<L_3$,
\be
\label{coorsp2L}
\Omega_2= \begin{pmatrix} L_1+L_2 & -L_2 \\
-L_2 & L_2 + L_3 
\end{pmatrix}\ .
\ee
The variables $L_1,L_2,L_3$ can be interpreted as Schwinger time parameters for the three edges of the two-loop `sunset' Feynman diagram which describes the maximal non-separating degeneration of a genus two Riemann surface (see Figure \ref{fig_gen2}). The integration measure is normalized as in \cite{Pioline:2014bra}, 
\be
\label{defmu2}
\de\mu_2(\Omega) = \frac{\prod_{I\leq J} \de\, \Re(\Omega_{IJ})\,  \de\, \Im(\Omega_{IJ})}{|\Omega_2|^3} = \frac{\de L_1 \de L_2 \de L_3}{(L_1 L_2 + L_2 L_3 + L_3 L_1)^3}\ \prod_{I\leq J} \de\, \Re(\Omega_{IJ})\ ,
\ee
so that the volume of the fundamental domain $\cF_2$ is $\cV_2=2\zetastar(2)\zetastar(4)=\frac{\pi^3}{270}$.

The first source of divergences is the region I where $\Im(\Omega_{22})$ is scaled to infinity, keeping
the other entries in $\Omega$ fixed. In this region, it is convenient to parametrize
\be
\label{coorsp2sp1}
\Omega=\begin{pmatrix} \rho & \rho u_2 - u_1 \\
\rho u_2 - u_1 & t_1+\I (t+ \rho_2 u_2^2)
\end{pmatrix}\ ,
\ee
where $t\in\IR^+, \rho\in \cH_1, (u_1,u_2,t_1)\in \IR^3$, so that the integration measure becomes
\be
\de\mu_2(\Omega) = \frac{\de t}{t^3} \de\mu_1(\rho)\, \de u_1 \, \de u_2 \, \de t_1 \ ,
\ee
where $\de\mu_1(\rho)=\de\rho_1\de\rho_2/\rho_2^2$. The region of interest is $t\to \infty$, keeping $\rho,u_1,u_2,t_1$ finite. Since the stabilizer of the cusp $t=\infty$ is $Sp(2,\IZ)\ltimes \IZ^2 \ltimes \IZ$, the fundamental  domain $\cF_2$ simplifies in this limit to $\IR^+ \times \cF_1 \times [-\tfrac12,\tfrac12]^2/\IZ_2 \times [-\tfrac12,\tfrac12]$, where the center $\IZ_2$ of $Sp(2,\IZ)=SL(2,\IZ)$ acts by flipping the sign of $(u_1,u_2)$.  In string theory, this region is responsible for one-loop infrared subdivergences, described by a one-loop diagram in supergravity, with an insertion of a one-loop counterterm.
The parameter $t$ is interpreted as the Schwinger time parameter for the propagation of massless supergravity states around the loop, while $\rho$ is the complexified Schwinger parameter for the
counterterm.

The second source of divergences is the region II where the whole matrix $\Omega_2$ scales to infinity. In the language of string theory, this region is responsible for primitive two-loop infrared divergences. In this region, it is convenient to choose a different parametrization for the imaginary part of $\Omega$,
\be
\Omega_2=\frac{1}{V \tau_2} \begin{pmatrix} 1 & \tau_1 \\
\tau_1 & |\tau|^2 
\end{pmatrix}\ ,
\ee
where $V\in \IR^+$ and $\tau\in \cH_1$, so that the integration measure in coordinates $V,\tau,\Omega_1$ becomes
\be
 \de\mu_2(\Omega) = 2 V^2 \de V \frac{\de\tau_1\de\tau_2}{\tau_2^2}\, \prod_{I<J} 
 \de\Re(\Omega_{IJ})\ .
\ee
 The region of interest is then $V\to 0$ keeping $\tau$ and $\Omega_1$
fixed. Since the stabilizer of the cusp $V=0$ is $\Gamma_{2,\infty} = PGL(2,\IZ)\ltimes \IZ^3$, the fundamental domain $\cF_2$ simplifies in this limit to $\IR^+ \times (\cF_1/\IZ_2) \times [-\tfrac12,\tfrac12]^3$, where the involution 
$\IZ_2$ (corresponding to the element ${\rm diag}(1,-1)$ in $PGL(2,\IZ)$) acts by $\tau\mapsto -\bar\tau$. In string theory, this region is responsible for primitive two-loop divergences. Indeed, as indicated in \eqref{coorsp2L}, the fundamental domain $\IR^+\times \cF_1/\IZ_2$ is isomorphic to the space $(\IR^+)^3/\sigma_3$ parametrized by the ordered Schwinger parameters
$0<L_2<L_1<L_3$, so the  divergence can be cast into the form of a two-loop 
amplitude in  supergravity. The fact that two-loop supergravity amplitudes have a hidden 
modular invariance was first noticed in \cite{Green:1999pu}, and it becomes manifest when these amplitudes are obtained as field theory limits of string amplitudes. 

It is worth noting that the locus $\tau_1=0$
inside $\cF_1/\IZ_2$, fixed under the involution $\tau\mapsto -\bar\tau$, corresponds to
the intersection of the non-separating and separating degeneration loci $t=\infty$ and 
$\rho u_2-u_1=\infty$. Since we assume
that the integrand is regular in the separating degeneration limit, there is no divergence originating
from this region.

In order to regularize the integral \eqref{DefRS2}, we shall define, for $\Lambda$ sufficiently large,
\be
\label{DefRS2reg}
\cR_2^\star(F,s) = \int_{\cF_2^{\Lambda}} \de\mu_2 \, \cE_2^\star(s,\Omega)\, F(\Omega)\ ,
\ee
where
\be
\cF_2^{\Lambda} = \cF_2 \cap \{ t<\Lambda \} \,,
\ee
is the `truncated fundamental domain'. The conditions \eqref{defF2} require that
$0\leq u_2\leq \tfrac12$ and $\rho_2<t/(1-u_2^2)\leq \tfrac43 t$ in $\cF_2$, so the cut-off $t<\Lambda$
ensures that the domain $\cF_2^{\Lambda}$ is compact. Thus, for any Siegel modular function
$F(\Omega)$, smooth in $\cH_2$, $\cR_2^\star(F,s)$ inherits the analytic structure of $\cE_2^\star(s,\Omega)$ as a function of $s$. Our goal is to determine its dependence on the cut-off $\Lambda$ as $\Lambda\to\infty$, up to exponentially suppressed terms in $\Lambda$, and define the renormalized integral \eqref{DefRS2} by subtracting these contributions and taking the limit $\Lambda\to\infty$.

A significant complication in extracting the large $\Lambda$ behavior of \eqref{DefRS2reg} is that
the regions I and II overlap, corresponding to overlapping divergences in supergravity. 
In order to disentangle their contributions, it is useful to introduce an 
auxiliary cut-off $\Lambda_1$ such that  $1\ll \Lambda_1\ll \Lambda$, and split $\cF_2^\Lambda$ into three domains \cite{Pioline:2015nfa}: 
\be
\label{defF2split}
\begin{split}
\cF_2^{0}=& \cF_2 \cap \{ \rho_2 < t+u_2^2 \rho_2 < \Lambda_1\ , \quad  t<\Lambda\}  \, ,\\	
\cF_2^{I}=& \cF_2 \cap \{ \rho_2 < \Lambda_1 < t+u_2^2 \rho_2\ , \quad t < \Lambda\} \, ,\\	
\cF_2^{II}=& \cF_2 \cap \{  \Lambda_1 < \rho_2 <  t+u_2^2 \rho_2\ , \quad t < \Lambda \} \,.	
\end{split}
\ee
The integral over $\cF_2^{0}$ gives a finite result, independent of $\Lambda$.  The integrals
over $\cF_2^{I}$ and $\cF_2^{II}$ will have power-like dependence on $\Lambda$ and $\Lambda_1$, but mixed terms depending on both $\Lambda$ and $\Lambda_1$ will cancel
in the sum, since the union of the three regions is independent of $\Lambda_1$.

\subsection{Renormalizing the integral}

Our second task is to understand the behavior of the Eisenstein series 
$\cE^\star_2(s,\Omega)$ in regions I and II. Recall that  $\cE^\star_2(s,\Omega)$ is defined for 
$s>\tfrac32$ by the sum over images in \eqref{defEstarh}, and has a meromorphic continuation
to the $s$-plane, invariant under $s\mapsto \tfrac32-s$, with simple poles at $s=0, \tfrac12, 1, \tfrac32$.
The residue at $s=\tfrac32$ (or minus the residue at $s=0$) is a constant $r_2=\tfrac12 \zetastar(3)$,
while the residue at $s=1$ is a non-trivial real-analytic Siegel modular form. The behavior of $\cE^\star_2(s,\Omega)$ in the regions I and II is given by the Langlands constant term formula, 
\be
\label{decsp4sp2}
\cE^\star_2(s,\Omega) = t^s\, \zetastar(4s-2)\, \cE^\star_1(s,\rho) + 
t^{\tfrac32-s}\, \zetastar(4s-3)\, \cE^\star_1(s-\tfrac12,\rho) + \cO(t^{-N})\ ,\qquad \forall N>0 \,,
\ee
in the limit $t\to\infty$, and 
\be
\label{decsp4gl2}
\begin{split}
\cE_2^\star(s,\Omega) = & \zetastar(2s)\, \zetastar(4s-2) \, V^{-2s}
+ \zetastar(2s-2) \zetastar(4s-3)\, V^{2s-3}\\
& +\zetastar(2s-1) \, V^{-1}\,  \cE_1^\star(2s-1,\tau) + \cO(V^N) \ , \qquad \forall N>0\,,
\end{split}
\ee
in the limit $V\to 0$. Using \eqref{kron1} it is easily checked that the two expansions agree in their common domain of validity. 

In order to control the divergences of the integrand, we shall require that the function $F$ behaves in the limit $t\to\infty$ as 
\be
\label{Fgrowth21}
F(\Omega) = \varphi+ \cO(t^{-N})\qquad \forall N>0\ , \quad
\varphi=\sum_{i=1}^{\ell} t^{\sigma_i}\, \varphi_i(\rho)  \, ,
\ee
where  $\{\sigma_i\}$ is a set of distinct complex exponents and $\varphi_i(\rho)$ are real-analytic modular forms of weight zero. More generally, we could allow $\varphi_i$ to be a real analytic Jacobi form
in  $\rho, u_1,u_2,t_1$, of zero weight and index. In the analysis below, $\varphi_i(\rho)$ should then be understood as the average
$\int_{[-1/2,1/2]^3} \varphi_i(\rho, u_1,u_2,t_1)\, \de u_1 \de u_2 \de t_1$.

Similarly, in the limit $V\to 0$, we require that 
 $F(\Omega)$ grows as 
\be
\label{Fgrowth22}
F(\Omega) = \tilde\varphi+ \cO(V^N)\qquad \forall N>0\ , \quad
\tilde\varphi=\sum_{j=1}^{\tilde \ell} \, V^{\alpha_j} \tilde\varphi_j(\tau)  \,,
\ee
where $\alpha_j$ is a set of distinct complex exponents, and $\tilde\varphi_j(\tau)$ are
Maass forms of weight 0 under $GL(2,\IZ)$. We assume that $\tilde\varphi_j(\tau)$
is smooth at $\tau_1=0$, corresponding to the intersection of the non-separating and
separating degeneration loci. 
More generally, we could allow $\tilde\varphi_j$ 
to be functions of $\tau,\Omega_1$, invariant under $\Gamma_{2,\infty}$. In this case, in the analysis below, 
one should again interpret $\tilde\varphi_j(\tau)$ as the average
$\int_{[-1/2,1/2]^3} \tilde\varphi_j(\tau,\Omega_1) \de\Omega_1$.

Using $V=1/\sqrt{t\rho_2}, \tau_2=\sqrt{t/\rho_2}$, the compatibility of \eqref{Fgrowth21} and \eqref{Fgrowth22} requires
that 
\be
\label{Fgrowthcompatible}
\varphi_i(\rho) \stackrel{\rho_2\to\infty}{\sim} \sum_j \,c_{i,j}\, \rho_2^{\eta_{i,j}}  \ ,\quad
\tilde\varphi_j(\tau)\stackrel{\tau_2\to\infty}{\sim} \sum_i\, \tilde c_{j,i} \,\tau_2^{\beta_{j,i}}\,,
\ee
with $c_{i,j}=\tilde c_{j',i'}$ whenever $\sigma_i=\tfrac12(\beta_{j',i'}-\alpha_{j'}), \eta_{i,j}=-\tfrac12(\beta_{j',i'}+\alpha_{j'})$.

Under these assumptions, in region I we can approximate $F$ by \eqref{Fgrowth21} and 
$\cE_2^\star(s,\Omega)$ by \eqref{decsp4sp2}, so that, after integrating over 
$t$ from its (irrelevant) lower bound $t(\rho,u_1,u_2,t_1)$ to $t=\Lambda$, and then over
 $(u_1,u_2,t_1)\in [-\frac12,\tfrac12]^2/\IZ_2\times [-\frac12,\tfrac12]$,
\be
\label{DefRS2reg1}
\begin{split}
\int_{\cF_2^{I}} \de\mu_2 \, \cE_2^\star(s,\Omega)\, F(\Omega) =&
\frac12 \sum_{i=1}^{\ell} \frac{ \zetastar(4s-2)\,  \Lambda^{s+\sigma_i-2}}{s+\sigma_i-2}
\int_{\cF_1^{\Lambda_1}} \de\mu_1 \,\varphi_i(\rho) \,  \cE^\star_1(s,\rho) 
\\
+& \frac12 \sum_{i=1}^{\ell}\frac{ \zetastar(4s-3)\,  \Lambda^{\sigma_i-\tfrac12-s}}{\sigma_i-\tfrac12-s}
\int_{\cF_1^{\Lambda_1}} \de\mu_1  \, \varphi_i(\rho)  \, \cE^\star_1(s-\tfrac12,\rho) 
\\
 +& \dots
\end{split}
\ee
where
 the dots denote exponentially suppressed terms in $\Lambda$. On the other hand, using \eqref{RN1},
 one can express the integral over $\cF_1^{\Lambda_1}$ in terms of the renormalized integral over
 $\cF_1$,
 \be
\label{DefRS2reg12}
\begin{split}
\int_{\cF_2^{I}} \de\mu_2& \, \cE_2^\star(s,\Omega)\, F(\Omega) \sim
\frac12 \sum_i\,  \frac{ \zetastar(4s-2)\,  \Lambda^{s+\sigma_i-2}}{s+\sigma_i-2}
\RN \int_{\cF_1} \de\mu_1(\rho) \,\varphi_i(\rho)\,  \cE^\star_1(s,\rho)
\\
+& \frac12 \sum_i\,  \frac{ \zetastar(4s-3)\,  \Lambda^{\sigma_i-\tfrac12-s}}{\sigma_i-\tfrac12-s}
\RN \int_{\cF_1} \de\mu_1(\rho)  \, \varphi_i(\rho)  \, \cE^\star_1(s-\tfrac12,\rho) 
\\
+& \frac12  \sum_{i,j} c_{i,j} 
\left( 
\tfrac{\zetastar(2s)\,\zetastar(4s-2)\,  \Lambda^{s+\sigma_i-2}\, \Lambda_1^{\eta_{i,j}+s-1}}{(s+\sigma_i-2)(\eta_{i,j}+s-1)}
+\tfrac{\zetastar(2s-1)\,\zetastar(4s-2)\,  \Lambda^{s+\sigma_i-2}\, \Lambda_1^{\eta_{i,j}-s}}{(s+\sigma_i-2)(\eta_{i,j}-s)}
 \right) 
\\
+ & \frac12  \sum_{i,j} c_{i,j} 
\left( 
\tfrac{\zetastar(2s-1)\,\zetastar(4s-3)\,  
\Lambda^{\sigma_i-\tfrac12-s}\, \Lambda_1^{\eta_{i,j}+s-\tfrac32}}
{(\sigma_i-\tfrac12 -s)(\eta_{i,j}+s-\tfrac32)}
+\tfrac{\zetastar(2s-2)\,\zetastar(4s-3)\,  
\Lambda^{\sigma_i-\tfrac12-s}\, \Lambda_1^{\eta_{i,j}-s+\tfrac12}}
{(\sigma_i-\tfrac12 -s)(\eta_{i,j}-s+\tfrac12)}
 \right)  \,.
\end{split}
\ee
In region II,  we can instead approximate $F$ by \eqref{Fgrowth22} and 
$\cE_2^\star(s,\Omega)$ by \eqref{decsp4gl2}. In terms of the variables $V,\tau_2$, 
the integration domain $\cF_2^{II}$ corresponds to
\be
\tau\in \cF_1^{\sqrt{{\Lambda}/{\Lambda_1}}}/\IZ_2\ ,\quad \frac{\tau_2}{\Lambda} < V < \frac{1}{\tau_2 \Lambda_1} \,.
\ee
After integrating over $V$, one finds
 \be
\label{DefRS2reg21}
\begin{split}
\int_{\cF_2^{II}} \de\mu_2& \, \cE_2^\star(s,\Omega)\, F(\Omega) \sim
\sum_{j} \int_{\cF_1^{\sqrt{\frac{\Lambda}{\Lambda_1}}}} \de\mu_1(\tau)
\\
& 
\left\{ \frac{\zetastar(2s)\,\zetastar(4s-2)}{\alpha_j-2s+3}
 \left[ (\tau_2\Lambda_1)^{2s-\alpha_j-3} - \left(\frac{\tau_2}{\Lambda}\right)^{\alpha_j+3-2s} \right]
 \, \tilde\varphi_j(\tau)
\right. \\
& +\frac{\zetastar(2s-2)\,\zetastar(4s-3)}{2s+\alpha_j}
 \left[ (\tau_2\Lambda_1)^{-2s-\alpha_j} - \left(\frac{\tau_2}{\Lambda}\right)^{\alpha_j+2s} \right]
 \, \tilde\varphi_j(\tau)
  \\
& +\frac{\zetastar(2s-1)}{\alpha_j+2}
 \left[ (\tau_2\Lambda_1)^{-2-\alpha_j} - \left(\frac{\tau_2}{\Lambda}\right)^{\alpha_j+2} \right]\,
 \cE^\star_1(2s-1,\tau)\, \tilde\varphi_j(\tau)
 \left. \right\} \,.
\end{split}
\ee
Again, we can replace  in this expression\footnote{We denote by $\tau_2^{\gamma}$ the modular invariant (but not smooth) function equal to $\tau_2^{\gamma}$ inside the standard fundamental domain $\cF$.}
\be
\int_{\cF_1^{\sqrt{\frac{\Lambda}{\Lambda_1}}}} \de\mu_1(\tau)\,\tau_2^{\gamma}\, \tilde\varphi_j(\tau)
\mapsto 
\RN \int_{\cF_1} \de\mu_1\,\tau_2^{\gamma}\, \tilde\varphi_j(\tau) 
+ \sum_{i} \frac{\tilde c_{j,i} \left(\frac{\Lambda}{\Lambda_1}\right)^{\tfrac12(\beta_{j,i}+\gamma-1)} }
{\beta_{j,i}+\gamma-1}\,,
\ee
and similarly for the integral $\int_{\cF_1^{\sqrt{\Lambda/\Lambda_1}}} \de\mu_1(\tau)\,\tau_2^{\gamma}\, \cE^\star_1(2s-1,\tau)\, \varphi_j(\tau)$. By construction, the mixed terms 
depending on both $\Lambda$ and $\Lambda_1$ cancel between \eqref{DefRS2reg12} and \eqref{DefRS2reg21}, thanks to the relation between $c_{i,j}$ and $\tilde c_{j,i}$ mentioned below
\eqref{Fgrowthcompatible}. Thus, the relevant contributions from region II are obtained by keeping
only the power of $\tau_2/\Lambda$ in the square brackets, and replacing the integral over 
$\cF_1^{\sqrt{{\Lambda}/{\Lambda_1}}}$ by the renormalized integral.

It follows that, up to exponentially suppressed terms in $\Lambda$, the regularized integral
\eqref{DefRS2reg} depends on $\Lambda$ as 
\be
\label{DefRS2regfull}
\begin{split}
\int_{\cF_2^{\Lambda}} \de\mu_2 & \, \cE_2^\star(s,\Omega)\, F(\Omega) \sim 
\frac12 \sum_i\, \frac{ \zetastar(4s-2)\,  \Lambda^{s+\sigma_i-2}}{s+\sigma_i-2}
\RN \int_{\cF_1} \de\mu_1(\rho) \,\varphi_i(\rho)\,  \cE^\star_1(s,\rho)
\\
+& \frac12 \sum_i\,  \frac{ \zetastar(4s-3)\,  \Lambda^{\sigma_i-\tfrac12-s}}{\sigma_i-\tfrac12-s}
\RN \int_{\cF_1} \de\mu_1(\rho)  \, \varphi_i(\rho)  \, \cE^\star_1(s-\tfrac12,\rho) 
\\
-& 
\sum_{j}  \frac{\zetastar(2s)\,\zetastar(4s-2)\, \Lambda^{2s-\alpha_j-3}}{\alpha_j-2s+3}
\RN \int_{\cF_1} \de\mu_1(\tau)\, \tau_2^{\alpha_j+3-2s}\, \tilde\varphi_j(\tau)
\\
- & \sum_{j} \frac{\zetastar(2s-2)\,\zetastar(4s-3)\,\Lambda^{-2s-\alpha_j}}{2s+\alpha_j}
\RN \int_{\cF_1} \de\mu_1(\tau)\, \tau_2^{\alpha_j+2s}
 \, \tilde\varphi_j(\tau)
  \\
- & \sum_{j} \frac{\zetastar(2s-1)\, \Lambda^{-2-\alpha_j} }{\alpha_j+2}
 \RN \int_{\cF_1} \de\mu_1(\tau)\, \tau_2^{\alpha_j+2}
 \cE^\star_1(2s-1,\tau)\, \tilde\varphi_j(\tau)\,.
\end{split}
\ee
We therefore define the renormalized integral \eqref{DefRS2} by subtracting these cut-off dependent terms before taking the limit $\Lambda\to\infty$,
\be
\label{DefRS2exp}
\begin{split}
\cR_2^\star(F,s) \equiv & \RN \int_{\cF_2} \de\mu_2 \, \cE_2^\star(s,\Omega)\, F(\Omega)
\equiv \lim_{\Lambda\to\infty} \left\{ \int_{\cF_2^{\Lambda}} \de\mu_2 \, \cE_2^\star(s,\Omega)\, F(\Omega) \right. 
\\
-&\frac12 \sum_i\, \frac{ \zetastar(4s-2)\,  \Lambda^{s+\sigma_i-2}}{s+\sigma_i-2}
\RN \int_{\cF_1} \de\mu_1(\rho) \,\varphi_i(\rho)\,  \cE^\star_1(s,\rho)
\\
-& \frac12 \sum_i\,  \frac{ \zetastar(4s-3)\,  \Lambda^{\sigma_i-\tfrac12-s}}{\sigma_i-\tfrac12-s}
\RN \int_{\cF_1} \de\mu_1(\rho)  \, \varphi_i(\rho)  \, \cE^\star_1(s-\tfrac12,\rho) 
\\
+& 
\sum_{j}  \frac{\zetastar(2s)\,\zetastar(4s-2)\, \Lambda^{2s-\alpha_j-3}}{\alpha_j-2s+3}
\RN \int_{\cF_1} \de\mu_1(\tau)\, \tau_2^{\alpha_j+3-2s}\, \tilde\varphi_j(\tau)
\\
+& \sum_{j} \frac{\zetastar(2s-2)\,\zetastar(4s-3)\,\Lambda^{-2s-\alpha_j}}{2s+\alpha_j}
\RN \int_{\cF_1} \de\mu_1(\tau)\, \tau_2^{\alpha_j+2s}
 \, \tilde\varphi_j(\tau)
  \\
+& \sum_{j} \frac{\zetastar(2s-1)\, \Lambda^{-2-\alpha_j}}{\alpha_j+2}
 \RN \int_{\cF_1} \de\mu_1(\tau)\, \tau_2^{\alpha_j+2}
 \cE^\star_1(2s-1,\tau)\, \tilde\varphi_j(\tau)
 \Biggr\}\,.
\end{split}
\ee
We note that the last 5 lines have poles, respectively, at 
\be
\begin{split}
s\in&\{2-\sigma_i, \eta_{ij}, 1-\eta_{ij}, 0, \tfrac12, \tfrac34, 1\}, \\ 
s\in&\{\sigma_i-\tfrac12, \eta_{ij}+\tfrac12, \tfrac32-\eta_{ij}, \tfrac12, \tfrac34, 1, \tfrac32\}, \\
s\in&\{\tfrac{\alpha_j+3}{2},1-\eta_{ij},0,\tfrac12, \tfrac34\}, \\
s\in&\{-\tfrac{\alpha_j}{2}, \eta_{ij}+\tfrac12, \tfrac32,1, \tfrac34\}, \\
s\in&\{\eta_{ij}, \tfrac32-\eta_{ij}, \tfrac12, 1\} \ .
\end{split}
\ee
However, for generic values of $\sigma_i, \alpha_j, \eta_{ij}$, only the poles at $s\in \{ 
2-\sigma_i, \sigma_i-\tfrac12, \tfrac{\alpha_j+3}{2}, -\tfrac{\alpha_j}{2}\}$  have 
$\Lambda$-independent residues. As we shall see, the other poles (except those at $s\in \{0, \tfrac12, \tfrac32, 2\}$) must cancel against those in $ \int_{\cF_2^{\Lambda}} \de\mu_2 \, \cE_2^\star(s,\Omega)\, F(\Omega)$ on the first line.

\subsection{Constructing $\Diamond_2$}
Having defined the renormalized integral \eqref{DefRS2} for the class of functions $F$ satisfying the growth conditions \eqref{Fgrowth21}, \eqref{Fgrowth22}, we now turn to the problem of relating it to 
a suitably regularized Mellin transform of the constant term $F^{(0)}$. For this purpose, following the strategy laid out in the introduction, we consider the invariant differential operators
\be
\label{Diamond2sigma}
\Diamond_2(\sigma)=\Delta^{(4)}_{Sp(4)} - (\sigma-\tfrac12)(\sigma-\tfrac32) \Delta_{Sp(4)} + 
\sigma(\sigma-\tfrac12)(\sigma-\tfrac32)(\sigma-2)\ .
\ee
and
\be
\label{Diamond2alpha}
\tilde\Diamond_2(\alpha)=\Delta^{(4)}_{Sp(4)}-\tfrac{1}{4}(\Delta_{Sp(4)})^2 +(\tfrac{3}{4}+\tfrac{1}{2}\alpha(\alpha+3))\Delta_{Sp(4)} - 
\tfrac{1}{4}\alpha(\alpha+1)(\alpha+2)(\alpha+3)\ .
\ee
Here, $\Delta_{Sp(4)}$ is the usual Laplace-Beltrami operator, proportional to the quadratic Casimir of the action of $\Gamma_2$ on $L^2(\cH_2)$, while $\Delta^{(4)}_{Sp(4)}$ is the invariant quartic differential operator proportional to the product of Maass' raising and lowering operators (see
Appendix \ref{app_gendegree} for further details). These operators are 
normalized such that
\be
\label{DeltaE2}
\begin{split}
\Delta_{Sp(4)}\, \cE_2^\star(s,\Omega) =& 2s (s-\tfrac32) \, \cE_2^\star(s,\Omega)\ ,
\\
\Delta^{(4)}_{Sp(4)} 
\, \cE_2^\star(s,\Omega) = &
 s (s-\tfrac12) (s-1) (s-\tfrac32)\, \cE_2^\star(s,\Omega)\ .
\end{split}
\ee
The operator $\Diamond_2(\sigma)$ is designed to 
annihilate $t^\sigma \varphi(\rho)$ for any function $\varphi(\rho)$. To see this, 
it suffices to write the operators  $\Delta_{Sp(4)}, \Delta^{(2)}_{Sp(4)}$ in the coordinates appropriate to region I,  
\be
\label{DeltaSp4toSp2}
\begin{split}
\Delta_{Sp(4)} \to & \Delta_t+ \Delta_\rho  \\
\Delta^{(4)}_{Sp(4)}  \to &  (\Delta_t+\frac{3}{4})\,  \Delta_{\rho} \ ,
\end{split}
\ee
where 
\be
\Delta_t = t^2 \partial_{t^2} - t \partial _t\ ,\quad \Delta_{\rho}= \rho_2^2 ( \partial_{\rho_1}^2 +\partial_{\rho_2}^2 )  \ .
\ee
The formulae  \eqref{DeltaSp4toSp2} hold up to  terms which annihilate functions of $t,\rho$, independent of $u_1,u_2,t_1$. Using these relations, it is straightforward to check that $\Diamond_2(\sigma)$ annihilates $t^\sigma \varphi(\rho)$ for any function $\varphi(\rho)$. 

Similarly, the operator  
$\tilde\Diamond_2(\alpha)$  is designed to 
annihilate $V^\alpha \tilde\varphi(\tau)$ for any function $\tilde\varphi(\tau)$. This is easily seen
by writing $\Delta_{Sp(4)}, \Delta^{(4)}_{Sp(4)}$ in the coordinates appropriate  to region II,
\be
\label{DeltaSp4toGL2}
\begin{split}
\Delta_{Sp(4)} \to & \frac12 ( \Delta_V + \Delta_\tau)  \ ,\quad \\
\Delta^{(4)}_{Sp(4)}  \to &  \frac{1}{4} \left[ \frac14 (\Delta_V - \Delta_\tau)^2 + \frac12 \Delta_V - \frac32 \Delta_\tau \right] \ ,
\end{split}
\ee
where 
\be
\Delta_V = V^2 \partial_{V^2} +4 V \partial _V\ ,\quad \Delta_{\tau}= \tau_2^2 ( \partial_{\tau_1}^2 +\partial_{\tau_2}^2 )  \,,
\ee
and the formulae \eqref{DeltaSp4toGL2} similarly hold up to terms annihilating functions of $V,\tau$, independent of $\Omega_1$. 

Thus, in order to construct an operator which annihilates the non-decaying part of $F$ in all regions, we may consider the product $\Diamond=\prod_{i=1}^{\ell} \Diamond_2(\sigma_i) \prod_{j=1}^{\tilde \ell} \tilde\Diamond_2(\alpha_j)$. However, in some circumstances, this may  not be the most economical choice. Indeed, since $\Diamond_2(\sigma)=\Diamond_2(2-\sigma)$ and 
$\tilde\Diamond_2(\alpha)= \tilde\Diamond_2(-3-\alpha)$, we may keep only one element in each pair $(\sigma_i,2-\sigma_i)$ or $(\alpha,-3-\alpha)$, in cases where the  two elements of the pair occur in the expansion. It may also be the case that an operator 
$\Diamond_2(\sigma_i)$ annihilates a term $V^{\alpha_j} \tilde\varphi_j(\tau)$ in the expansion 
\eqref{Fgrowth22}, or conversely, that $\tilde\Diamond_2(\alpha_j)$ annihilates a term 
$t^\sigma \varphi_i(\rho)$ in \eqref{Fgrowth21}. To see when this can happen, we compute
\be
\begin{split}
\Diamond_{2}(\sigma) \cdot V^{\alpha} \tilde\varphi(\tau) =& \frac14 V^{\alpha}\,
\left[ \Delta_\tau - (\alpha+2\sigma)(\alpha+2\sigma-1)\right]\, 
\left[ \Delta_\tau - (\alpha-2\sigma+3)(\alpha-2\sigma+4)\right] \cdot \tilde\varphi(\tau)\ ,
\\
\tilde\Diamond_2(\alpha)\cdot t^\sigma \varphi(\rho) =& -\frac14
\left[ \Delta_\rho - (\alpha+\sigma)(\alpha+\sigma+1)\right]\,
\left[ \Delta_\rho - (2+\alpha-\sigma)(3+\alpha-\sigma)\right]\cdot \varphi(\rho)\ .
\end{split}
\ee
Thus, this phenomenon may take place whenever $\varphi_i(\rho)$ and $\tilde\varphi_j(\tau)$ are
eigenmodes of the respective Laplace operators with a suitable eigenvalue. This happens, for example, when $F=\cE_2^\star(s',\Omega)$, although this  is a very special case since it is an eigenmode of 
$\Diamond_2(\sigma)$ and $\tilde\Diamond_2(\alpha)$ for any $\sigma$ and $\alpha$:
\be
\label{DiamondE2}
\begin{split}
\Diamond_2(\sigma)\,\cE_2^\star(s,\Omega) = &
(s-\sigma)(s-\sigma+\tfrac12)(s+\sigma-\tfrac32)(s+\sigma-2)\, \cE_2^\star(s,\Omega)\ ,\\
\tilde\Diamond_2(\alpha)\,\cE_2^\star(s,\Omega) =& 
(\alpha+1)(\alpha+2)(s+\tfrac12 \alpha)(s-\tfrac 12 (\alpha+3))\, \cE_2^\star(s,\Omega)\ .
\end{split}
\ee
As we shall see in \S\ref{sec_latdd2}, another example is the Narain lattice partition function $\Gamma_{d,d,2}$, whose non-decaying part is annihilated by $\Diamond_2(\tfrac{d}{2})$ in both regions I and II.

To take advantage of these possible simplifications, we shall take
\be
\label{defDiamond2}
\Diamond=\prod_{i\in I} \Diamond_2(\sigma_i) \prod_{j\in J} \tilde\Diamond_2(\alpha_j)
\ee
where $I$ and $J$ are suitable subsets of $1\dots \ell$ and $1\dots \tilde\ell$ such that 
$\Diamond$ annihilates both non-decaying
parts \eqref{Fgrowth21} and \eqref{Fgrowth22} in regions I and II.

We now consider the regularized integral $\cR_2^{\star,\Lambda}(\Diamond F)$. On the one hand,
$\Diamond F$ is of fast decay in regions  I and II, so the integral is finite as $\Lambda\to\infty$, and 
the standard Rankin-Selberg method reviewed in
the introduction produces
\be
\begin{split}
\cR_2^{\star}(\Diamond F) =&  \lim_{\Lambda\to\infty} \cR_2^{\star,\Lambda}(\Diamond F)\\
=& 
\zetastar(2s) \,  \zetastar(4s-2)\, 
\int_{\Gamma_{\infty,2}\backslash \cH_2} \de\mu_2 \, |\Omega_2|^{s}\, \Diamond F(\Omega)\\
=& 
\zetastar(2s) \,  \zetastar(4s-2)\, 
\int_0^{\infty} V^{2-2s} \de V\, \int_{\cF_1} \frac{\de\tau_1\de\tau_2}{\tau_2^2} \, \Diamond (F^{(0)}
-\tilde\varphi) \,,
\end{split}
\ee
where, in going from the second to the third line, we integrated over $\Omega_1$ and used the fact
that $\Diamond\tilde\varphi=0$. Integrating by parts and using \eqref{DiamondE2}, we find
\be
\cR_2^{\star}(\Diamond F) = \zetastar(2s) \,  \zetastar(4s-2)\, 
D_2(s)\, 
\int_0^{\infty} V^{2-2s} \de V\, \int_{\cF_1} \frac{\de\tau_1\de\tau_2}{\tau_2^2} \, (F^{(0)}-\tilde\varphi)\,,
\ee
where
\be
\begin{split}
D_2(s)&=\prod_{i\in I} (s-\sigma_i)(s-\sigma_i-\tfrac12)(s+\sigma_i-\tfrac32)(s+\sigma_i-2)\\
		& \,\times\prod_{j\in J}
		(\alpha_j+1)(\alpha_j+2)(s+\tfrac 12 \alpha_j)(s-\tfrac 12  (\alpha_j+3)) \ .
\end{split}
\ee
Since $\Diamond F$ is of rapid decay, $\cR_2^{\star}(\Diamond F)$ has a meromorphic continuation in $s$, invariant under $s\mapsto \tfrac32-s$, with simple poles only at $s=0,\tfrac12, 1,\tfrac32$.

On the other hand, by integrating by parts in the regularized integral $\cR_2^{\star,\Lambda}(\Diamond F)$ and using \eqref{DiamondE2}, we find after a tedious computation that, for finite $\Lambda$, 
\be
\label{DefRS2regfullDiamond}
\begin{split}
\int_{\cF_2^{\Lambda}} \de\mu_2 & \, \cE_2^\star(s,\Omega)\, \Diamond F(\Omega) =
D_2(s)\, \int_{\cF_2^{\Lambda}} \de\mu_2 \, \cE_2^\star(s,\Omega)\, F(\Omega)
\\
-&\frac12 \sum_i\, \frac{D_2(s)\, \zetastar(4s-2)\,  \Lambda^{s+\sigma_i-2}}{s+\sigma_i-2}
\RN \int_{\cF_1} \de\mu_1(\rho) \,\varphi_i(\rho)\,  \cE^\star_1(s,\rho)
\\
-& \frac12 \sum_i\,  \frac{D_2(s)\, \zetastar(4s-3)\,  \Lambda^{\sigma_i-\tfrac12-s}}{\sigma_i-\tfrac12-s}
\RN \int_{\cF_1} \de\mu_1(\rho)  \, \varphi_i(\rho)  \, \cE^\star_1(s-\tfrac12,\rho) 
\\
+& 
\sum_{j}  \frac{D_2(s)\,\zetastar(2s)\,\zetastar(4s-2)\, \Lambda^{2s-\alpha_j-3}}{\alpha_j-2s+3}
\RN \int_{\cF_1} \de\mu_1(\tau)\, \tau_2^{\alpha_j+3-2s}\, \tilde\varphi_j(\tau)
\\
+ & \sum_{j} \frac{D_2(s)\,\zetastar(2s-2)\,\zetastar(4s-3)\,\Lambda^{-2s-\alpha_j}}{2s+\alpha_j}
\RN \int_{\cF_1} \de\mu_1(\tau)\, \tau_2^{\alpha_j+2s}
 \, \tilde\varphi_j(\tau)
  \\
+ & \sum_{j} \frac{D_2(s)\,\zetastar(2s-1)\, \Lambda^{-2-\alpha_j}}{\alpha_j+2}
 \RN \int_{\cF_1} \de\mu_1(\tau)\, \tau_2^{\alpha_j+2}
 \cE^\star_1(2s-1,\tau)\, \tilde\varphi_j(\tau)\,.
\end{split}
\ee
Dividing out by $D_2(s)$ and using the definition \eqref{DefRS2exp}, we arrive at the desired 
relation between the renormalized integral and the renormalized Mellin transform,
\be
\label{RNMellin2}
\boxed{\cR_2^\star(F,s) = \zetastar(2s) \,  \zetastar(4s-2)\, 
\int_0^{\infty} V^{2-2s} \de V\, \int_{\cF_1} \frac{\de\tau_1\de\tau_2}{\tau_2^2} \, (F^{(0)}-\tilde\varphi)
=\frac{\cR_2^{\star}(\Diamond F)}{D_2(s)}\ .}
\ee
Since $\cR_2^{\star}(\Diamond F)$ has a meromorphic continuation in $s$ with simple poles only at $s\in\{0,\tfrac12, 1,\tfrac32\}$, it follows that similarly, $\cR_2^\star(F,s)$ has a meromorphic continuation in $s$, invariant under $s\mapsto \tfrac32-s$, with poles located at most at $s\in \{ 0,\tfrac12, 1,\tfrac32,
\sigma_i, \sigma_i-\tfrac12, 2-\sigma_i,\tfrac32-\sigma_i, -\tfrac{\alpha_j}{2}, \tfrac{\alpha_j+3}{2}\}$
with $i\in I, j\in J$.

For now, we assume that none of these values collected in curly brackets collide, so that 
the poles are simple. The simple pole at $s=\sigma\notin \{ 0,\tfrac12, 1,\tfrac32\}$ originates from the subtractions in \eqref{DefRS2exp}
\be
\begin{split}
\Res_{s=\sigma} \cR_2^\star(F,s) =& 
\frac 12\left(\sum_{i\atop \sigma=\sigma_i-\frac12}-\sum_{i\atop \sigma=2-\sigma_i} \right)\zetastar(4\sigma_i-5)\, \RN \int_{\cF_1} \de\mu_1(\rho) \,\varphi_i(\rho)\,  \cE^\star_1(\sigma_i-1,\rho)\\
+&\frac12 
\left(\sum_{j\atop \sigma=-\frac{\alpha_j}{2}}-\sum_{j\atop \sigma=\frac{\alpha_j+3}{2}} \right)
\zetastar(\alpha_j+3)\, \zetastar(2\alpha_j+4)\, \RN \int_{\cF_1} \de\mu_1(\tau) \,\tilde\varphi_j(\tau)\,  ,
\end{split}
\ee
while the simple poles at $s=\sigma\in \{ 0,\tfrac12, 1,\tfrac32\}$ originate from the poles
in $\cE^\star_2(s)$. We define the renormalized integral of $F$ by subtracting the power-like
terms in $\Lambda$,
\be
\begin{split}
\RN \, \int_{\cF_2} \de\mu_2\, F(\Omega) \equiv& \lim_{\Lambda\to\infty} \biggr\{ \int_{\cF_2^{\Lambda}} \de\mu_2 \, F(\Omega) 
-\frac12 \sum_{i\atop \sigma_i\neq 2}\, \frac{\Lambda^{\sigma_i-2}}{\sigma_i-2}
\RN \int_{\cF_1} \de\mu_1(\rho) \,\varphi_i(\rho)\, 
\\
-& \frac{1}{2}\log\Lambda\sum_{i\atop \sigma_i=2}\RN\int_{\mathcal F_1}\de\mu_1(\rho)\,\varphi_i(\rho)+\sum_{j\atop \alpha_j\neq -3}  \frac{\Lambda^{-\alpha_j-3}}{\alpha_j+3}
\RN \int_{\cF_1} \de\mu_1(\tau)\, \tau_2^{\alpha_j+3}\, \tilde\varphi_j(\tau) \,
\\
& -\log\Lambda \sum_{j\atop\alpha_j=-3}\RN \int_{\cF_1} \de\mu_1(\tau)\, \tilde\varphi_j(\tau) \biggr\}\ .
\end{split}
\ee
With this definition, assuming that $\tfrac32\notin \{\sigma_i-\tfrac12, 2-\sigma_i, -\tfrac{\alpha_j}{2}, \tfrac{\alpha_j+3}{2}\}$,   the renormalized integral of $F$  equals $1/r_2$ times the residue of $\cR_2^\star(F,s)$ at the simple pole $s=\tfrac32$,
\be
\begin{split}
\RN \, \int_{\cF_2} \de\mu_2\, F(\Omega) = \frac{2}{\zetastar(3)} \Res_{s=\tfrac32} \cR_2^\star(F,s) \ .
\end{split}
\ee
When the pole is of order greater than one, the two differ by a finite contribution,
\be
\begin{split}
\RN \, \int_{\cF_2} \de\mu_2\, F(\Omega) =& \frac{2}{\zetastar(3)} \Res_{s=\tfrac32} \cR_2^\star(F,s)
+\frac{\zetastar(4)}{\zetastar(3)} \sum_{\sigma_i=\tfrac12} 
\RN\, \int_{\cF_1} \de\mu_1(\rho)\, \varphi_i(\rho)\, \cE_1^\star(\tfrac32,\rho) \\
& +\frac12\sum_{\sigma_i=2}
\RN\, \int_{\cF_1} \de\mu_1(\rho)\, \varphi_i(\rho)\, \left[\log(64\pi^3 e^{\gamma-4} \rho_2 |\eta(\rho)|^4)-4\,\frac{\zeta'(3)}{\zeta(3)}\right]\\
& +\zetastar(4) \sum_{\alpha_j=0} \RN\, \int_{\cF_1} \de\mu_1(\tau)\,\tilde\varphi_j(\tau) \\
& - \sum_{\alpha_j=-3} \RN\, \int_{\cF_1} \de\mu_1(\tau) \,\tilde\varphi_j(\tau)\,\left[\log\left(\frac{e^{2-\gamma/2}}{8\pi^{3/2}}\,\tau_2\right)+2\,\frac{\zeta'(3)}{\zeta(3)}\right]\,
\ .
\end{split}
\ee
For the example discussed in the next subsection, this difference was denoted by $\delta$ in \cite[(4.30)]{Pioline:2014bra}, but was left implicit in that early attempt.

\subsection{Lattice partition function \label{sec_latdd2}}

We now apply the previous result to the special case when $F$ is the lattice partition 
function defined in \eqref{defGammaddh}. In this case, the asymptotic behavior in regions I and II
is given by 
\be
\begin{split}
\varphi=& t^{\tfrac{d}{2}} \Gamma_{d,d,1}(\rho)\ ,\\
\tilde\varphi=& V^{-d} +V^{-d}  \sum_{(p,q)=1} \sum'_{(m_i,n^i)\in\IZ^{2d}\atop
m_i n^i=0} e^{-\frac{\pi |p-q\tau|^2}{\tau_2 V}\cM^2(m_i,n^i)}\,,
\end{split}
\ee
so that the conditions of the previous subsection are obeyed with 
\be
\sigma_i=\tfrac{d}{2}, \quad \varphi_i=\Gamma_{d,d,1}, \quad \alpha_j=-d,\quad
\tilde\varphi_j(\tau)=1\ ,\quad \Diamond = \Diamond_2(\tfrac{d}{2})\ .
\ee
Note that the second term in $\tilde\varphi$ (corresponding to the $\Omega_1$-independent 
terms in \eqref{defGammaddh} where 
the matrix $\cL^{IJ}$ has rank one) is not of the form $V^\alpha \tilde\varphi(\tau)$, but it is 
obtained from $\int_0^1 \de\rho_1 \varphi$ by a modular transformation $SL(2,\IZ)\subset Sp(4,\IZ)$, so it is also annihilated by $\Diamond_2(\tfrac{d}{2})$.

The renormalized integral $\RN \int_{\cF_2}\,\de\mu_2 \,\cE^\star_2(s,\Omega)\, \Gamma_{d,d,2}$
is obtained by applying the general prescription \eqref{DefRS2exp}. The integrals $\RN \int_{\cF_1} \de\mu_1(\rho) \Gamma_{d,d,1}\, \cE^\star_1(s',\rho)$ appearing on the second and third line of this expression were evaluated in \eqref{RN1Gdd}. The integrals appearing on the fourth and sixth line
can be evaluated explicitly \cite[(2.34)]{Angelantonj:2011br},
\be
\label{intF1t2gamma}
c_2(\gamma) \equiv \RN\,  \int_{\cF_1} \de\mu_1(\tau) \,\tau_2^\gamma 
= \frac{2}{1-\gamma} \int_{0}^{\tfrac12} \de\tau_1\, (1-\tau_1^2)^{\tfrac{\gamma-1}{2}}
=\frac{2\, c(\tfrac{1-\gamma}{2})}{1-\gamma}\ ,
\ee
where $c(\alpha)=\int_0^{\tfrac12} (1-x^2)^{-\alpha}=\tfrac12 \,_2F_1(\alpha,\tfrac12;\tfrac32;\tfrac14)$.
The pole at $\gamma=1$ arises from the logarithmic divergence of the integral over $\tau_2$. We note the special values $c(0)=\tfrac12$ and $c(\tfrac12)=\tfrac{\pi}{6}$, consistent with the volume of the fundamental domain $\cV_1=\frac{\pi}{3}$. 

Finally, the integral $c_2(\gamma,s')\equiv \RN \int_{\cF_1} \de\mu_1\, \tau_2^{\gamma}\, \cE^\star_1(s',\tau)$ appearing on the last line of  \eqref{DefRS2exp} does not seem to be computable in closed form, although by using the fact that both factors in the integrand are eigenmodes of 
the Laplacian $\Delta$, it can be reduced to an integral $\int_{-1/2}^{1/2} \de\tau_1\, \tau_2^{\gamma-1} \cE^\star_1(s,\tau)$, where the integrand is evaluated at $|\tau|=1$. Its analytic structure as a function of $s$ and $\gamma$ is determined by the constant terms in $\cE^\star_1(s,\tau)$, since the non-zero Fourier modes are exponentially suppressed as $\tau_2\to\infty$. In particular, it has simple poles at $\gamma+s=1$ and $\gamma=s$. Furthermore it vanishes at $\gamma=0$, since the renormalized integral of $\cE^\star_1(s,\tau)$ vanishes.

The renormalized integral is
then defined as the limit
\be
\label{RNRstarG}
\begin{split}
\RN &\int_{\cF_2}\,\de\mu_2 \, 
\cE^\star_2(s,\Omega)\, \Gamma_{d,d,2}  \equiv  \cR_2^\star(\Gamma_{d,d,2},s)  =
\lim_{\Lambda\to\infty}  \left[ \int_{\cF_2^{\Lambda}}\,\de\mu_2 \, 
\cE^\star_2(s,\Omega)\, \Gamma_{d,d,2}  \right. 
\\
& -\frac12 \frac{ \zetastar(4s-2)  \Lambda^{s+\tfrac{d}{2}-2}}{s+\tfrac{d}{2}-2}
\cR^\star_1(\Gamma_{d,d,1},s) 
-\frac12  \frac{\zeta^\star(4s-3) \Lambda^{-s+\tfrac{d-1}{2}}}{\tfrac{d-1}{2}-s}
  \cR^\star_1(\Gamma_{d,d,1},s-\tfrac12) 
  \\
 & -  
2\frac{\zetastar(2s) \zetastar(4s-2) \, c(s+\tfrac{d}{2}-1) \Lambda^{2s+d-3}}{(2s+d-2)(2s+d-3)} 
- 
2\frac{\zetastar(2s-2) \zetastar(4s-3)  \, c(\tfrac{d+1}{2}-s) \Lambda^{d-2s}}{(2s-d)(2s-d-1)} 
\\
& \left. - 
\frac{\zetastar(2s-1)\, c_2(2-d;2s-1) \Lambda^{d-2}}{d-2}\, \right]\,.
\end{split}
\ee

On the other hand, using \eqref{GenEuler} the renormalized Mellin transform \eqref{RNMellin2} evaluates to 
\be
\label{RN2Gdd}
\cR_2^\star(\Gamma_{d,d,2},s) = \frac{\zetastar(2s) \,  \zetastar(4s-2)\, 
\Gamma(s+\tfrac{d-3}{2})\, \Gamma(s+\tfrac{d-4}{2})}{\pi^{2s+d-\tfrac72}}
\sum_{\substack{(m_i^I, n^{i,I})\in\IZ^{4d} / GL(2,\IZ)
\\ m_i^I n^{i J}+m_i^J n^{i I}=0 \\
{\rm rank}(m_i^I, n^{i,I})=2}}
|\cL|^{-s+\tfrac{3-d}{2}}\,,
\ee
where $\cL$ is the $2\times 2$ matrix defined in \eqref{defcL}. This is recognized as the completed
Langlands-Eisenstein series attached to the two-index 
antisymmetric representation of $SO(d,d,\IZ)$,\footnote{This relation is
discussed in more detail in Appendix \ref{sec_lath}.}
\be
\cR_2^\star(\Gamma_{d,d,2},s) = 2\, \cE^{SO(d,d),\star}_{\Lambda^2V}(s+\tfrac{d-3}{2})\,.
\ee
Indeed, it is known from the Langlands constant term formula that $\cE^{SO(d,d),\star}_{\Lambda^2V}(s')$ has, for generic dimension $d$, simple poles at
\be
\label{polesE2}
s'=0,\tfrac12, \tfrac{d-3}{2}, \tfrac{d-2}{2},\tfrac{d-1}{2},\tfrac{d}{2}, d-2, d-\tfrac32\ ,
\ee
which translate into poles at $s= \tfrac{3-d}{2}, \tfrac{4-d}{2}, 0, \tfrac12, 1, \tfrac32, \tfrac{d-1}{2}, \tfrac{d}{2}$ as predicted by the Rankin-Selberg method (in particular,  the apparent pole at $s=\tfrac{d+1}{2}$ in \eqref{RN2Gdd} cancels). Moreover, from the fact that 
the regularized integral $\int_{\cF_2^{\Lambda}}\,\de\mu_2 \, \cE^\star_2(s)\, \Gamma_{d,d,2} $ has only simple poles at $s=0,\tfrac12,1,\tfrac32$, we deduce from \eqref{RNRstarG} the value of the
residues of $\cR_2^\star(\Gamma_{d,d,2},s)$ at the poles,
\be
\label{resE2}
\begin{split}
{\rm Res}_{s=\tfrac{d}{2}} \cR^\star_2(\Gamma_{d,d,2},s) = &\zetastar(2)\,\zeta^\star(d-2)\, \zeta^\star(2d-3)\,,
\\
{\rm Res}_{s=\tfrac{d-1}{2}} \cR^\star_2(\Gamma_{d,d,2},s) = &  \zeta^\star(2d-5)\, \cE^{SO(d,d),\star}_{V}(d-2)\,,
\\
{\rm Res}_{s=\tfrac{3}{2}} \cR^\star_2(\Gamma_{d,d,2},s) =& 
\frac12 \zetastar(3) \,\RN \int_{\cF_2}\,\de\mu_2 \, \Gamma_{d,d,2}\,,
\\
{\rm Res}_{s=1} \cR^\star_2(\Gamma_{d,d,2},s) =&
\frac12 \zetastar(3) \,
\RN \int_{\cF_2^{\Lambda}}\,\de\mu_2  \left[ {\rm Res}_{s=1} \cE^\star_2(s) \right] \,\Gamma_{d,d,2} \,,
\end{split}
\ee
where the renormalized integrals are defined for generic $d$ by 
\be
\label{defRN2Gdd}
\begin{split}
&\RN \int_{\cF_2}\,\de\mu_2 \, \Gamma_{d,d,2}  =  \lim_{\Lambda\to\infty} \left[ 
\int_{\cF_2^{\Lambda}}\,\de\mu_2 \, \Gamma_{d,d,2} 
 -\frac{2\Lambda^{d-3} c(\tfrac{d}{2}-1)}{(d-2)(d-3)} 
-\frac{\Lambda^{\tfrac{d}{2}-2}}{d-4}\, \RN 
\int_{\cF_1}\,\de\mu_1 \, \Gamma_{d,d,1} \right] \,,
\\
&\RN \int_{\cF_2^{\Lambda}}\,\de\mu_2  \left[ {\rm Res}_{s=1} \cE^\star_2(s) \right] \Gamma_{d,d,2} 
= \lim_{\Lambda\to\infty}  \left[ 
\int_{\cF_2^{\Lambda}}\,\de\mu_2 \, \left[ {\rm Res}_{s=1} \cE^\star_2(s) \right] \Gamma_{d,d,2} \right. \\
& \left. - \frac{\zetastar(2) \, \Lambda^{\tfrac{d}{2}-1}}{\zetastar(3)(d-2)}\, \RN 
\int_{\cF_1}\,\de\mu_1 \, \Gamma_{d,d,1}  
-\frac{2\Lambda^{d-2} c(\tfrac{d-1}{2})\, \log \Lambda}{(d-1)(d-2)\zetastar(3)}
-\frac{\Lambda^{\tfrac{d-3}{2}}}{2(d-3)\zetastar(3)} \cR_1^\star(\Gamma_{d,d,1},\tfrac12) 
\right] \,.
\end{split}
\ee
For the values of $d$ where the subtracted terms in these equations are singular ($d=2,3,4$ for the first line and $d=1,2,3$ in the second line), we define the renormalized integrals by  subtracting the corresponding logarithmic divergences, as in   \eqref{defRN1Gdd}. In these cases, the Rankin-Selberg transform \eqref{RN2Gdd} has a pole of higher degree, and the residue at the pole 
differs from the renormalized integral by a computable term. Focussing on the residue at $s=\tfrac32$,
\be
\begin{split}
\int_{\cF_2^{\Lambda}}\,\de\mu_2 \Gamma_{3,3,2} = &
\frac{4\pi}{\zeta(3)} {\rm Res}_{s=3/2}  \cR^\star_2(\Gamma_{3,3,2},s)
+\frac{\pi}{3} (\log \Lambda + {\rm cte}) - \frac{1}{\sqrt{\Lambda}}  
\RN \int_{\cF_1}\,\de\mu_1 \Gamma_{3,3,1}\,,
\\
\int_{\cF_2^{\Lambda}}\,\de\mu_2 \Gamma_{4,4,2} = &
\frac{4\pi}{\zeta(3)} {\rm Res}_{s=3/2}  \cR^\star_2(\Gamma_{4,4,2},s)
+\frac{1}{2} \RN \int_{\cF_1}\,\de\mu_1 \Gamma_{4,4,1}\, (\log \Lambda+{\rm cte}) \\
&-2 {\widehat\cE}^{SO(4,4),\star}_V(2) +c(1)\, \Lambda\, ,
\end{split}
\ee
where ${\widehat\cE}^{SO(4,4),\star}_V(2)=\lim_{s\to 2}\left[ {\cE}^{SO(4,4),\star}_V(s) -
\frac{1}{4(s-2)} \RN \int_{\cF_1}\,\de\mu_1 \Gamma_{4,4,1}\right]$.

For $d=2$, it was shown in \cite{Pioline:2014bra} that the Langlands-Eisenstein series $\cE^{SO(2,2),\star}_{\Lambda^2V}$ can be written as a sum of $SL(2)$ Eisenstein series
\be
 \cE^{SO(2,2),\star}_{\Lambda^2V} = \zetastar(2s+1)\, \zetastar(2s)\, \zetastar(2s-1)
 \left[ \cE_1^\star(2s,T) + \cE_1^\star(2s,U)\right]\ .
\ee
The residue at the simple pole $s=\tfrac32$ yields
\be
\RN\int_{\cF_2}\,\de\mu_2 \Gamma_{2,2,2} = 2\zetastar(2)  
\left[ \cE_1^\star(3,T) + \cE_1^\star(3,U)\right]\ .
\ee
For $d<2$, the Rankin-Selberg transform \eqref{RN2Gdd} vanishes, but the renormalized integral
is still non-trivial,
\be
\begin{split}
\int_{\cF_2^{\Lambda}}\,\de\mu_2 = &2 \zetastar(2) \left( \zetastar(4) - \frac{1}{4\Lambda^2} \right)\,,
\\
\int_{\cF_2^{\Lambda}}\,\de\mu_2 \Gamma_{1,1,2}
= & 2\zetastar(2) \zetastar(4) (R^2+1/R^2) -\frac{\pi}{9\Lambda^{3/2}} (R+1/R) 
+ \frac{\sqrt{3}}{8\Lambda^2}+\frac{\pi}{12\Lambda^2}\ ,
\end{split}
\ee
reproducing the result of \cite[(4.46)]{Pioline:2014bra} in the limit $\Lambda\to\infty$. 

\subsection{Product of two Eisenstein series}

We now apply our general result to the case $F=\cE_2^\star(s';\Omega)$. The asymptotic behavior
in region I and II can be read off from \eqref{decsp4sp2} and \eqref{decsp4gl2}, with $s\mapsto s'$.
In this case, rather than using the operator \eqref{defDiamond2}, it is easiest to exploit the fact 
that $F$ is an eigenmode of the Laplacian, see \eqref{DeltaE2}. Taking 
$\Diamond= \Delta_{Sp(4)}-2s' (s'-\tfrac32)$, we trivially get $\cR^\star_2(\Diamond F)=0$,
and therefore $\cR^\star_2(F)=0$. This reflects the general fact that the renormalized integral (or more generally, the renormalized Rankin-Selberg transform) of an eigenmode of the Laplacian vanishes.
The  regularized integral is therefore purely given by the boundary terms in \eqref{DefRS2regfull}. Since the $\varphi_i(\rho)$ are $SL(2,\IZ)$ Eisenstein series,
and since the renormalized integral of the product of two Eisenstein series vanishes \cite{MR656029},
the contributions from region I vanish, leaving only the last three lines in \eqref{DefRS2regfull}:
\be
\begin{split}
\cR_2^{\star,\Lambda}(\cE_2(s');s)= 
&  \frac{\zetastar(2 s) \zetastar(4 s-2) \zetastar(2 s')
   \zetastar(4 s'-2)\, c(s+s'-1) \Lambda^{2 s+2
   s'-3}}{(s+s'-1) (2 s+2 s'-3)} \\
+& \frac{\zetastar(2 s) \zetastar(4 s-2) \zetastar(2 s'-2)
   \zetastar(4 s'-3)\, c\left(s-s'+\frac{1}{2}\right) \Lambda^{2s-2s'}}{(2 s-2 s'+1) (s-s')} \\
+& \frac{\zetastar(2 s-2) \zetastar(4 s-3) \zetastar(2 s')
   \zetastar(4 s'-2) \,c\left(-s+s'+\frac{1}{2}\right) \Lambda^{2
   s'-2 s}}{(2 s-2 s'-1) (s-s')} \\
+& \frac{\zetastar(2 s-2) \zetastar(4 s-3) \zetastar(2 s'-2)
   \zetastar(4 s'-3) \,c(-s-s'+2) \Lambda^{-2 s-2
   s'+3}}{(s+s'-2) (2 s+2 s'-3)} 
  \\
- & \frac{\zetastar(2s-1)\,\zetastar(2s')\,\zetastar(4s'-2)\, c_2(2-2s',2s-1)\, \Lambda^{2s'-2} }{2-2s'}\,
 \\
- & \frac{\zetastar(2s-1)\, \zetastar(2s'-2)\,\zetastar(4s'-3)\, c_2(2s'-1,2s-1)\, \Lambda^{1-2s'} }{2s'-1}
\\
 -&
  \frac{\zetastar(2s)\,\zetastar(4s-2)\, \zetastar(2s'-1)\, c_2(2-2s,2s'-1)\, \Lambda^{2s-2}}{2-2s}
\\
- &\frac{\zetastar(2s-2)\,\zetastar(4s-3)\,\zetastar(2s'-1)\,c_2(2s-1,2s'-1)\,\Lambda^{1-2s}}{2s-1}
  \\
- & \zetastar(2s-1)\, \zetastar(2s'-1)\,\Lambda^{-1} \,
c_2(1,2s-1,2s'-1)
\end{split}
\ee
where
\be
c_2(\gamma,s,s') = \RN \int_{\cF_1} \de\mu_1\, \tau_2^\gamma\,\cE^\star_1(s,\tau)\, 
\cE^\star_1(s',\tau)\ .
\ee
Although the functions $c_2(\gamma,s)$ and $c_2(\gamma,s,s')$ do not seem to be computable in closed form, one may check that the apparent poles at $s'=s$, $s'=1-s$ and images under the symmetries $s\mapsto \tfrac32-s$ and $s\leftrightarrow s'$ cancel, leaving only the poles at $s=0,\tfrac12,1,\tfrac32$ and   $s'=0,\tfrac12,1,\tfrac32$. The above gives an explicit example of Maass-Selberg relations at genus 2.

\section{Degree three\label{sec_three}}

We now turn to the case of Siegel modular forms of degree three, with at most polynomial growth at the cusp. As before, our aim is to define the 
renormalized integral
\be
\label{DefRS3}
\cR_3^\star(F,s) = \RN \int_{\cF_3} \de\mu_3 \, \cE_3^\star(s,\Omega)\, F(\Omega)\ ,
\ee
and relate it to the generalized Mellin transform of the zero-th Fourier coefficient $F^{(0)}(\Omega_2)$, defined with a suitable subtraction. 

\subsection{Regularizing the divergences}

\begin{figure}
\begin{center}
\begin{picture}(0,0)%
\includegraphics{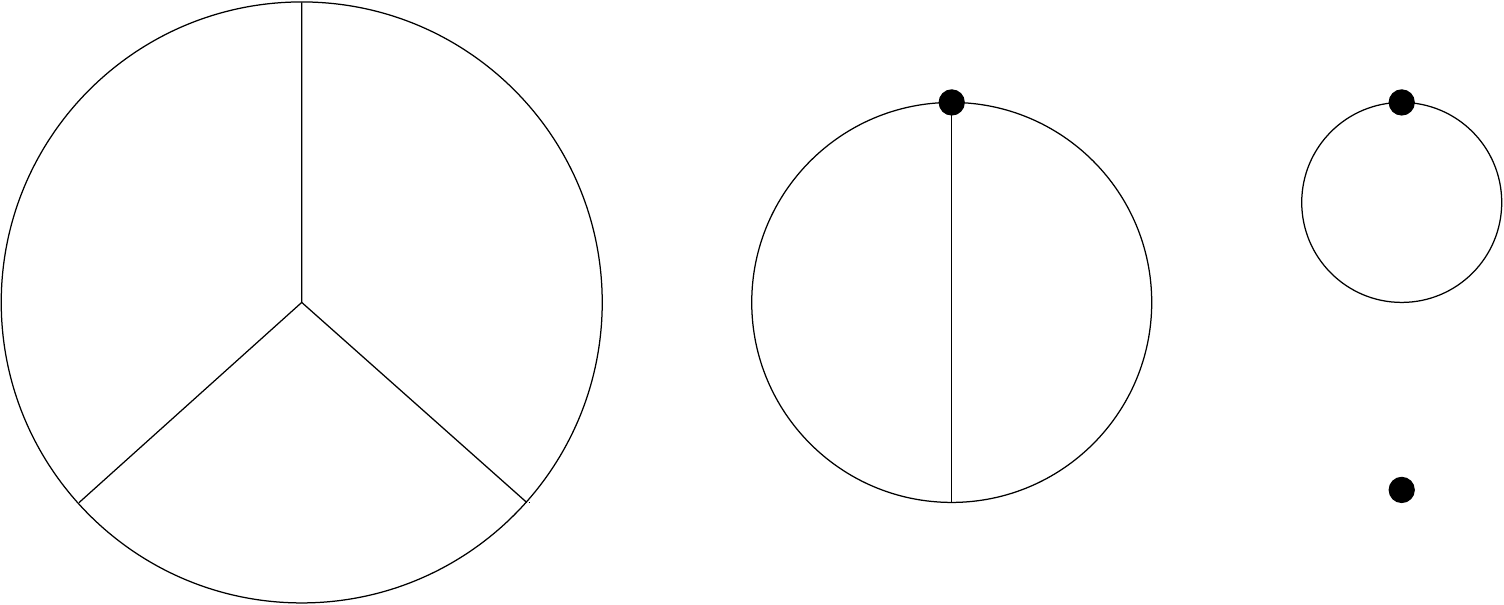}%
\end{picture}%
\setlength{\unitlength}{3158sp}%
\begingroup\makeatletter\ifx\SetFigFont\undefined%
\gdef\SetFigFont#1#2#3#4#5{%
  \reset@font\fontsize{#1}{#2pt}%
  \fontfamily{#3}\fontseries{#4}\fontshape{#5}%
  \selectfont}%
\fi\endgroup%
\begin{picture}(9019,3623)(1490,-4272)
\put(7351,-1561){\makebox(0,0)[lb]{\smash{{\SetFigFont{10}{12.0}{\familydefault}{\mddefault}{\updefault}{\color[rgb]{0,0,0}$\rho$}%
}}}}
\put(9826,-961){\makebox(0,0)[lb]{\smash{{\SetFigFont{10}{12.0}{\familydefault}{\mddefault}{\updefault}{\color[rgb]{0,0,0}2}%
}}}}
\put(10276,-1861){\makebox(0,0)[lb]{\smash{{\SetFigFont{10}{12.0}{\familydefault}{\mddefault}{\updefault}{\color[rgb]{0,0,0}t}%
}}}}
\put(9751,-1561){\makebox(0,0)[lb]{\smash{{\SetFigFont{10}{12.0}{\familydefault}{\mddefault}{\updefault}{\color[rgb]{0,0,0}$\tilde\Omega$}%
}}}}
\put(9826,-3286){\makebox(0,0)[lb]{\smash{{\SetFigFont{10}{12.0}{\familydefault}{\mddefault}{\updefault}{\color[rgb]{0,0,0}3}%
}}}}
\put(9826,-3961){\makebox(0,0)[lb]{\smash{{\SetFigFont{10}{12.0}{\familydefault}{\mddefault}{\updefault}{\color[rgb]{0,0,0}$\Omega$}%
}}}}
\put(3481,-1666){\makebox(0,0)[lb]{\smash{{\SetFigFont{10}{12.0}{\familydefault}{\mddefault}{\updefault}{\color[rgb]{0,0,0}$L_1$}%
}}}}
\put(4201,-2911){\makebox(0,0)[lb]{\smash{{\SetFigFont{10}{12.0}{\familydefault}{\mddefault}{\updefault}{\color[rgb]{0,0,0}$L_2$}%
}}}}
\put(2101,-2911){\makebox(0,0)[lb]{\smash{{\SetFigFont{10}{12.0}{\familydefault}{\mddefault}{\updefault}{\color[rgb]{0,0,0}$L_3$}%
}}}}
\put(1801,-1861){\makebox(0,0)[lb]{\smash{{\SetFigFont{10}{12.0}{\familydefault}{\mddefault}{\updefault}{\color[rgb]{0,0,0}$R_2$}%
}}}}
\put(4501,-1861){\makebox(0,0)[lb]{\smash{{\SetFigFont{10}{12.0}{\familydefault}{\mddefault}{\updefault}{\color[rgb]{0,0,0}$R_3$}%
}}}}
\put(3196,-3886){\makebox(0,0)[lb]{\smash{{\SetFigFont{10}{12.0}{\familydefault}{\mddefault}{\updefault}{\color[rgb]{0,0,0}$R_1$}%
}}}}
\put(7126,-961){\makebox(0,0)[lb]{\smash{{\SetFigFont{10}{12.0}{\familydefault}{\mddefault}{\updefault}{\color[rgb]{0,0,0}1}%
}}}}
\put(6076,-2461){\makebox(0,0)[lb]{\smash{{\SetFigFont{10}{12.0}{\familydefault}{\mddefault}{\updefault}{\color[rgb]{0,0,0}$L'_1$}%
}}}}
\put(7276,-2461){\makebox(0,0)[lb]{\smash{{\SetFigFont{10}{12.0}{\familydefault}{\mddefault}{\updefault}{\color[rgb]{0,0,0}$L'_2$}%
}}}}
\put(8101,-2461){\makebox(0,0)[lb]{\smash{{\SetFigFont{10}{12.0}{\familydefault}{\mddefault}{\updefault}{\color[rgb]{0,0,0}$L'_3$}%
}}}}
\end{picture}%
\end{center}
\caption{Tetrahedron three-loop diagram, and its one-loop, two-loop and three-loop counterterms, corresponding to the regions III, II, I, 0 defined in \eqref{defF3split}.
\label{fig_gen3}}
\end{figure}
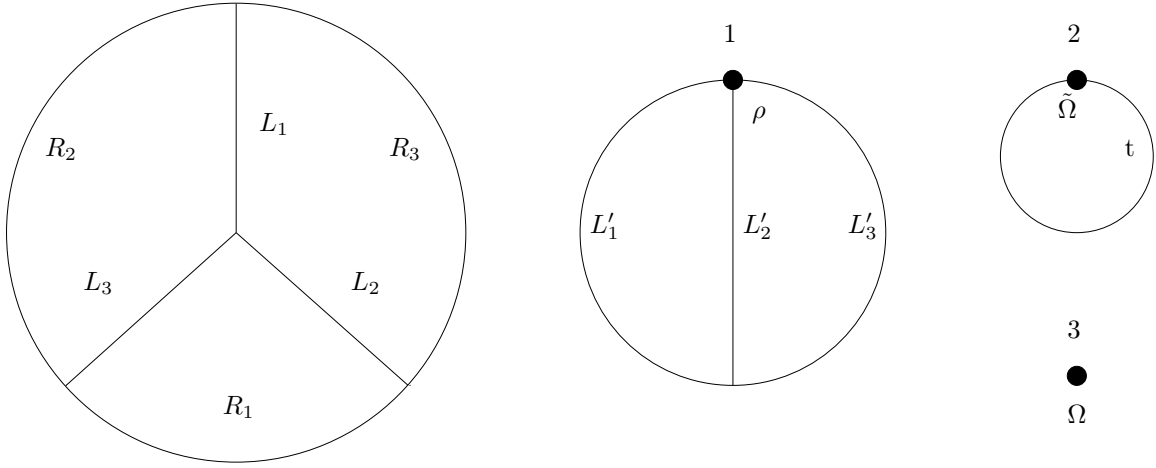

Our first task is to understand the possible sources of divergence in the integral \eqref{DefRS2}. For this we choose a fundamental domain $\cF_3$ defined by 
\be
\label{defF3}
\begin{split}
(1) & \qquad -\frac12 < \Re(\Omega_{IJ}) \leq \frac12 \\
(2) & \qquad \Im(\Omega) \in \cF_{PGL(3,\IZ)}   \\
(3) & \qquad  |C\Omega+D|>1 \ \mbox{for all}\  \begin{pmatrix} A & B \\ C& D \end{pmatrix} \in Sp(6,\IZ)
\end{split}
\ee
where $\cF_{PGL(3,\IZ)}$ is a fundamental domain for the action of $PGL(3,\IZ)$ on the space of positive definite symmetric matrices of rank 3. Various distinct fundamental domains are discussed in the literature \cite{zbMATH02628371,zbMATH04037921,zbMATH00572465}, however we shall find it convenient to use the one which appears in the maximal degeneration described by the tetrahedron three-loop diagram (see Figure \ref{fig_gen3}). Indeed, it follows from the Torelli and Schottky theorems for metric graphs in \cite{zbMATH05717813,Brannetti:2009} that any generic positive definite rank 3 matrix 
can be uniquely conjugated by an element of $GL(3,\IZ)$ into the period matrix of  the tetrahedron
diagram,\footnote{The tetrahedron diagram is one of the five diagrams which appear in the 
maximal degenerations of a genus 3 Riemann surface (see e.g. Figure 1 in \cite{zbMATH06126971}). The period matrix of the other four lie in lower dimensional cells of the moduli space of tropical
Abelian varieties of dimension 3. See \cite{Tourkine:2013rda} for a discussion of the relevance
of tropical geometry for string amplitudes.}
\be
\label{OmMercedes}
 \Omega_2 = \begin{pmatrix} L_1+L_2+R_3 & -L_2 & -L_1 \\ 
 -L_2 & L_2+L_3+R_1 & -L_3 \\
 -L_1 & -L_3 & L_1+L_3+R_2 \end{pmatrix}\,,
\ee
up to an automorphism of this diagram. Since the symmetry group of the tetrahedron is $\sigma_4$,
which acts by permuting the 4 faces, we can fix this symmetry by requiring that the sum of length of the edges of each face be ordered,  
\be
\label{LRorder}
L_1+L_2+R_3 < L_2+L_3+R_1 < L_1+L_3+R_2 < R_1 + R_2 + R_3 \ .
\ee
Thus, we choose for $\cF_{PGL(3,\IZ)}$ the space of all matrices \eqref{OmMercedes} where 
$L_i, R_i$ are positive real  variables such that \eqref{LRorder} is obeyed.
 The integration measure is normalized as in \cite{Pioline:2014bra}, 
\be
\de\mu_3(\Omega) = \frac{\prod_{I\leq J} \de\, \Re(\Omega_{IJ})\,  \de\, \Im(\Omega_{IJ})}{|\Omega_2|^4}\ ,
\ee
so that the volume of the fundamental domain $\cF_3$ is 
$\cV_3=2\zetastar(2)\zetastar(4)\zetastar(6)=\frac{\pi^6}{127575}$.

The first source of divergences is the region I where $\Im(\Omega_{33})$  is scaled to infinity, keeping
the other entries in $\Omega$ fixed (or equivalently, $R_2\to \infty$ keeping $R_1,R_3,L_1,L_2,L_3$ fixed). In this region, it is convenient to parametrize
\be
\label{coorsp3sp2}
\Omega = \begin{pmatrix} \tilde\Omega & \tilde\Omega \tilde u_2 + \tilde u_1 \\ 
\tilde u_2^t \tilde\Omega +  \tilde u_1^t & t_1+\I(t+  \tilde u_2^t \tilde\Omega_2  \tilde u_2) 
\end{pmatrix}\ ,
\ee
where $t\in\IR^+, \tilde\Omega\in \cH_2, (\tilde u_1,\tilde u_2,t_1)\in \IR^5$, so that the region of interest is $t\to \infty$, keeping $\tilde\Omega,\tilde u_1,\tilde u_2,t_1$ finite. In the language of string theory, the region $t\to\infty$ is responsible for one-loop infrared subdivergences, with the parameter $t$ being interpreted as the Schwinger time parameter around the loop. 
Since the stabilizer of the cusp $t=\infty$ is $Sp(4,\IZ)\ltimes \IZ^4 \ltimes \IZ$, the fundamental 
domain $\cF_3$ simplifies in this limit to $\IR^+ \times \cF_2 \times [-\tfrac12,\tfrac12]^4/\IZ_2 \times [-\tfrac12,\tfrac12]$, where 
the center $\IZ_2$ of $Sp(4,\IZ)$ acts by flipping the sign of $(\tilde u_1,\tilde u_2)$. 
In the coordinates \eqref{coorsp2sp1}, 
the integration measure becomes
\be
\label{mu3asI}
\de\mu_3=\frac{\de t}{t^4}\, \de\mu_2(\tilde\Omega)\, \de^2  \tilde u_1\, \de^2  \tilde u_2\, \de t_1\ ,
\ee
where $\de\mu_2(\tilde\Omega)$ is the measure defined in \eqref{defmu2}.

The second source of divergences is the region II where the $2\times 2$ submatrix $
{\Im \scriptsize \begin{pmatrix} \Omega_{22} & \Omega_{23} \\ \Omega_{23} & \Omega_{33}
\end{pmatrix}}$ scales to infinity (or equivalently, $R_1,R_2,L_3\to \infty$, keeping $L_1,L_2,R_3$ fixed). In the language of string theory, this region is responsible for  two-loop infrared subdivergences. In this region, it is useful to parametrize
\be
\label{coorsp2gl2}
\Omega = \begin{pmatrix} \rho & \rho u_2 - u_1 \\ 
u_2^t \rho - u_1^t & t_1 + \I(t_2+ u_2^t \rho_2 u_2) 
\end{pmatrix} \ ,\quad
t_2= \frac{1}{V \tau_2} \begin{pmatrix} 1 & \tau_1 \\ \tau_1 & |\tau|^2 \end{pmatrix}\ ,
\ee
with $\rho\in \cH_1$, $\tau\in\cH_1$, $V\in\IR^+$, $(u_1,u_2)\in  \IR^4$,  and 
$t_1$ is a two-by-two symmetric matrix. The stabilizer of the cusp $V\to 0$ is 
$[Sp(2,\IZ)\times GL(2,\IZ)]/\IZ_2 \ltimes \IZ^4\ltimes \IZ^3$, where the first two factors
act by fractional linear transformations on $\rho$ and $\tau$, respectively, and the 
last two factors by integer translations of $(u_1,u_2,t_1)$. Therefore,  in the region $V\to 0$ the fundamental
domain $\cF_3$ reduces to $\IR^+\times \cF_1 \times (\cF_1/\IZ_2) \times 
[-\tfrac12,\frac12]^4/\IZ_2 \times 
[-\tfrac12,\frac12]^3$. In this domain, the matrix $t_2$ can be understood
as the period matrix of a two-loop sunset diagram (see Figure \ref{fig_gen3}),
\be
t_2 = \begin{pmatrix} L'_1+L'_2 & - L'_2 \\ - L'_2 & L'_3+ L'_2 \end{pmatrix}\ ,\quad 0< L_2' < L_1' < L_3'\ .
\ee
 The locus $\tau_1=0$ corresponds to the separating degeneration limit for the two-loop
sunset subdiagram.
In the coordinates \eqref{coorsp2gl2}, the integration measure
becomes
\be
\label{mu3asII}
\de\mu_3(\Omega)= 2\, V^4 \de V \,\de\mu_1(\tau)\, \de\mu_1(\rho)\, \de^3 t_1\, \de^2 u_1 \de^2 u_2\ .
\ee

Finally, the third source of divergences is the region III where all entries in  $\Omega_2$ scale to infinity, keeping $\Omega_1$ fixed, corresponding to primitive three-loop divergences. In this region it is useful to parametrize
\be
\Omega = \Omega_1 + \I \cV^{-1}\, \hat\Omega_2\,,
\ee
where $\hat\Omega_2$ is an element of $SL(3)/SO(3)$ in Iwasawa parametrization, 
\be
\Omega_2 =  \cV^{-1}\,
\begin{pmatrix} \frac{ 1}{  L^2} & \frac{A_1}{L^2} & \frac{A_2}{L^2}\\ 
\frac{A_1}{L^2} &\frac{L}{\tau_2} + \frac{A_1^2}{L^2} &\frac{L\tau_1}{\tau_2}+\frac{A_1 A_2}{L^2} \\ 
\frac{A_2}{L^2} &\frac{L\tau_1}{\tau_2}+\frac{A_1 A_2}{L^2}  &\frac{L |\tau|^2}{\tau_2} + \frac{A_2^2}{L^2}
\end{pmatrix}  \,.
\ee
The stabilizer of the cusp is the parabolic subgroup $PGL(3,\IZ)\ltimes \IZ^6$, where the second factor acts by integer translation of the entries in $\Omega_1$.  Therefore, in the region $\cV\to 0$,
 the fundamental domain $\cF_3$ reduces to $\IR^+\times \hat\cF_{PGL(3,\IZ)} \times [-\tfrac12,\tfrac12]^6 $, where $\hat\cF_{PGL(3,\IZ)}=\cF_{PGL(3,\IZ)}\cap \{|\Omega_2|=1\}$ is a
fundamental domain of the action of $PGL(3,\IZ)$ on the space of unimodular positive definite matrices. 
The integration measure in these variables is 
\be
\label{mu3asIII}
\de\mu_3(\Omega) = 6\, \cV^5 \de \cV\,\de\hat\mu_3\,  \de^6\Omega_1\ ,
\ee
where 
\be
\label{defhmu3}
\de\hat\mu_3=
\frac{\de L}{ L^4} \frac{\de \tau_1 \de \tau_2}{\tau_2^2}\ \de A_1 \de A_2 \,,
\ee
is the invariant measure on $\hat\cF_{PGL(3,\IZ)}$, normalized so that 
$\int_{\hat\cF_{PGL(3,\IZ)}} \de\hat\mu_3 = \frac12 \zetastar(2) \zetastar(3)$ \cite{zbMATH02628371}. 

The region III itself admits two higher-codimension cusps, corresponding to i) $Y_{11}\ll Y_{22} \sim Y_{33}$, and ii) $Y_{11}\sim Y_{22} \ll Y_{33}$. The first cusp corresponds to $L\to\infty$ keeping $\tau$ fixed. In this limit, $PGL(3,\IZ)$ is broken to $GL(2,\IZ)\ltimes \IZ^2$, so $\hat\cF_{PGL(3,\IZ)}$ reduces to 
$\IR^+ \times (\cF_1/\IZ_2) \times [0,1]^2/\IZ_2$, where the three factors correspond to
the variables $L, \tau, (A_1,A_2)$. In order to study
the cusp ii), it is useful to change variables to
\be
\label{LtoLp}
L' = \sqrt{L\tau_2}\ ,\quad \tau'=-A_1+\I \sqrt{\frac{L^3}{\tau_2}}\ ,\quad 
A_1'=-\tau_1\ ,\quad A_2'=A_1 \tau_1-A_2\ .
\ee
The measure \eqref{defhmu3} takes the same form in primed coordinates, reflecting the fact that \eqref{LtoLp} acts as an outer automorphism $\hat\Omega_2 \mapsto \sigma \hat\Omega_2^{-1} \sigma^T$, with $\sigma$ being the permutation matrix ${\scriptsize \begin{pmatrix} 0 & 0 & 1 \\
0 & 1 & 0 \\ 1 & 0 & 0\end{pmatrix}}$. The second cusp then corresponds to $L'\to\infty$ keeping 
$\tau'$ fixed, so that $\hat\cF_{PGL(3,\IZ)}$ reduces again to 
$\IR^+ \times (\cF_1/\IZ_2) \times [0,1]^2/\IZ_2$, where the three factors correspond to
the variables $L', \tau', (A_1',A_2')$. These two cusps intersect when 
$Y_{11}\ll Y_{22} \ll Y_{33}$, corresponding to $L\gg \tau_2^{1/3}\gg 1$, or equivalently
$L'\gg \tau'^{1/3}_2\gg 1$.

 In addition, region III contains loci where one  the Schwinger parameters $L_i, R_i$ is scaled to zero:
\be
\label{LRshrink}
\begin{array}{ll}
L_1\to 0:  \, A_2\to 0\, , \qquad\qquad &
R_1\to 0: \, A_1(1+A_1+A_2) + \frac{L^3}{\tau_2}(1+\tau_1)=0\\
L_2\to 0: \, A_1\to 0\, ,\qquad\qquad & 
R_2\to 0:  \, A_2(1+A_1+A_2) + \frac{L^3}{\tau_2}(|\tau|^2+\tau_1)=0\\
L_3\to 0: \, A_1 A_2 + \frac{L^3 \tau_1}{\tau_2}\to 0\, , \qquad\qquad &
R_3\to 0: \, A_1+A_2=1
\end{array}
\ee
These loci all correspond to the descendent of the tetrahedron diagram  in Figure \ref{figdeg}. They
are also obtained by degenerating the three-loop ladder diagram on the top row. The latter being two-particle reducible, it corresponds to a double separating degeneration of the Riemann
surface. According to our assumptions about the integrand, the loci \eqref{LRshrink} do not generate further
infrared divergences. Intersections of these loci generate further degenerations
of the  tetrahedron diagram, which do not contribute to infrared divergences.

In order to regulate all divergences, it is therefore be sufficient to enforce a cut-off on the largest element in \eqref{LRorder}. More conveniently, we shall define the truncated fundamental domain as 
\be
\cF_3^{\Lambda} = \cF_3 \cap \{ t<\Lambda \}\ ,
\ee
where $t$ is the variable defined in \eqref{coorsp3sp2}, and define the regularized integral by
\be
\label{defR3reg}
\cR^{\star,\Lambda}_3(s) = \int_{\cF_3^{\Lambda}} \de\mu_3\, \cE^\star_3(s,\Omega) \, F(\Omega)\ .
\ee
In order to disentangle overlapping divergences and extract the polynomial dependence on $\Lambda$, we shall further split the integration domain into 4 regions, schematically
\be
\label{defF3split}
\begin{split}
\cF_3^{0} =& \cF_3 \cap \{ Y_{11} \leq Y_{22} \leq Y_{33} \leq \Lambda_1\} \,, \\
\cF_3^{I} =& \cF_3 \cap \{ Y_{11} \leq Y_{22} \leq \Lambda_1 \leq Y_{33} \leq \Lambda\} \,, \\
\cF_3^{II} =& \cF_3 \cap \{ Y_{11} \leq\Lambda_1  \leq Y_{22}  \leq Y_{33} \leq \Lambda\} \,,\\
\cF_3^{III} =& \cF_3 \cap \{ \Lambda_1 \leq  Y_{11} \leq Y_{22} \leq Y_{33} \leq \Lambda\} \,.
\end{split} 
\ee
In region I, $t$ is therefore bounded by $\Lambda$, while $\tilde\Omega$ lies in $\cF_2^{\Lambda_1}$. In region II, 
\be
\rho\in \cF_1^{\Lambda_1}\ ,\quad \tau\in\cF_1^{\sqrt{\Lambda/\Lambda_1}}\ ,\quad
\frac{\tau_2}{\Lambda} < V < \frac{1}{\tau_2\Lambda_1}\ .
\ee
Finally, in region III, for a purely imaginary diagonal period matrix, the range of the variables $\cV, L,\tau_2$ is given by 
\be
1\leq \tau_2 \leq \sqrt{\frac{\Lambda}{\Lambda_1}}\ ,\quad 
\tau_2 < L^3 < \frac{\Lambda}{\tau_2\Lambda_1}\ ,\quad 
\frac{L\tau_2}{\Lambda} < \cV < \frac{1}{\Lambda_1 L^2}\ .
\ee
Clearly, the dependence on $\Lambda_1$  cancels when summing over all regions 0, I, II, III. We shall, henceforth, explicitly display only $\Lambda$ dependent terms.

\subsection{Renormalizing the integral}

Having defined the regularized integral \eqref{defR3reg}, we shall now define the renormalized integral \eqref{DefRS3} by subtracting the terms which diverge as the cut-off is removed. For this we need to make assumptions on the behavior of the integrand near the cusps. First, we recall the behavior of
the Eisenstein series $\cE^\star_3(s,\Omega)$ in region I,
\be
\begin{split}
\cE^\star_3(s,\Omega) \to t^s\, \cE^\star_2(s,\tilde\Omega) + t^{2-s}\, \cE^\star_2(s-\tfrac12,\tilde\Omega)\ ,
\end{split} 
\ee
in region II, 
\be
\begin{split}
\cE^\star_3(s,\Omega) \to &  V^{-2s}\, \zetastar(4s-2)\, \cE^\star_1(s,\rho) 
+ V^{2s-4}\, \zetastar(4s-5)\, \cE^\star_1(s-1,\rho) \\
& +V^{-3/2}\, \cE^{\star}_1(2s-\tfrac32,\tau)\, \cE^\star_1(s-\tfrac12,\rho)\,,
\end{split}
\ee
and finally in region III,
\be
\begin{split}
\cE^\star_3(s,\Omega) \to & \zetastar(2s) \zetastar(4s-2) \, \cV^{-3s}
+ \zetastar(2s-3) \zetastar(4s-5) \,\cV^{3s-6}\\
& +\zetastar(2s-1)\, \cV^{-s-1}
\cE^{\star;SL(3,\IZ)}_{V}\left( 2s-1,\hat\Omega_2 \right) \\
&+\zetastar(2s-2)\, \cV^{s-3}
\cE^{\star;SL(3,\IZ)}_{\Lambda^2 V}\left( 2s-\tfrac32 ,\hat\Omega_2\right) \,,
\end{split}
\ee
where $\cE^{\star;SL(3,\IZ)}_{V}\left(s', \hat\Omega_2\right)$ and $\cE^{\star;SL(3,\IZ)}_{\Lambda^2 V}\left( s',\hat\Omega_2\right)$ are Eisenstein series for $SL(3,\IZ)$, attached to the fundamental and
anti-fundamental representations, respectively. They are meromorphic functions of $s'$, with a simple pole at $s'=0$ and $s'=\tfrac32$, and are exchanged under $s'\mapsto \tfrac32-s'$.

We shall assume that similarly to the Eisenstein series $\cE^\star_3$, the function $F(\Omega)$ is regular in the bulk of the Siegel upper-half plane,
but has at most polynomial growth near the cusps. Namely,  we assume that $F$ behaves in region I as
\be
\label{Fgrowth31}
F(\Omega) = \varphi+ \cO(t^{-N})\qquad \forall N>0\ , \quad
\varphi=\sum_{i=1}^{\ell} t^{\sigma_i}\, \varphi_i(\tilde\Omega)\ , 
\ee
where $\varphi_i(\tilde\Omega)$ is a Siegel modular function of degree 2, satisfying the same assumptions as $F(\Omega)$ in \eqref{Fgrowth21} and \eqref{Fgrowth22}; 
in region II, as
\be
\label{Fgrowth32}
F(\Omega) = \tilde\varphi+ \cO(V^N)\qquad \forall N>0\ , \quad
\tilde\varphi=\sum_{j=1}^{\tilde \ell} \, V^{\alpha_j} \tilde\varphi_j(\tau)\, \tilde\varphi'_j(\rho) \,,
\ee
where $ \tilde\varphi_j(\tau),  \tilde\varphi'_j(\rho)$ are modular functions under $PGL(2,\IZ)$ and  $SL(2,\IZ)$ with polynomial growth at the cusps $\tau\to\I\infty$ and $\rho\to\I\infty$, 
with $\tilde\varphi_j(\tau)$ smooth on the separating degeneration locus $\tau_1=0$ and on its images under $PGL(2,\IZ)$;
and finally in region III as
\be
\label{Fgrowth33}
F(\Omega) = \hat\varphi+ \cO(\cV^N)\qquad \forall N>0\ , \quad
\hat\varphi=\sum_{k=1}^{\hat \ell} \, \cV^{\gamma_k} \,\hat\varphi_k(\hat\Omega_2) \,,
\ee
where $\hat\varphi_k(\hat\Omega_2)$ is a modular function  on $\cF_{PGL(3,\IZ)}$, with polynomial growth at the cusps $L\to\infty$ and $L'\to\infty$, and regular at the loci \eqref{LRshrink} and their images under $PGL(3,\IZ)$. Of course, the expansions of $\varphi_i, \tilde\varphi_j$ and 
$\hat\varphi_k$ at the respective cusps must agree whenever the regimes of validity overlap.

Under these assumptions on the asymptotic behavior of $F$, it is now straightforward to extract
the leading dependence of the regularized integral \eqref{defR3reg} on $\Lambda$ as $\Lambda\to\infty$. From region I, we find
\be
\label{RS3LambdaregI}
\begin{split}
 \int_{\cF_3^{\Lambda,I}}\de\mu_3\, \cE^\star_3(s) \, F(\Omega) \sim & \sum_{i=1}^{\ell}
  \frac12\int_{\Lambda_1}^{\Lambda} \frac{\de t}{t^4} 
 \int_{\cF_{2}^{\Lambda_1}} \de\mu_2
 \left[ t^{s+\sigma_i}\, \cE^\star_2(s,\tilde\Omega) + t^{2-s+\sigma_i}\, \cE^\star_2(s-\tfrac12,
 \tilde\Omega) \right] \varphi_i(\tilde\Omega) \\
 =&  \frac12 \sum_{i=1}^{\ell}  \frac{\Lambda^{s+\sigma_i-3}}{s+\sigma_i-3} \, 
 \left( \RN
  \int_{\cF_{2}} \de\mu_2\, \cE^\star_2(s,\tilde\Omega)  \, \varphi_i(\tilde\Omega)+\dots \right) \\
   +&   \frac12 \sum_{i=1}^{\ell}  \frac{ \Lambda^{\sigma_i-1-s}}{\sigma_i-1-s} \, \left( \RN
  \int_{\cF_{2}} \de\mu_2\, \cE^\star_2(s-\tfrac12,\tilde\Omega)  
  \, \varphi_i(\tilde\Omega) + \dots\right)\,,
 \end{split} 
\ee
where the dots stand for $\Lambda_1$-dependent terms, which will cancel against contributions from region II and III. From region II, we find
\be
\label{RS3LambdaregII}
\begin{split}
 \int_{\cF_3^{\Lambda,II}} & \de\mu_3\, \cE^\star_3(s) \,F(\Omega) \sim \frac{1}{2}\sum_{j=1}^{\tilde \ell}
 \int_{\cF_1^{\Lambda_1}} \de\mu_1(\rho)\,
  \int_{\cF_1^{\sqrt{\Lambda/\Lambda_1}}} \de\mu_1(\tau)\
  \int_{\tau_2/\Lambda}^{1/(\tau_2 \Lambda_1)} \de V \, V^{4+\alpha_j} \, 
  \\ &\tilde\varphi_j(\tau)\, \tilde\varphi'_j(\rho) \,
  \left[ V^{-2s}\, \zetastar(4s-2)\, \cE^\star_1(s,\rho) 
+ V^{2s-4}\, \zetastar(4s-5)\, \cE^\star_1(s-1,\rho) \right.
\\ & \left.
+V^{-\tfrac32}\, \cE^{\star}_1(2s-\tfrac32,\tau) \, \cE^\star_1(s-\tfrac12,\rho) \right]\,  
\\
& =  \frac{1}{2}\sum_{j=1}^{\tilde \ell} \left[ 
\frac{\zetastar(4s-2)\, \Lambda^{2s-5-\alpha_j}}
{2s-5-\alpha_j} 
\left( \RN\,  \int_{\cF_1} \de\mu_1 \, \tau_2^{5+\alpha_j-2s}\tilde\varphi_j(\tau)\right)\, 
\left( \RN\,  \int_{\cF_1} \de\mu_1 \, \cE^\star_1(s)\, \tilde\varphi'_j(\rho)   \right)
\right.
 \\
 & -
 \frac{ \zetastar(4s-5)\, \Lambda^{-\alpha_j-1-2s}}{2s+\alpha_j+1}
 \left( \RN\,  \int_{\cF_1} \de\mu_1 \, \tau_2^{1+\alpha_j+2s}\tilde\varphi_j(\tau)\right)\, 
\left( \RN\,  \int_{\cF_1} \de\mu_1 \, \cE^\star_1(s-1)\tilde\varphi'_j(\rho)   \right)
 \\
 &
 \left. -
 \frac{\Lambda^{-\alpha_j-\tfrac72}}{\alpha_j+\tfrac72} \,
\left( \RN\,  \int_{\cF_1} \de\mu_1 \, \tau_2^{\alpha_j+\tfrac72} \cE^\star_1(2s-\tfrac32 )
\, \tilde\varphi_j(\tau)   \right)
\left( \RN\,  \int_{\cF_1} \de\mu_1 \, \cE^\star_1(s-\tfrac12) \, \tilde\varphi'_j(\rho)   \right) \right]+\dots
 \end{split} 
\ee
Finally, in region III,
\be
\label{RS3LambdaregIII}
\begin{split}
 \int_{\cF_3^{\Lambda,III}} & \de\mu_3\, \cE^\star_3(s) \,F(\Omega) \sim 6\sum_{k=1}^{\hat \ell}
 \int_{\hat \cF_{PGL(3,\IZ)}} \de\hat\mu_3\, \int_{L\tau_2/\Lambda}^{\infty} \cV^{5+\gamma_k} \de \cV \,  \hat\varphi_k(\hat\Omega_2)  \\
 & 
 \left[  \zetastar(2s) \zetastar(4s-2) \, \cV^{-3s}  
+ \zetastar(2s-3) \zetastar(4s-5) \,\cV^{3s-6}\right. \\
& \left. +\zetastar(2s-1)\, \cV^{-s-1}
\cE^{\star;SL(3,\IZ)}_{V}\left( 2s-1 \right) +\zetastar(2s-2)\, \cV^{s-3}
\cE^{\star;SL(3,\IZ)}_{\Lambda^2 V}\left( 2s-\tfrac32 \right) \right]
\\
=& 6\sum_{k=1}^{\hat \ell} 
\left[ \frac{ \zetastar(2s)\, \zetastar(4s-2)\, \Lambda^{3s-\gamma_k-6}}{3s-\gamma_k-6}
\RN \int_{\hat \cF_{PGL(3,\IZ)}} \de\hat\mu_3\, (L\tau_2)^{6+\gamma_k-3s}\, \hat\varphi_k(\hat\Omega_2)
\right.\\
& -  \frac{ \zetastar(2s-3)\, \zetastar(4s-5)\, \Lambda^{-3s-\gamma_k}}{3s+\gamma_k}
\RN \int_{\hat \cF_{PGL(3,\IZ)}} \de\hat\mu_3\, (L\tau_2)^{\gamma_k+3s}\, \hat\varphi_k(\hat\Omega_2)
\\
& +  \frac{ \zetastar(2s-1)\, \Lambda^{s-5-\gamma_k}}{s-5-\gamma_k}
\RN \int_{\hat \cF_{PGL(3,\IZ)}} \de\hat\mu_3\, (L\tau_2)^{5+\gamma_k-s}\, 
\cE^{\star;SL(3,\IZ)}_{V}\left( 2s-1 \right) \, \hat\varphi_k(\hat\Omega_2)\,
\\
& \left. -  \frac{ \zetastar(2s-2)\, \Lambda^{-s-3-\gamma_k}}{s+3+\gamma_k}
\RN \int_{\hat \cF_{PGL(3,\IZ)}} \de\hat\mu_3\, (L\tau_2)^{s+3+\gamma_k}\, 
\cE^{\star;SL(3,\IZ)}_{\Lambda^2 V}\left( 2s-\tfrac32 \right) \, \hat\varphi_k(\hat\Omega_2) \right] + \dots
\end{split}
\ee
We therefore define the renormalized integral by subtracting the divergent terms,
\be
\label{RS3Lambdaregtot}
\begin{split}
&\cR_3^\star(F,s) = \lim_{\Lambda\to\infty} \Biggr\{ 
\int_{\cF_3^{\Lambda}} \de\mu_3\, \cE^\star_3(s) \,F(\Omega) 
 \\
&
 - \frac12 \sum_{i=1}^{\ell} 
\left[  
  \frac{\Lambda^{s+\sigma_i-3}}{s+\sigma_i-3} \, 
 \RN  \int_{\cF_{2}} \de\mu_2\, \cE^\star_2(s)  \, \varphi_i(\tilde\Omega)+  \frac{ \Lambda^{\sigma_i-1-s}}{\sigma_i-1-s} \,  
   \RN  \int_{\cF_{2}} \de\mu_2\, \cE^\star_2(s-\tfrac12)  \, \varphi_i(\tilde\Omega) \right] 
\\
- &  \frac{1}{2}\sum_{j=1}^{\tilde \ell} \left[ 
\frac{\zetastar(4s-2)\, \Lambda^{2s-5-\alpha_j}}
{2s-5-\alpha_j} 
\left( \RN\,  \int_{\cF_1} \de\mu_1 \, \tau_2^{5+\alpha_j-2s}\tilde\varphi_j(\tau)\right)\, 
\left( \RN\,  \int_{\cF_1} \de\mu_1 \, \cE^\star_1(s)\tilde\varphi'_j(\rho)   \right)
\right.
 \\
 & -
 \frac{ \zetastar(4s-5)\, \Lambda^{-\alpha_j-1-2s}}{2s+\alpha_j+1}
 \left( \RN\,  \int_{\cF_1} \de\mu_1 \, \tau_2^{1+\alpha_j+2s}\tilde\varphi_j(\tau)\right)\, 
\left( \RN\,  \int_{\cF_1} \de\mu_1 \, \cE^\star_1(s-1)\tilde\varphi'_j(\rho)   \right)
 \\
 &
 \left. -
 \frac{\Lambda^{-\alpha_j-\tfrac72}}{\alpha_j+\tfrac72} \,
\left( \RN\,  \int_{\cF_1} \de\mu_1 \, \tau_2^{\alpha_j+\tfrac72} \cE^\star_1(2s-\tfrac32 )
\, \tilde\varphi_j(\tau)   \right)
\left( \RN\,  \int_{\cF_1} \de\mu_1 \, \cE^\star_1(s-\tfrac12) \, \tilde\varphi'_j(\rho)   \right) \right]
\\
-& 6\sum_{k=1}^{\hat \ell} 
\left[ \frac{ \zetastar(2s)\, \zetastar(4s-2)\, \Lambda^{3s-\gamma_k-6}}{3s-\gamma_k-6}
\RN \int_{\hat \cF_{PGL(3,\IZ)}} \de\hat\mu_3\, (L\tau_2)^{6+\gamma_k-3s}\, \hat\varphi_k(\hat\Omega_2)
\right.\\
& -  \frac{ \zetastar(2s-3)\, \zetastar(4s-5)\, \Lambda^{-3s-\gamma_k}}{3s+\gamma_k}
\RN \int_{\hat \cF_{PGL(3,\IZ)}} \de\hat\mu_3\, (L\tau_2)^{\gamma_k+3s}\, \hat\varphi_k(\hat\Omega_2)
\\
& +  \frac{ \zetastar(2s-1)\, \Lambda^{s-5-\gamma_k}}{s-5-\gamma_k}
\RN \int_{\hat \cF_{PGL(3,\IZ)}} \de\hat\mu_3\, (L\tau_2)^{5+\gamma_k-s}\, 
\cE^{\star;SL(3,\IZ)}_{V}\left( 2s-1 \right) \, \hat\varphi_k(\hat\Omega_2)\,
\\
& \left. -  \frac{ \zetastar(2s-2)\, \Lambda^{-s-3-\gamma_k}}{s+3+\gamma_k}
\RN \int_{\hat \cF_{PGL(3,\IZ)}} \de\hat\mu_3\, (L\tau_2)^{s+3+\gamma_k}\, 
\cE^{\star;SL(3,\IZ)}_{\Lambda^2 V}\left( 2s-\tfrac32 \right) \, \hat\varphi_k(\hat\Omega_2) \right]  \Biggr\}\,.
\end{split}
\ee

\subsection{Constructing $\Diamond_3$}

In this subsection, we construct an invariant differential operator annihilating
the non-decaying part of $F$, in such a way that the right-hand side of \eqref{RS3LambdaregIII},
at finite value of $\Lambda$, is recognized as $\int_{\cF_3^{\Lambda}} \de\mu_3\, \cE^\star_3(s)\,\Diamond F$, up to a polynomial in $s$. 

\subsubsection*{Region I}

In analogy with \eqref{Diamond2sigma} we introduce the invariant differential operator
\be
\label{Diamond3sigma}
\begin{split}
\Diamond_3(\sigma)=& \Delta^{(6)}_{Sp(6)} - (\sigma-1)(\sigma-2) \Delta^{(4)}_{Sp(6)} 
+(\sigma-\tfrac12)(\sigma-1)(\sigma-2)(\sigma-\tfrac52) \Delta_{Sp(6)} 
\\ &-
\sigma(\sigma-\tfrac12)(\sigma-1)(\sigma-2)(\sigma-\tfrac52)(\sigma-3)\ .
\end{split}
\ee
where $ \Delta_{Sp(6)}, \Delta^{(4)}_{Sp(6)},  \Delta^{(6)}_{Sp(6)}$ are the quadratic, quartic and sextic Casimir operators of $Sp(6)$ (see Appendix \ref{app_gendegree}), normalized such that
\be
\begin{split}
\Delta_{Sp(6)}\, \cE^\star_3(s) =& 3s(s-2)\,  \cE^\star_3(s)\ ,
\\
\Delta^{(4)}_{Sp(6)} \, \cE^\star_3(s) = &3 s(s-\tfrac12)(s-\tfrac32)(s-2)\,  \cE^\star_3(s)\ ,
\\
\Delta^{(6)}_{Sp(6)} \, \cE^\star_3(s) =  &
s(s-\tfrac12)(s-1)^2(s-\tfrac32)(s-2)\, \cE^\star_3(s) \ .
\end{split} 
\ee
In terms of the coordinates $t,\tilde\Omega, \tilde u_1, \tilde u_2, t_1$ appropriate to region I,
\be
\label{DeltaSp6asI}
\begin{split}
\Delta_{Sp(6)} \to & \Delta_t + \Delta_{Sp(4)}\ ,
\\
\Delta^{(4)}_{Sp(6)} \to &\Delta^{(4)}_{Sp(4)} + ( \Delta_t + \tfrac{5}{4})\, \Delta_{Sp(4)}\ ,
\\
\Delta^{(6)}_{Sp(6)} \to & (\Delta_t+2)\, \Delta^{(4)}_{Sp(4)} \ .
\end{split}
\ee
where $\Delta_t=t^2 \pa^2_t -2 t \pa_t$. Using this, one may check that the operator 
$\Diamond_3(\sigma)$ annihilates $t^\sigma\, \varphi(\tilde\Omega)$ for any 
function $\varphi(\tilde\Omega)$, independent of $\tilde u_1, \tilde u_2, t_1$. The
Eisenstein series $\cE^\star_3(s)$ is an eigenmode of $\Diamond_3(\sigma)$
for any $\sigma$,
\be
\label{DiamondE31}
\Diamond_3(\sigma)\, \cE^\star_3(s) =  (s-\sigma)\,(s-\sigma+\tfrac{1}{2})\, (s-\sigma+1)\, 
(s+\sigma-2)\, (s+\sigma-\tfrac{5}{2})\, (s+\sigma-3)\,  \cE^\star_3(s) \ .
\ee

\subsubsection*{Region II}

In the coordinates appropriate to region II, we have instead
\be
\label{DeltaSp6asII}
\begin{split}
\Delta_{Sp(6)}\to  & \frac12 \Delta_V + \Delta_\rho + \frac12\Delta_\tau \ ,
\\
\Delta^{(4)}_{Sp(6)}\to  & \frac{1}{16}\left[ 
8 \Delta_{\rho} (\Delta_{\tau}+ \Delta_{V})+32 \Delta_{\rho}
+\Delta_{\tau}^2-2\Delta_{\tau} \Delta_{V}-20 \Delta_{\tau}+\Delta_{V}^2+4 \Delta_{V}\right]\ ,
\\
\Delta^{(6)}_{Sp(6)}\to &  \frac{1}{16}\left[-2 \Delta_\rho \Delta_\tau \Delta_V + 
(\Delta_\tau^2+\Delta_V^2)\Delta_\rho +  (10\Delta_V-7\Delta_\tau+24)\Delta_\rho \right]\ ,
\end{split}
\ee
with $\Delta_V =  V^2 \pa_V^2 + 6 V \pa_V$. Using these formulae, one can check that the
following degree 12 invariant operator
\be
\begin{split}
\widetilde{\Diamond}_3(\alpha) =&
   \left(-4 \Delta_{Sp(6)}^3 {\Delta^{(6)}_{Sp(6)}}+\Delta_{Sp(6)}^2 [\Delta^{(4)}_{Sp(6)}]^2
   +18 \Delta_{Sp(6)} {\Delta^{(4)}_{Sp(6)}}
   {\Delta^{(6)}_{Sp(6)}}-4 [\Delta^{(4)}_{Sp(6)}]^3-27 [\Delta^{(6)}_{Sp(6)}]^2\right)\\
   &+ \left(-(2 \alpha  (\alpha +5)+11)
   \Delta_{Sp(6)}^3 {\Delta^{(4)}_{Sp(6)}}+3 (6 \alpha  (\alpha +5)+31) \Delta_{Sp(6)}^2 {\Delta^{(6)}_{Sp(6)}}
   \right. \\
   &\left. +2 (3 \alpha   (\alpha +5)+17) \Delta_{Sp(6)} [\Delta^{(4)}_{Sp(6)}]^2 
   -18 (3 \alpha  (\alpha +5)+16) {\Delta^{(4)}_{Sp(6)}}
   {\Delta^{(6)}_{Sp(6)}}\right)\\
   &+\frac{1}{4} \left(4 (\alpha +1) (\alpha +2) (\alpha +3) (\alpha +4)
   \Delta_{Sp(6)}^4 \right.\\
   &\left. 
   +(2 \alpha  (\alpha +5) (4 \alpha  (\alpha +5)+47)+293) \Delta_{Sp(6)}^2 {\Delta^{(4)}_{Sp(6)}}\right.\\
   &\left. -6
   (\alpha  (\alpha +5) (16 \alpha  (\alpha +5)+161)+399) \Delta_{Sp(6)} {\Delta^{(6)}_{Sp(6)}}\right.\\
   &\left. -4 (\alpha 
   (\alpha +5) (7 \alpha  (\alpha +5)+80)+237) [\Delta^{(4)}_{Sp(6)}]^2\right) \nn
   \end{split}
   \ee
   \be
   \begin{split}
   &+\frac{1}{2}
   \left(-(\alpha +1) (\alpha +2) (\alpha +3) (\alpha +4) (8 \alpha  (\alpha +5)+37)
   \Delta_{Sp(6)}^3\right.\\
   &\left. +(\alpha  (\alpha +5) (4 \alpha  (\alpha +5) (\alpha  (\alpha +5)+14)+265)+447)
   \Delta_{Sp(6)} {\Delta^{(4)}_{Sp(6)}}\right.\\
   &\left. +(\alpha +2) (\alpha +3) (\alpha  (\alpha +5) (52 \alpha  (\alpha
   +5)+495)+1134) {\Delta^{(6)}_{Sp(6)}}\right)\\
   &+\frac{1}{16} (\alpha +2) (\alpha +3)  \left((\alpha +1)
   (\alpha +4) (8 \alpha  (\alpha +5) (12 \alpha  (\alpha +5)+101)+1549) \Delta_{Sp(6)}^2\right.\\
   &\left. -8 (\alpha
    (\alpha +5) (\alpha  (\alpha +5) (4 \alpha  (\alpha +5)+49)+216)+378)
   {\Delta^{(4)}_{Sp(6)}}\right)\\
   &-\frac{1}{8} ((\alpha +1) (\alpha +2) (\alpha +3) (\alpha +4) (\alpha 
   (\alpha +5) (4 \alpha  (\alpha +5) (8 \alpha  (\alpha +5)+91)+1197)+945)
   \Delta_{Sp(6)})\\
   &+\frac{1}{16} \alpha  (\alpha +1) (\alpha +2)^2 (\alpha +3)^2 (\alpha +4) (\alpha
   +5) (2 \alpha +1) (2 \alpha +3) (2 \alpha +7) (2 \alpha
   +9) \,,
\end{split}
\ee
annihilates $V^{\alpha} \tilde\varphi(\tau)\, \tilde\varphi'(\rho)$ for arbitrary  
$\tilde\varphi(\tau), \tilde\varphi'(\rho)$. The
Eisenstein series $\cE^\star_3(s)$ is an eigenmode of $\widetilde{\Diamond}_3(\alpha)$
for all $\alpha$,
\be
\label{DiamondE32}
\begin{split}
\widetilde{\Diamond}_3(\alpha) \cE^\star_3(s) =& \frac{1}{16} (\alpha +2)^2 (\alpha +3)^2 
(2 \alpha +3) (2 \alpha +7) (2 \alpha -4 s+9) (2 \alpha +4 s+1) \\
& \times (\alpha -2 s+5) (\alpha -2 s+4) (\alpha +2 s) 
(\alpha +2 s+1) \,\cE^\star_3(s)\ .
\end{split}
\ee

\subsubsection*{Region III}

Finally, in the coordinates appropriate to region III,
\be
\label{DeltaSp6asIII}
\begin{split}
\Delta_{Sp(6)} \to & \frac13 \Delta_{\cV} +\frac12 \Delta_{SL(3)}\ ,
\\
\Delta^{(4)}_{Sp(6)} \to & \frac1{27} \Delta_{\cV}^2 +\frac1{4}  \Delta_{\cV}+
\frac{1}{16}(\Delta_{SL(3)})^2+\frac34(1+\cV\partial_\cV) \Delta^{(3)}_{SL(3)} 
-\frac12 \Delta_{SL(3)}\ ,
\\
\Delta^{(6)}_{Sp(6)} \to &
\frac{1}{46656}\left( 27  \Delta^{(3)}_{SL(3)}  + 9  (3+2\cV \partial_\cV)\Delta_{SL(3)} 
-4 \cV \partial_\cV (3+\cV \partial_\cV)(3+2\cV \partial_\cV)\right)\\
&\times\left( 27  \Delta^{(3)}_{SL(3)}  + 9  (9+2\cV \partial_\cV)\Delta_{SL(3)} 
-4 (3+\cV \partial_\cV) (6+\cV \partial_\cV)(9+2\cV \partial_\cV)\right) \ ,
\end{split}
\ee
where $\Delta_\cV=\cV^2 \pa^2_{\cV} + 7 \cV\pa_{\cV}$, and $\Delta_{SL(3)} , \Delta^{(3)}_{SL(3)}$
denote the quadratic and cubic Casimirs on $SL(3)/SO(3)$, normalized such that 
the two-parameter Langlands-Eisenstein series 
satisfies \cite{Pioline:2009qt}
\be
\begin{split}
{\Delta_{SL(3)} \cE^{\star;SL(3,\IZ)}(s_1,s_2)} =&\left[ \frac43(s_1^2+s_2^2+s_1 s_2)-2(s_1+s_2) \right]\, {\cE^{\star;SL(3,\IZ)}(s_1,s_2)}
\\
{\Delta^{(3)}_{SL(3)} \cE^{\star;SL(3,\IZ)}(s_1,s_2)} 
=&-\frac{2}{27}(s_1-s_2)(2s_1+4s_2-3)(2s_2+4s_1-3)\, {\cE^{\star;SL(3,\IZ)}(s_1,s_2)} \,.
\end{split}
\ee
Using these formulae, one can check that the
following degree 8 invariant operator
\be
\begin{split}
\widehat{\Diamond}_3(\gamma) =&
\frac{1}{16} \left[\Delta_{Sp(6)}^2-4 {\Delta^{(4)}_{Sp(6)}}\right]^2\\
&+ \left(-\frac{1}{4} (\gamma +2)
   (\gamma +4) \Delta_{Sp(6)}^3+(\gamma +2) (\gamma +4) \Delta_{Sp(6)} {\Delta^{(4)}_{Sp(6)}}-4 (\gamma +3)^2
   {\Delta^{(6)}_{Sp(6)}}\right)\\
&+\frac{1}{8}  \left((\gamma  (\gamma +6) (3 \gamma  (\gamma +6)+43)+146)
   \Delta_{Sp(6)}^2-4 (\gamma +3)^2 (\gamma  (\gamma +6)+4) {\Delta^{(4)}_{Sp(6)}}\right)\\
&-\frac{1}{4} 
  (\gamma +3)^2 (\gamma  (\gamma +6) (\gamma  (\gamma +6)+10)+20)
   \Delta_{Sp(6)}\\
   & +\frac{1}{16} \gamma  (\gamma +1) (\gamma +2) (\gamma +3)^2 (\gamma +4)
   (\gamma +5) (\gamma +6) \,,
\end{split}
\ee
 annihilates $\cV^\gamma \hat\varphi(\hat\Omega_2)$ for any  $\hat\varphi(\hat\Omega_2)$.
 The Eisenstein series $\cE^\star_3(s)$ is an eigenmode of $\widehat{\Diamond}_3(\gamma)$
for all $\alpha$,
\be
\label{DiamondE33}
\begin{split}
\widehat{\Diamond}_3(\gamma)  \cE^\star_3(s) =& \frac{1}{16} (\gamma -3 s+6) (\gamma -s+3) (\gamma -s+4) (\gamma -s+5) \\
&\times (\gamma +s+1) (\gamma
   +s+2) (\gamma +s+3) (\gamma +3 s)
 \,\cE^\star_3(s)\ .
\end{split}
\ee

\subsubsection*{All regions}

Combining these results, we see that $\Diamond=\prod_{i=1}^{\ell} 
\Diamond_3(\sigma_i) \prod_{j=1}^{\tilde \ell} \widetilde{\Diamond}_3(\alpha_j)
\prod_{k=1}^{\hat \ell} \widehat{\Diamond}_3(\gamma_k)$ annihilates the decaying part
of $F$ in all regions. However, this may not be the most economical choice. Indeed,
the symmetry properties $\Diamond_3(\sigma)=\Diamond_3(3-\sigma)$, 
$ \widetilde{\Diamond}_3(\alpha)= \widetilde{\Diamond}_3(-5-\alpha)$,
$\widehat{\Diamond}_3(\gamma)=\widehat{\Diamond}_3(-6-\gamma)$ allow to
keep only one element in each pair, in case the two elements are present. 
Second, the operator $\Diamond_3(\sigma)$
in region II factorizes into 
\be
\begin{split}
\Diamond_3(\sigma)=&
	\tfrac{1}{16}[\Delta_\rho-(\sigma-1)(\sigma-2)]\, \\
	& [\Delta_\tau-\alpha(\alpha-1)+2\sigma(1-2\sigma-2\alpha)]\, [\Delta_\tau-\alpha(\alpha-1)+2(\sigma-3)(5+2\sigma-2\alpha)] \,,
\end{split}
\ee
so it annihilates $V^{\alpha} \tilde\varphi(\tau)\, \tilde\varphi'(\rho)$ whenever one of these
factors vanishes. Similarly, in region III it factorizes into
\be
\begin{split}
\Diamond_3(\sigma)=&
\frac{1}{46656} \left[-4 (2 \gamma -6 \sigma +15) (\gamma -3 \sigma +6) (\gamma -3 \sigma +9)+9 \Delta_{SL(3)} (2
   \gamma -6 \sigma +15)+27 {\Delta^{(3)}_{SL(3)}}\right]\\
   &\times \left[-4 (\gamma +3 \sigma -3) (\gamma +3 \sigma ) (2
   \gamma +6 \sigma -3)+9 \Delta_{SL(3)} (2 \gamma +6 \sigma -3)+27 {\Delta^{(3)}_{SL(3)}}\right] \,,
\end{split}
\ee
so it annihilates $\cV^\gamma \hat\varphi(\hat\Omega_2)$ whenever either of these factors vanishes.

As for the operator  $\widetilde{\Diamond}_3(\alpha)$, it factorizes in region I as 
\be
\begin{split}
&\widetilde{\Diamond}_3(\alpha) = 
\frac{1}{16} \left((\alpha +1) (\alpha +2) (\alpha +3) (\alpha +4)+\Delta_{Sp(4)}^2-(2 \alpha 
   (\alpha +5)+11) {\Delta_{Sp(4)}}-4 {\Delta^{(4)}_{Sp(4)}}\right)\\
  & \times  \left[(2 \alpha -2 \sigma +7) (2 \alpha -2 \sigma
   +9) (\alpha -\sigma +3) (\alpha -\sigma +5)-{\Delta_{Sp(4)}} (2 \alpha -2 \sigma +7) (2 \alpha -2
   \sigma +9)+4 {\Delta^{(4)}_{Sp(4)}}\right] \\
  & \times \left[(\alpha +\sigma ) (\alpha +\sigma +2) (2 \alpha +2 \sigma +1) (2
   \alpha +2 \sigma +3)-{\Delta_{Sp(4)}} (2 \alpha +2 \sigma +1) (2 \alpha +2 \sigma +3)+4
   {\Delta^{(4)}_{Sp(4)}}\right]\,,
   \end{split}
   \ee
and in region III as 
\be
\begin{split}
\widetilde{\Diamond}_3(\alpha) = &
-\frac{1}{2985984}\left(-16 (\alpha +2)^2 (\alpha +3)^2 (2 \alpha +3) (2 \alpha +7)+4 \Delta_{SL(3)}^3-3 (12
   \alpha  (\alpha +5)+71) \Delta_{SL(3)}^2 
   \right.\\
   &\left. +24 (\alpha +2) (\alpha +3) (4 \alpha  (\alpha +5)+23)
   \Delta_{SL(3)}-27 [\Delta^{(3)}_{SL(3)}]^2\right)\\
   &\times  (-4 (6 \alpha -4 \gamma +3) (3 \alpha -2 \gamma ) (3
   \alpha -2 \gamma +3)+9 \Delta_{SL(3)} (6 \alpha -4 \gamma +3)+27 {\Delta^{(3)}_{SL(3)}})\\
   &\times 
   (-4 (3 \alpha +2
   \gamma +12) (3 \alpha +2 \gamma +15) (6 \alpha +4 \gamma +27)+9 \Delta_{SL(3)} (6 \alpha +4
   \gamma +27)-27 {\Delta^{(3)}_{SL(3)}}) \,.
   \end{split}
   \ee
Finally, the operator $\widehat{\Diamond}_3(\gamma)$ factorizes in region I as 
\be
\begin{split}
\widehat{\Diamond}_3(\gamma) =&
\frac{1}{16} \left((\gamma -\sigma +3) (\gamma -\sigma +4) (\gamma -\sigma +5) (\gamma -\sigma
   +6)+[\Delta_{Sp(4)}]^2 
   \right.\\
   &\qquad \left.
  -{\Delta_{Sp(4)}} \left(2 \gamma ^2+\gamma  (18-4 \sigma )+2 (\sigma -9) \sigma
   +39\right)-4 {\Delta^{(4)}_{Sp(4)}}\right)\\
   &\times  \left((\gamma +\sigma ) (\gamma +\sigma +1) (\gamma +\sigma
   +2) (\gamma +\sigma +3)+[\Delta_{Sp(4)}]^2\right.\\
   &\qquad \left. -{\Delta_{Sp(4)}} \left(2 \gamma ^2+\gamma  (4 \sigma +6)+2
   \sigma  (\sigma +3)+3\right)-4 {\Delta^{(4)}_{Sp(4)}}\right) \,,
   \end{split}
   \ee
and in region II as
   \be
   \begin{split}
\widehat{\Diamond}_3(\gamma) =&\frac{1}{16} 
\left[\Delta_{\rho}-(\alpha -\gamma -1) (\alpha -\gamma )\right]\, 
\left[\Delta_{\rho}-(\alpha +\gamma +5) (\alpha +\gamma +6)\right]\\
   &\times \left[\Delta_{\rho}^2-2 \Delta_{\rho} \left((\gamma
   +3)^2+\Delta_{\tau}\right)+(\Delta_{\tau}-(\gamma +2) (\gamma +3)) (\Delta_{\tau}-(\gamma +3)
   (\gamma +4))\right] \,.
   \end{split} 
\ee
To take advantage of these possible simplifications, we shall choose 
\be
\Diamond=\prod_{i\in I}
\Diamond_3(\sigma_i) \prod_{j\in J} \widetilde{\Diamond}_3(\alpha_j)
\prod_{k\in K} \widehat{\Diamond}_3(\gamma_k)
\ee
where $I,J,K$ are suitable subsets of $1\dots \ell$, $1\dots \tilde\ell$, $1\dots \hat\ell$ such that 
$\Diamond$ annihilates the non-decaying parts \eqref{Fgrowth31},\eqref{Fgrowth31},\eqref{Fgrowth33} of $F$ in all regions I, II, III.

Having constructed an operator $\Diamond$ which annihilates the non-decaying part of $F$
for all degenerations, we can now compute the integral $\cR_3^{\Lambda}(\Diamond F)$
using the standard Rankin-Selberg method,
\be
\begin{split}
\cR_3^{\star}(\Diamond F) =&  \lim_{\Lambda\to\infty} \cR_3^{\star,\Lambda}(\Diamond F)\\
=& 
\zetastar(2s) \,  \zetastar(4s-2)\, 
\int_{\Gamma_{\infty,3}\backslash \cH_3} \de\mu_3 \, |\Omega_2|^{s}\, \Diamond F(\Omega)\\
=& 
6\zetastar(2s) \,  \zetastar(4s-2)\, 
\int_0^{\infty} \cV^{5-3s} \de \cV\, \int_{\hat\cF_{PGL(3,\IZ)}} \de\hat\mu_3 \, \Diamond (F^{(0)}
-\tilde\varphi)
\end{split}
\ee
where, in going from the second to the last line, we integrated over $\Omega_1$ and used the fact
that $\Diamond\tilde\varphi=0$. Integrating by parts and using \eqref{DiamondE31},  \eqref{DiamondE32},  \eqref{DiamondE33}, we find
\be
\cR_3^{\star}(\Diamond F) = 6\zetastar(2s) \,  \zetastar(4s-2)\, 
D_3(s)\, 
\int_0^{\infty} \cV^{5-3s} \de \cV\, \int_{\hat\cF_{PGL(3,\IZ}} \de\hat\mu_3 \, (F^{(0)}-\tilde\varphi)
\ee
where
\be
\begin{split}
D_3(s)&=\prod_{i\in I}  (s-\sigma_i)\,(s-\sigma_i+\tfrac{1}{2})\, (s-\sigma_i+1)\, 
(s+\sigma_i-2)\, (s+\sigma_i-\tfrac{5}{2})\, (s+\sigma_i-3)
\\
& \times\prod_{j\in J} \frac{1}{16} (\alpha_j +2)^2 (\alpha_j +3)^2 
(2 \alpha_j +3) (2 \alpha_j +7) (2 \alpha_j -4 s+9) (2 \alpha_j +4 s+1) \\
& \quad (\alpha_j -2 s+5) (\alpha_j -2 s+4) (\alpha_j +2 s) (\alpha_j +2 s+1)
\\
& \times\prod_{k\in K} 
\frac{1}{16}  (\gamma+k -3 s+6) (\gamma_k -s+3) (\gamma_k -s+4) (\gamma_k -s+5) \\
&\quad (\gamma_k +s+1) (\gamma_k+s+2) (\gamma_k +s+3) (\gamma_k +3 s) \,.
\end{split}
\ee
Since $\Diamond F$ is of rapid decay, $\cR_3^{\star}(\Diamond F)$ has a meromorphic continuation in 
$s$ with simple poles at $s\in \{0,\tfrac12,\tfrac32,2\}$, invariant under $s\mapsto 2-s$. 
On the other hand, by integrating by parts in the regularized integral $\cR_3^{\star,\Lambda}(\Diamond F)$ and using \eqref{DiamondE31},\eqref{DiamondE32},\eqref{DiamondE33}, a very tedious computation shows that for finite $\Lambda$, 
\be
\label{DefRS3regfullDiamond}
\begin{split}
\int_{\cF_3^{\Lambda}} \de\mu_2 & \, \cE_3^\star(s,\Omega)\, \Diamond F(\Omega) =
D_3(s)\, \int_{\cF_3^{\Lambda}} \de\mu_3 \, \cE_3^\star(s,\Omega)\, F(\Omega)
\\
-&\frac12 \sum_i\, \frac{D_3(s)\,   \Lambda^{s+\sigma_i-3}}{s+\sigma_i-3}
\RN \int_{\cF_2} \de\mu_2(\tilde\Omega) \,\varphi_i(\tilde\Omega)\,  \cE^\star_2(s,\tilde\Omega) - \dots
\end{split}
\ee
where the dots stand for the divergent remaining terms in \eqref{RS3Lambdaregtot}, multiplied by $D_3(s)$. It follows that the renormalized integral \eqref{RS3Lambdaregtot} is equal to the
renormalized Mellin transform,
\be
\label{RNMellin3}
\boxed{\cR_3^\star(F,s) = 6\, \zetastar(2s) \,  \zetastar(4s-2)\, 
\int_0^{\infty} \cV^{5-3s} \de \cV\, \int_{\hat\cF_{PGL(3,\IZ)}} \de\hat\mu_3  (F^{(0)}-\tilde\varphi)
=\frac{\cR_3^{\star}(\Diamond F)}{D_3(s)}\ .}
\ee
Thus, $\cR_3^\star(F,s)$ has a meromorphic continuation in $s$, invariant under $s\mapsto 2-s$, with poles located at most at $s\in \{0,\tfrac12, \tfrac32,2\}$ and at the zeros of $D_3(s)$ . Assuming that $D_3(s)$ does not vanish at $s=2$, so that $\cR_3^\star(F,s)$ has a simple pole at $s=2$, its residue then produces the renormalized integral of $F$,
\be
\begin{split}
\RN \, \int_{\cF_3} \de\mu_3\, F(\Omega) = \frac{2}{\zetastar(3)} \Res_{s=2} \cR_3^\star(F,s) \ .
\end{split}
\ee
If the order of the pole at $s=2$ is greater than one, then the renormalized integral of $F$
(defined by minimally subtracting the divergent terms in $\int_{\cF_3^\Lambda} \de\mu_3\, F$) will differ from the residue of $\cR_3^\star(F,s)$ at $s=2$ by a finite term $\delta$, which is easily computed from \eqref{RS3Lambdaregtot}.

\subsection{Lattice partition function \label{sec_latdd3}}

We now apply the previous result to the special case $F$ is the lattice partition 
function defined in \eqref{defGammaddh}. In this case, the asymptotic behavior in regions I, II, III
is given by 
\be
\begin{split}
\varphi=& t^{\tfrac{d}{2}} \Gamma_{d,d,2}(\tilde\Omega)\ ,\quad
\tilde\varphi= V^{-d}\, \Gamma_{d,d,1}(\rho)\ ,\quad
\hat\varphi = \cV^{-3d/2} \ ,
\end{split}
\ee
so that we can use the operator $\Diamond = \Diamond_3(\tfrac{d}{2})$. 
The renormalized integral $\RN\, \int_{\cF_3} \de\mu_3\, \cE^\star_3(s) \,\Gamma_{d,d,3}$ is
obtained by applying the general prescription \eqref{RS3Lambdaregtot}. The integrals appearing
on the second to fifth line of this expression were evaluated in  \eqref{RN1Gdd}, \eqref{intF1t2gamma}.
\eqref{RN2Gdd}. The last four lines of \eqref{RS3Lambdaregtot} involve integrals over the fundamental domain of $PGL(3,\IZ)$
of the form
\be
\label{defc3}
\begin{split}
c_3(\alpha,\beta)= &  \RN\, \int_{\hat\cF_{PGL(3,\IZ)}} \de\hat\mu_3\,  L^\alpha\, \tau_2^\beta \,,\\
c_3(\alpha,\beta;s')= &  \RN\, \int_{\hat\cF_{PGL(3,\IZ)}} \de\hat\mu_3\,  L^\alpha\, \tau_2^\beta\,\cE^{\star;SL(3,\IZ)}_{V}(s') \,,\\
\tilde c_3(\alpha,\beta;s')= &  \RN\, \int_{\hat\cF_{PGL(3,\IZ)}} \de\hat\mu_3\,  L^\alpha\, \tau_2^\beta
\cE^{\star;SL(3,\IZ)}_{\Lambda^2 V}(s') \,.
\end{split}
\ee
Although  these integrals do not seem to be computable in closed form, it is easy to determine their analytic properties. As mentioned below \eqref{defhmu3}, the fundamental domain $\hat\cF_{PGL(3,\IZ)}$ admits two different cusps, corresponding to i) $L\to\infty$ keeping $\tau$
fixed, ii) $L'\to\infty$ keeping $\tau'$ fixed. In the first case,  $\hat\cF_{PGL(3,\IZ)}$ reduces to 
$\IR^+_V \times \hat\cF_{PGL(2,\IZ),\tau} \times [0,1]_{A_1,A_2}^2/\IZ^2$. The integral over $L$
in the first integral $c_3(\alpha,\beta)$ is then convergent if $\Re(\alpha)<3$, and has a pole at 
$\alpha=3$ with residue 
\be
\Res_{\alpha=3} \,c_3(\alpha,\beta) = \frac14 \RN \int_{\cF_1} \de\mu_1(\tau)\, \tau_2^\beta = 
\frac{c\left( \frac{1-\beta}{2}\right)}{2(1-\beta)}\ .
\ee
For what concerns the other two integrals in \eqref{defc3}, using the asymptotic behavior
of the Eisenstein series as $L\to \infty$,
\be
\label{limESL3}
\begin{split}
\cE^{\star;SL(3,\IZ)}_{V}\left( \hat\Omega_2,s' \right) \to &\,
\zetastar(2s')\, L^{2s'} + L^{\tfrac32-s'}\, \cE^\star_1(s'-\tfrac12,\tau)\,,
\\ 
\cE^{\star;SL(3,\IZ)}_{\Lambda^2 V}\left( \hat\Omega_2,s' \right)   \to &\,
\zetastar(2s'-2)\, L^{3-2s'} + L^{s'} \,\cE^\star_1(s',\tau) \,,
  \end{split} 
\ee
we similarly find that $c_3(\alpha,\beta;s')$ has poles at $\alpha+2s'=3$ 
and at $\alpha-s'=\tfrac32$, with residues
\be
\Res_{\alpha=3-2s'} \,c_3(\alpha,\beta;s') = 
 \frac{\zetastar(2s')\, c\left( \frac{1-\beta}{2}\right)}{2(1-\beta)}\ ,\quad 
\Res_{\alpha=s'+\tfrac32} \,c_3(\alpha,\beta;s') = 
\frac14 c_2(\beta;s'-\tfrac12)\,.
\ee
Similarly, $\tilde c_3(\alpha,\beta;s')$
 has poles at $\alpha=2s'$ 
and at $\alpha=3-s'$, with residues
\be
\Res_{\alpha=2s'} \,\tilde c_3(\alpha,\beta;s') = 
 \frac{\zetastar(2s'-2)\, c\left( \frac{1-\beta}{2}\right)}{2(1-\beta)}\ ,\quad 
\Res_{\alpha=3-s'} \,\tilde c_3(\alpha,\beta;s') = 
\frac14 c_2(\beta;s')\,.
\ee

In the vicinity of the cusp ii), rewriting $L^\alpha\, \tau_2^\beta = L'^{\frac{\alpha+3\beta}{2}}
\tau_2'^{\frac{\alpha-\beta}{2}}$ and using \eqref{limESL3p}, one finds that $c_3(\alpha,\beta)$ is convergent if $\Re(\alpha+3\beta)<6$, and has a pole at 
$\alpha+3\beta=6$ with residue 
\be
\Res_{\alpha+3\beta=6} \,c_3(\alpha,\beta) = 
\frac{c\left( \beta-1 \right)}{4(\beta-1)}\ .
\ee
Similarly, using \be
\label{limESL3p}
\begin{split}
\cE^{\star;SL(3,\IZ)}_{V}\left( \hat\Omega_2,s' \right)   \to &\,
\zetastar(2s'-2)\, L'^{3-2s'} + L'^{s'} \,\cE^\star_1(s',\tau) \,,
\\
\cE^{\star;SL(3,\IZ)}_{\Lambda^2 V}\left( \hat\Omega_2,s' \right) \to &\,
\zetastar(2s')\, L'^{2s'} + L'^{\tfrac32-s'}\, \cE^\star_1(s'-\tfrac12,\tau)\,,
  \end{split} 
\ee
as $L'\to\infty$, we see that $c_3(\alpha,\beta,s')$ has poles at $\alpha+3\beta-4s'=0$ 
and $\alpha+3\beta+2s'=6$,  with residues
\be
\Res_{\alpha+3\beta=4s'} \,c_3(\alpha,\beta;s') = 
 \frac{\zetastar(2s'-2)\, c\left( \tfrac{2+\alpha-\beta}{2}\right)}{2+\alpha-\beta}\ ,\quad 
\Res_{\alpha+3\beta+2s'=6} \,c_3(\alpha,\beta,s') = 
\frac14 c_2\left(\tfrac{\alpha-\beta}{2};s'\right)\,,
\ee
while $\tilde c_3(\alpha,\beta,s')$
 has poles at $\alpha+3\beta+4s'=6$ 
and at $\alpha+3\beta-2s'=3$, with residues
\be
\Res_{\alpha+3\beta+4s'=6} \,\tilde c_3(\alpha,\beta;s') = 
\frac{\zetastar(2s')\, c\left( \tfrac{2+\alpha-\beta}{2}\right)}{2+\alpha-\beta}\ ,\quad 
\Res_{\alpha+3\beta-2s'=3} \,\tilde c_3(\alpha,\beta;s') = 
\frac14 c_2(\beta;s'-\tfrac12)\,.
\ee
These equations are of course consistent  with the functional equation 
$\cE^{\star;SL(3,\IZ)}_{V}( \hat\Omega_2,s')=\cE^{\star;SL(3,\IZ)}_{\Lambda^2 V}( \hat\Omega_2,\tfrac32-s')$. For brevity, we shall denote $c_3(\alpha)=c_3(\alpha,\alpha), c_3(\alpha;s)=c_3(\alpha,\alpha;s)$ and similarly for $\tilde c_3$. Moreover, it is useful to note the special values
\be
c_3(0)=\frac{1}{24} \zeta(3)=\frac12\zetastar(2)\,\zetastar(3)\ ,\quad c_3(0;s) = \tilde c_3(0;s) = 0\ ,
\ee
where the first relation is the volume of the fundamental domain of
$PGL(3,\IZ)$ \cite{zbMATH02628371}, and the second relation follows from the fact that the renormalized integral of 
Eisentein series vanishes.

Returning to \eqref{RS3Lambdaregtot}, the renormalized integral is therefore given by
\be
\label{RS3Lambdareglat}
\begin{split}
\RN\, &\int_{\cF_3} \de\mu_3\, \cE^\star_3(s) \,\Gamma_{d,d,3}(\Omega) 
\equiv \cR_3^\star(\Gamma_{d,d,3},s) 
= \lim_{\Lambda\to\infty} \left\{
\int_{\cF_3^{\Lambda}} \de\mu_3\, \cE^\star_3(s) \,\Gamma_{d,d,3}(\Omega) 
\right. \\
 - & \frac12  
\left[  
  \frac{\Lambda^{s+\tfrac{d}{2}-3}}{s+\tfrac{d}{2}-3} \, 
 \RN  \int_{\cF_{2}} \de\mu_2\, \cE^\star_2(s)  \, \Gamma_{d,d,2}
 +  \frac{ \Lambda^{\tfrac{d-2}{2}-s}}{\tfrac{d-2}{2}-s} \,  
   \RN  \int_{\cF_{2}} \de\mu_2\, \cE^\star_2(s-\tfrac12)  \, \Gamma_{d,d,2}\right] 
\\
- & \frac{1}{2}\left[ 
\frac{\zetastar(4s-2)\, c_2(5-d-2s)\, \Lambda^{2s-5+d}}
{2s-5+d} 
\left( \RN\,  \int_{\cF_1} \de\mu_1 \, \cE^\star_1(s)\, \Gamma_{d,d,1}   \right)
\right.
 \\
 &
  -
 \frac{ \zetastar(4s-5)\,  c_2(1-d+2s)\, \Lambda^{d-1-2s}}{2s-d+1}
\left( \RN\,  \int_{\cF_1} \de\mu_1 \, \cE^\star_1(s-1)\, \Gamma_{d,d,1}    \right)
 \\
 &
 \left. -
 \frac{ c_2\left( \tfrac72-d; 2s-\tfrac32\right)\Lambda^{d-\tfrac72}}{\tfrac72-d} \,
\left( \RN\,  \int_{\cF_1} \de\mu_1 \, \cE^\star_1(s-\tfrac12) \, \Gamma_{d,d,1}  \right) \right]
\\
-& 
6\left[ \tfrac{ \zetastar(2s)\, \zetastar(4s-2)\, c_3\left(6-\tfrac{3d}{2}-3s\right)\, \Lambda^{3s+\tfrac{3d}{2}-6}}{3s+\tfrac{3d}{2}-6}
\right.
-  \tfrac{ \zetastar(2s-3)\, \zetastar(4s-5)\, c_3\left(-\tfrac{3d}{2}+3s\right)\, 
\Lambda^{-3s+\tfrac{3d}{2}}}{3s-\tfrac{3d}{2}}
\\& 
+  \tfrac{ \zetastar(2s-1)\, c_3\left(5-\tfrac{3d}{2}-s;2s-1\right)\, 
\Lambda^{s-5+\tfrac{3d}{2}}}{s-5+\tfrac{3d}{2}}
\left.\left. -  \tfrac{ \zetastar(2s-2)\, \tilde c_3\left( s+3-\tfrac{3d}{2}; 2s-\tfrac32\right)\, 
\Lambda^{-s-3+\tfrac{3d}{2}}}{s+3-\tfrac{3d}{2}}
\right]
\right\} \,.
\end{split}
\ee

On the other hand, using \eqref{GenEuler} the renormalized Mellin transform \eqref{RNMellin3} evaluates to (see Appendix \ref{sec_lath} for more details)
\be
\cR_3^\star(\Gamma_{d,d,3},s) =\frac{ \zetastar(2s) \,  \zetastar(4s-2)\, 
\Gamma(s+\tfrac{d-4}{2})\, \Gamma(s+\tfrac{d-5}{2})\, \Gamma(s+\tfrac{d-6}{2})}
{\pi^{3s+\tfrac32(d-5)}}\, \!\!\!
\sum_{\substack{(m_i^I, n^{i,I})\in\IZ^{6d} / GL(3,\IZ)
\\ m_i^I n^{i J}+m_i^J n^{i I}=0 \\
{\rm rank}(m_i^I, n^{i,I})=3}}\!\!\!
|\cL|^{-s+\tfrac{4-d}{2}}\,,
\ee
where $\cL$ is the $3\times 3$ matrix defined in \eqref{defcL}. This is recognized as the completed
Langlands-Eisenstein series attached to the three-index antisymmetric representation of 
$SO(d,d,\IZ)$ 
\be
\cR_3^\star(\Gamma_{d,d,3},s) = 2\, \cE^{SO(d,d),\star}_{\Lambda^3V}(s+\tfrac{d-4}{2})\,.
\ee
Indeed, it is known from the Langlands constant term formula that $\cE^{SO(d,d),\star}_{\Lambda^3V}(s')$ has, for generic dimension $d$, simple poles at
\be
\label{polesE3}
s'=0,\tfrac12,1, \tfrac{d-4}{2}, \tfrac{d-3}{2},\tfrac{d-1}{2},\tfrac{d}{2}, d-3, d-\tfrac52\ ,d-2\,,
\ee
which translate into poles at $s= \tfrac{4-d}{2}, \tfrac{5-d}{2}, \tfrac{6-d}{2}, 0, \tfrac12, \tfrac32, 2, \tfrac{d-2}{2}, \tfrac{d-1}{2}, \tfrac{d}{2}$ as predicted by the Rankin-Selberg method. Moreover, from the fact that 
the regularized integral $\int_{\cF_3^{\Lambda}}\,\de\mu_3 \, \cE^\star_3(s)\, \Gamma_{d,d,3} $ has only simple poles at $s=0,\tfrac12,\tfrac32,2$, we deduce from \eqref{RS3Lambdareglat} the value of the
residues of $\cR_3^\star(\Gamma_{d,d,3},s)$ at the poles for generic value of $d$,
\be
\label{resE3}
\begin{split}
\Res_{s=\frac{d}{2}}  \cR^{\star}_3(\Gamma_{d,d,3},s) =&  
\zetastar(2)\, \zetastar(3)\, \zetastar(d-3)\, \zetastar(2d-5)\,,\\
\Res_{s=\frac{d-1}{2}}  \cR^{\star}_3(\Gamma_{d,d,3},s) =&  
\zetastar(2)\, \zetastar(2d-7)\,  \cE^{\star,SO(d,d)}_{V}(d-\tfrac52) \,,\\
\Res_{s=\frac{d-2}{2}}   \cR^{\star}_3(\Gamma_{d,d,3},s) =&  
\cE^{\star,SO(d,d)}_{\Lambda^2 V}(d-3) \,,\\
\Res_{s=2}  \cR^{\star}_3(\Gamma_{d,d,3},s)  =&  
\frac12 \zetastar(3)\, \RN\int_{\cF_3} \de\mu_3\, \Gamma_{d,d,3}\,,
\end{split}
\ee
where the renormalized integral $\RN\int_{\cF_3} \de\mu_3\, \Gamma_{d,d,3}$ is defined, for all values of $d$, as 
\be
\label{defRNintlat3}
\begin{split}
\RN \int_{\cF_3} \de\mu_3\, \Gamma_{d,d,3}= & 
\lim_{\Lambda\to\infty} \left[ \int_{\cF_3^\Lambda} \de\mu_3\, \Gamma_{d,d,3} \right.\\
&
-\frac12 \left( \frac{\Lambda^{\tfrac{d}{2}-3}}{\tfrac{d}{2}-3}\Theta(d-6)  +  \log\Lambda\, \delta_{d,6}\right)
\RN \int_{\cF_2} \de\mu_2\, \Gamma_{d,d,2}\,
\\
&-\left ( \frac{c(\tfrac{d-4}{2}) \Lambda^{d-5}}{(d-4)(d-5)} \Theta(d-5) + \frac{\pi}{6} \log\Lambda
\, \delta_{d,5} \right)\, \RN \int_{\cF_1} \de\mu_1\, \Gamma_{d,d,1}
\\
&\left.  -\left(\frac{6\, c_3(6-\tfrac{3d}{2}) 
\Lambda^{\tfrac32(d-4)}}{\frac{3}{2}(d-4)} \Theta(d-4) +
\frac14\zeta(3)\, \log\Lambda\, \delta_{d,4} \right)
\right]\,.
\end{split} 
\ee

For $d=4,5,6$, $\cR^{\star}_3(\Gamma_{d,d,3},s)$ has a pole of higher order at $s=2$, in which case its residue at $s=2$ will differ from $\tfrac12 \zetastar(3)$ times the renormalized integral 
\eqref{defRNintlat3} by a term which can be easily computed. For $d=0,1,2$, the 
Rankin-Selberg transform $\cR^{\star}_3(\Gamma_{d,d,3},s)$ vanishes but the renormalized
integral $\RN \int_{\cF_3} \de\mu_3\, \Gamma_{d,d,3}$ is still non-zero. For example,
for $d=0$, the vanishing of \eqref{RS3Lambdareglat} at $s=2$ gives
\be
 \int_{\cF_3^{\Lambda}} \de\mu_3 = 
 2\zetastar(2)\zetastar(4)\zetastar(6) +\frac{\zeta^\star(4)\, c(1)}{6\Lambda^3}
 -\frac{\zetastar(2)\, c(-2)}{10 \Lambda^5}
 - \frac{c_3(6)}{\Lambda^6} + \dots
\ee
up to exponentially suppressed terms, reproducing the known value of $\cV_3$. 
Similarly, at $d=1$ one obtains
\be
\begin{split}
 \int_{\cF_3^\Lambda} \de\mu_3\, \Gamma_{1,1,3} = & 2\zetastar(2)\,\zetastar(4)\,\zetastar(6) \,
  (R^3 + 1/R^3) -\frac{4\zeta^\star(2)\,\zeta^\star(4)}{5\Lambda^{5/2}}(R^2+1/R^2) \\
	& - \frac{c_2(4)\,\zeta^\star(2)}{4\Lambda^4}\,(R+1/R)- \frac{4c_3(9/2)}{3\Lambda^{9/2}}+\ldots
  \end{split}
\ee
and at $d=2$,
\be
\begin{split}
 \int_{\cF_3^\Lambda} \de\mu_3\, \Gamma_{2,2,3} = & 2\zetastar(2)\zetastar(4)\left[ \cE_1^\star(3,T) + \cE_1^\star(3,U)\right] \\
 			&+ \frac{c_2(3)}{6\Lambda^3}\Bigr(\log\left[ T_2 U_2|\eta(T)\eta(U)|^4\right] + 5\log \Lambda+ {\rm cte}\Bigr)+\ldots
\end{split} 
\ee
reproducing the results of \cite[(4.46)]{Pioline:2014bra}.
For $d=3$, one may show using the Langlands constant term formula that 
\be
\cE^{SO(3,3),\star}_{\Lambda^3V}(s) = \zetastar(2s)\,\zetastar(2s-1)\,\zetastar(2s-2)\,\zetastar(2s-3)\,
\left[ \cE^{SO(3,3),\star}_{S}(2s-1) + \cE^{SO(3,3),\star}_{C}(2s-1) \right]\ , 
\ee
where $\cE^{SO(3,3),\star}_{S,C}$ are Eisenstein series attached to the two spinor representations of $SO(3,3,\IZ)$ (or equivalently, the fundamental and anti-fundamental representations of $SL(4,\IZ)$). The residue at $s=2$ yields
\be
 \RN \int_{\cF_3} \de\mu_3 \Gamma_{3,3,3} =2\zetastar(2)\zetastar(4)
 \left[ \cE^{SO(3,3),\star}_{S}(3) + \cE^{SO(3,3),\star}_{C}(3) \right]\,,
\ee
as announced in \cite[(4.59)]{Pioline:2014bra}.

In Appendix \S\ref{lap_gen3}, we use \eqref{defRNintlat3} as the starting point to show that 
the renormalized integral $\RN \int_{\cF_3} \de\mu_3\, \Gamma_{d,d,3}$ is an eigenmode of the 
Laplace operator on the Grassmannian $G_{d,d}$
parametrizing even selfdual lattices of signature $(d,d)$, up to anomalous source terms which originate from logarithmic divergences.

\appendix

\section{Siegel-Eisenstein series and invariant differential operators\label{app_gendegree}}

In this appendix, we collect various properties of Siegel-Eisenstein series which are known to hold for any degree. The completed Eisenstein series $\cE^\star_h(s,\Omega)$ is defined by
\be
\label{defEstarh2}
\cE^\star_h(s,\Omega) = \cN_h(s) \, 
\sum_{\gamma\in\Gamma_{h,\infty}\backslash \Gamma_h} |\Omega_2|^{s}\vert \gamma
\ee
for $\Re(s)>\tfrac{h+1}{2}$, and by analytic continuation in $s$ elsewhere. The normalization
factor
\be
\cN_h(s) = \zetastar(2s) \, \prod_{j=1}^{\lfloor h/2 \rfloor} \zetastar(4s-2j)\, 
\ee
is chosen such that $\cE^\star_h(s,\Omega)$ is invariant under $s\mapsto \tfrac{h+1}{2}-s$.
$\cE^\star_h(s,\Omega)$ has simple poles at most at $s=\tfrac{j}{4}$ where $j$ is an integer in the range $0\leq j \leq 2h+2$ \cite{0397.10021}. In particular, it has a simple pole at $s=\tfrac{h+1}{2}$, with constant residue $r_h= \frac12 \, \prod_{j=1}^{\lfloor h/2 \rfloor}\, \zetastar(2j+1)$.  

The Eisenstein series $\cE^\star_h(s,\Omega)$ is an eigenmode of all invariant differential operators, in particular of the Laplacian on the Siegel upper half plane,
\be
\label{eigenDelta2}
\Delta_{Sp(2h)}\, \cE^\star_h(s,\Omega) = h s (s-\tfrac{h+1}{2}) \, 
\cE^\star_h(s,\Omega)\ .
\ee
It will be convenient to choose a set of generators $\Delta^{(2r)}_{Sp(2h)}$, $1\leq r\leq h$ of the ring
of invariant differential operators, such that the eigenvalue of $\cE^\star_h(s,\Omega)$ is given by 
\be
\label{eigenDeltars}
\Delta^{(2r)}_{Sp(2h)}\, \cE^\star_h(s,\Omega) = \left( {h \atop r}\right)\, \prod_{k=0}^{r-1}
(s-\tfrac{k}{2}) (s-\tfrac{h+1-k}{2}) \, \cE^\star_h(s,\Omega)\,.
\ee
Up to normalization, these operators are defined by
\be
\label{defDeltar}
\Delta^{(2r)}_{Sp(2h)} \propto |Y|^{\tfrac{r-1}{2}}\, 
Y_{i_1 k_1}\dots Y_{i_ r k_r} Y_{j_1 l_1} \dots Y_{j_r l_r} 
\pa_{\bar\Omega_{i_1 j_1}} \dots \pa_{\bar\Omega_{i_{r} j_{r}}} 
\pa_{\Omega_{k_1 l_1}}\dots \pa_{\Omega_{k_r l_r}}  \, 
 |Y|^{-\tfrac{r-1}{2}}\,,
\ee
where $Y\equiv \Omega_2$ and the sum is completely antisymmetrized over $i_{1\dots r}$, $j_{1\dots r}$, $k_{1\dots r}$, $l_{1\dots r}$. For $r=1$, the operator $\Delta^{(2)}_{Sp(2h)}$ coincides
with the Laplacian $\Delta_{Sp(2h)}$. For $r=h$, the operator $\Delta^{(2h)}_{Sp(2h)}$
is proportional to the product $\overline\Box_{w+2}\, \Box_w$ of Maass' raising and lowering operators
\be
\begin{split}
\Box_w =&  |\Omega_2|^{\tfrac{h-1}{2}-w}\cdot \det\left( \tfrac{1+\delta_{IJ}}{2}  \partial_{\Omega_{IJ}} \right)\, |\Omega_2|^{w-\tfrac{h-1}{2}} \ ,\\
\overline\Box_w = &  |\Omega_2|^{2+\tfrac{h-1}{2}}\, 
 \det\left( \tfrac{1+\delta_{IJ}}{2}  \partial_{\bar\Omega_{IJ}} \right)\, \, |\Omega_2|^{-\tfrac{h-1}{2}}\ ,
\end{split}
\ee
which raise and lower the modular weight $w$ by two units, respectively \cite{zbMATH03355125}.
The eigenvalue of $\cE^\star_h(s,\Omega)$ under $\Delta^{(2r)}_{Sp(2h)}$ can be computed using lemma 9.1 in \cite{0519.10019}. The Fourier expansion of $\cE^\star_h(s,\Omega)$ can be found in \cite{0786.11024}.

We shall also consider the multi-parameter Eisenstein series
\be
\cE_h^\star(s_1,\dots s_h,\Omega) = \cN_h(s_1,\dots s_h)\, \sum_{\gamma\in B_h\backslash 
\Gamma_h} t_1^{s_1} t_2^{s_2}\dots t_h^{s_h}\vert_\gamma\,,
\ee
where $B_h$ is the Borel subgroup of $Sp(2h,\IZ)$. The series converges absolutely when
$\Re(x_i)>\tfrac12$ and $\Re(x_j-x_i)>\tfrac12$ whenever $i<j$. Upon choosing for the normalization factor 
\be
\cN_h(s_1,\dots s_h)  =\prod_{1\leq i<j\leq h} \zetastar(1+2 x_j-2x_i)\,  
\prod_{1\leq i<j\leq h}\zetastar(1+2 x_j+2x_i)\, \prod_{1\leq i\leq h} \zetastar(1+2 x_i)\,,
\ee
where  $x_i=s_i-\tfrac{i}{2}$ with $i=1\dots h$, the Eisenstein series $\cE^\star(s_1,\dots s_h)$
is invariant under the Weyl group, which permutes the $x_i$'s and takes one $x_i$ to minus itself.
The Eisenstein series $\cE^\star_h(s,\Omega)$ is proportional to the multi-residue of $\cE^\star(s_1,\dots s_h)$ at $s_1=s_2=\dots =s$.
$\cE^\star(s_1,\dots s_h)$ is an eigenmode of the ring of invariant differential operators, with eigenvalues
\be
\label{eigenDeltar}
\begin{split}
\Delta^{(2)}_{Sp(2h)} =& \sum_{1\leq i\leq h} x_i^2 -\tfrac{1}{24}h(h+1)(2h+1) \,,\\
\Delta^{(4)}_{Sp(2h)} =&  \sum_{1\leq i<j\leq h} x_i^2 x_j^2 -\tfrac{1}{24}h(h+1)(2h-1)  
\sum_{1\leq i\leq h} x_i^2  \\
&+\tfrac{1}{5760} h(h+1)(h-1)(2h+1)(2h-1)(5h-6) \,, \\
\Delta^{(6)}_{Sp(2h)} =&  \sum_{1\leq i<j<k\leq h} x_i^2 x_j^2 x_k^2 
-\tfrac{1}{24} (h-2) (h-1) (2 h-3) \sum_{1\leq i<j\leq h} x_i^2 x_j^2   
\\
&+ \tfrac{1}{5760}(h-2) (h-1) h (2 h-3) (2h-1) (5 h-11) \sum_{1\leq i\leq h} x_i^2  
\\&
   -\tfrac{1}{2903040}
   (h-2) (h-1) h (h+1) (2 h-3) (2 h-1) (2 h+1)
   (7 h (5 h-23)+186)\,,
\end{split}
\ee
under the operators $\Delta^{(2r)}_{Sp(2h)}$ with $1\leq r\leq 3$.
More generally, $\Delta^{(2r)}_{Sp(2h)}$ involves a sum of elementary symmetric functions
of the $x_i^2$, whose coefficients of the  are fixed by requiring that $\Delta^{(r)}_{Sp(2h)}$ vanishes 
at $s_i=s=0,\tfrac12,\dots, \tfrac{r-1}{2}$. 

The boundaries of the Siegel fundamental domain correspond to regions where the imaginary part of period matrix acquires very large values in a diagonal block of size $h_2$,
\be
\label{degh2}
\Omega \to \begin{pmatrix} \Omega' & \Omega' u_2-u_1 \\ u_2^t \Omega' - u_1^t 
& \I\, V^{-1} \hat\omega +\omega_1 \end{pmatrix} \,,
\ee
where  $\Omega'$ is a period matrix of degree $h_1=h-h_2$, $\hat\omega$ is a positive definite matrix of size $h_2$ with unit determinant, $u_1, u_2$ are two $h_1\times h_2$ real matrices, $\omega_1$ is a real $h_2\times h_2$ symmetric matrix, 
and $V$ is scaled towards 0. In this limit, the Siegel domain of degree $h$ decomposes according to
\be
\cH_h \to \cH_{h_1} \times \IR^+ \times \frac{SL(h_2)}{SO(h_2)} \times \IR^{2h_1h_2} \times \IR^{\tfrac12 h_2(h_2+1)}\ ,
\ee
while the integration measure factorizes into
\be
\label{degmesh}
\de\mu_h(\Omega) \propto \frac{\de V}{V}\, V^{h_1h_2+\tfrac12 h_2(h_2+1)} \, \de\mu_{h_1}(\Omega') \, \de\hat\omega\, \de u\ ,
\ee
up to a constant normalization factor.
In the limit $V\to 0$, the fundamental domain $\cF_h$ simplifies to 
\be
\cF_h \to \cF_{h_1} \times \IR^+ \times \hat\cF_{PGL(h_2,\IZ)} \times \IZ^{2h_1h_2}/\IZ_2 \times \IZ^{\tfrac12 h_2(h_2+1)}\ ,
\ee
where $\hat\cF_{PGL(h_2,\IZ)}$ denotes a fundamental domain of the action of $PGL(h_2,\IZ)$ on 
positive definite matrices of unit determinant and $\IZ_2$ acts by flipping the sign of $(u_1,u_2)$.

For $(h_1,h_2)=(h-1,1)$, setting $t=1/V$, the Eisenstein series $ \cE^\star_{h}(s,\Omega)$ decomposes into
\be
\begin{split}
\label{Eisdeg1}
 \cE^\star_{h}(s,\Omega)\to & t^s\,  \cE^\star_{h-1}(s,\Omega')+t^{\tfrac{h+1}{2}-s}\,  
 \cE^\star_{h-1}(s-\tfrac12,\Omega')\quad (h\ \mbox{odd}) \,,\\
  \cE^\star_{h}(s,\Omega)\to & t^s\,  \zetastar(4s-h)\, \cE^\star_{h-1}(s,\Omega')+
  t^{\tfrac{h+1}{2}-s}\,  \zetastar(4s-h-1)\,
 \cE^\star_{h-1}(s-\tfrac12,\Omega')\quad (h\ \mbox{even})\,.
\end{split}
\ee
More generally, the  Eisenstein series $\cE^\star_{h}(s_1,\dots, s_h,\Omega)$ 
decomposes into a sum of terms, one of which proportional to 
$t^{s_h} \cE^\star_{h-1}(s_1,\dots, s_{h-1},\Omega')$.
Consistency of these decompositions with with \eqref{eigenDeltars} and \eqref{eigenDeltar} 
implies that the operators $\Delta^{(r)}_{Sp(2h)}$ acting on functions independent of $u_1,u_2,\omega_1$ reduce to linear combinations of $\Delta^{(r)}_{Sp(2h-2)}$
and $\Delta^{(r-1)}_{Sp(2h-2)}$,
\be
\label{degDeltar}
\Delta^{(r)}_{Sp(2h)}\to \Delta^{(r)}_{Sp(2h-2)}+ \left[ t^2 \pa^2_t - (h-1) t\pa_t
+\tfrac{r-1}{2} (h-\tfrac{r-1}{2}) \right] \, \Delta^{(r-1)}_{Sp(2h-2)}\ .
\ee
It follows from \eqref{degDeltar} that the invariant operator
\be
\Diamond_h(\sigma)=\sum_{r=0}^h (-1)^r\,  \, \prod_{k=1}^r (\sigma-\frac{h-k}{2})
(\sigma-\frac{h+k}{2})\,
\Delta^{(h-r)}_{Sp(2h)}\,,
\ee
annihilates any function of the form $t^\sigma \varphi(\Omega')$. The Eisenstein series 
$\cE^\star_h(s,\Omega)$ is an eigenmode of $\Diamond_h(\sigma)$ with eigenvalue
\be
\Diamond_h(\sigma)\,\cE^\star_h(s,\Omega)=\left[\prod_{k=0}^{h-1}(s-\sigma+\tfrac{k}{2}) (s+\sigma-\tfrac{h+1+k}{2})\right]\, \cE^\star_h(s,\Omega)\ .
\ee
These relations generalize \eqref{Diamond2sigma}, \eqref{DiamondE2}, \eqref{Diamond3sigma}, \eqref{DiamondE31} to arbitrary degree.

More generally, for $h_2>1$,   the Eisenstein series $ \cE^\star_{h}(s,\Omega)$ decomposes into
\be
\label{Eisdeghr}
 \cE^\star_{h}(s,\Omega)\to \sum_{r=0}^{h_2} V^{-s(h_2-2r)-\frac{r(h_1+r+1)}{2}}\,
 c_{h,h_1,r}(s)\, 
  \cE^\star_{h_1}(s-\tfrac{r}{2},\Omega')\, \cE^{\star,SL(h_2)}_{\Lambda^r V}
  (2s-\tfrac{h_1+r+1}{2},\hat\omega)\,,
\ee
where $ c_{h,h_1,r}(s)$ is a product of 
zeta factors. This is consistent with \eqref{eigenDelta2}, since
it follows from \eqref{degmesh} that the Laplace operator decomposes into
\be
\Delta_{Sp(2h)}\to \Delta_{Sp(2h_1)}+\frac12 \Delta_{SL(h_2)} +
\frac{1}{h_2} \left( V^2 \pa^2_V + \left[1+h_1 h_2+\tfrac12 h_2(h_2+1)\right] V\pa_V\right)\ .
\ee
The decomposition of  $\Delta_{Sp(2h)}^{(2r)}$ with $r>1$ can be worked out on a case by case basis, by requiring that the multi-parameter Eisenstein series are eigenmodes with the
eigenvalues displayed in \eqref{eigenDeltar}.

\section{Lattice partition function and Langlands-Eisenstein series \label{sec_lath}}

Here we consider the renormalized Rankin-Selberg transform of the Siegel-Narain partition function
 \eqref{defGammaddh} in arbitrary degree:
\be \label{RSNarain}
\begin{split}
\cR_h^\star(\Gamma_{d,d,h},s) =\cN_h(s) \, 
\int_{\Gamma_{\infty,h}\backslash \cH_h} \de\mu_h \, |\Omega_2|^{s+\tfrac{d}{2}}\,  
\sum_{\substack{(m_i^I, n^{i,I})\in\IZ^{2dh}\\ {\rm Rk}(m_i^I, n^{i,I})=h}}\, 
e^{-\pi\cL^{IJ}\Omega_{2,IJ}+2\pi \I m_i^I n^{i,J}\Omega_{1,IJ}}
\end{split}
\ee
where the renormalization prescription amounts to keeping only the terms where the 
matrix $M=(m_i^I, n^{i,I})$ (of size $2d\times h$) has rank $h$. The integral over $\Omega_1$ enforces the condition $m_i^I n^{i,J}+m_i^J n^{i,I}=0$, while the integral over $\Omega_2\in \cP_h/PGL(h,\IZ)$ can be unfolded onto an integral over $\cP_h$, at the expense of restricting the sum to one representative in each orbit of the action of  $PGL(h,\IZ)$ on $M$.

The integral over $\Omega_1$ is straightforward and simply imposes the BPS condition $m_i^I n^{iJ}+m_i^J n^{i I}=0$. The integral over $\Omega_2$ may be performed using the identity  \cite{0066.32002,zbMATH03355125} 
\be
\label{GenEuler}
I=\int_{\cP_h} \, \de \Omega_2\, |\Omega_2|^{\delta-\frac{h+1}{2}} \, 
e^{-{\rm Tr}(Q \Omega_2)}  = \pi^{\frac{h(h-1)}{4}} |Q|^{-\delta}\,  
\prod_{k=0}^{h-1} \Gamma(\delta-\tfrac{k}{2})\, ,
\ee
where $Q$ being a real symmetric positive definite matrix. To prove this identity, notice that $Q$  can be uniquely decomposed into the product of a lower triangular matrix $L$ and its transpose, $Q=L^T L$. Changing variables to $\Omega_2' = L\Omega_2$ and noting that $d\Omega_2' = |Q|^{\frac{h+1}{2}} d\Omega_2$, one arrives at
\be
I= |Q|^{-\delta}\int_{\cP_h} \, \de \Omega_2'\, |\Omega_2'|^{\delta-\frac{h+1}{2}} \, 
e^{-{\rm Tr}(\Omega_2')}  \,.
\ee
The Cholesky decomposition may again be used to uniquely decompose $\Omega_2' = XX^T$, for some lower triangular matrix $X$ with positive elements along the diagonal, $x_{jj}>0$ for all $j=1,\ldots, h$. Since the Jacobian for this change of variables is $|J|=2^h \prod_{j=1}^h x_{jj}^{h+1-j}$, one obtains
\be
	I =2^h |Q|^{-\delta} \left(\prod_{k=1}^h \int_0^\infty \de x_{kk}\  x_{kk}^{2\delta-k}\,e^{-x_{kk}^2}\right)\left( \prod_{i>j=1}^h \int_{-\infty}^{\infty} \de x_{ij}\ e^{-x_{ij}^2}\right) \,,
\ee
from which one immediately recovers  (\ref{GenEuler}). 

Applying \eqref{GenEuler} result to
  the integral (\ref{RSNarain}) over $\Omega_2\in \cP_h$, we arrive at
\be
\begin{split}
\cR_h^\star(\Gamma_{d,d,h},s) =\frac{\cN_h(s) \, \prod_{j=1}^h \Gamma(s-\tfrac{h+j-d}{2})}
{\pi^{\tfrac{h}{4}(1-2d+3h-4s)}}\, 
\sum_{\substack{(m_i^I, n^{i,I})\in\IZ^{2hd} / PGL(h,\IZ)
\\ m_i^I n^{i J}+m_i^J n^{i I}=0 \\
{\rm Rk}(m_i^I, n^{i,I})=h}}
|\cL|^{-s+\tfrac{h+1-d}{2}}\, .
\end{split}
\ee
This is recognized as twice the completed
Langlands-Eisenstein series attached to the $h$-index antisymmetric representation of $SO(d,d,\IZ)$,
\be
\label{RhvsLE}
\cR_h^\star(\Gamma_{d,d,h},s) = 2\, \cE^{SO(d,d),\star}_{\Lambda^hV}(s+\tfrac{d-h-1}{2})\,.
\ee
To justify this identification, recall that the completed
Langlands-Eisenstein series is defined by 
\be
\cE^{SO(d,d),\star}_{\Lambda^hV}(s') = L_h(s')\, \sum_{\gamma\in P_h(\IZ)\backslash SO(d,d,\IZ)}
e^{2s' \langle \lambda_h, H(\gamma e)\rangle}
\ee
where $\lambda_h$ is the fundamental weight associated to the representation $\Lambda^hV$, 
$P_h$ is the  parabolic subgroup of $SO(d,d,\IZ)$ obtained by deleting the simple root dual
to $\lambda_h$ , 
 $H(e)$ is the Abelian component in the Iwasawa decomposition of a coset representative $e$ in
 $G_{d,d}$, and the normalization
 factor
\be
\label{LSOh}
L_h(s')=\zeta^\star(2s'+h+1-d)
\prod_{k=0}^{h-1}\zeta^\star(2s'-k)
\prod_{j=1}^{\lfloor h/2 \rfloor} \zeta^\star(4s'+2h+2-2d-2j) \,,
\ee
is chosen such that $\cE^{SO(d,d),\star}_{\Lambda^hV}(s')$ is invariant under $s'\mapsto d-\tfrac{h+1}{2}-s'$ (see e.g. \cite[A.3]{Pioline:2015yea}). The equality \eqref{RhvsLE} is thus equivalent to
\be
\sum_{\substack{(m_i^I, n^{i,I})\in\IZ^{2hd} / PGL(h,\IZ)
\\ m_i^I n^{i J}+m_i^J n^{i I}=0 \\
{\rm Rk}(m_i^I, n^{i,I})=h}}
|\cL|^{-s'} =2 \prod_{j=1}^h \zeta(2s'-j+1)\, \sum_{\gamma\in P_h(\IZ)\backslash SO(d,d,\IZ)}
e^{2s' \langle \lambda_h, H(\gamma e)\rangle} \,,
\ee
which, in turn, follows from the identity \cite{0013.24901}
\be
\sum_{M\in \IZ^{h\times h}/GL(h,\IZ) \atop {\rm Rk}(M)=h}\, |M|^{-2s'} = \prod_{j=1}^h \zeta(2s'-j+1)\ .
\ee
Finally, it may be shown from Langlands' constant term formula that the Langlands-Eisenstein
series $\cE^{SO(d,d),\star}_{\Lambda^hV}$ behaves in the limit where the radius $R$ of one circle in the torus $T^d$ becomes very large as
\be
\label{decompsohev}
\begin{split}
\cE^{\star,SO(d,d)}_{\Lambda^hV}(s) & \to   R^h\, \cE^{\star,SO(d-1,d-1)}_{\Lambda^hV}(s-\tfrac12) \\
&+ \zetastar(2s+1-h)\, \zetastar(4s+h+2-2d)\, R^{2s}\, \cE^{\star,SO(d-1,d-1)}_{\Lambda^{h-1}V}(s) 
\\ 
& +  \zeta^\star(2s+2h+1-2d)\,  \zetastar(4s+h+1-2d)\, R^{2d-h-1-2s}\,  
\cE^{\star,SO(d-1,d-1)}_{\Lambda^{h-1}V}(s-\tfrac12) 
\end{split}
\ee
for $h$ even, and 
\be
\label{decompsohodd}
\begin{split}
\cE^{\star,SO(d,d)}_{\Lambda^hV}(s) & \to R^h\, \cE^{\star,SO(d-1,d-1)}_{\Lambda^hV}(s-\tfrac12) \\
&+ \zetastar(2s+1-h)\, R^{2s}\, \cE^{\star,SO(d-1,d-1)}_{\Lambda^{h-1}V}(s) 
\\ 
& +  \zeta^\star(2s+2h+1-2d)\,  R^{2d-h-1-2s}\,  
\cE^{\star,SO(d-1,d-1)}_{\Lambda^{h-1}V}(s-\tfrac12) 
\end{split}
\ee
for $h$ odd. Using these formulae, one can establish recursively the analytic structure 
stated in \eqref{polesE2} and  \eqref{polesE3}, and  the formulae \eqref{resE1},
\eqref{resE2}, \eqref{resE3} for the residues at the poles for generic value of $d$.

\section{Laplace equation at genus 3\label{lap_gen3}}

In this section, we establish that the renormalized integral
$\RN\int_{\cF_3} \de\mu_3 \Gamma_{d,d,3}$ studied in \S\ref{sec_latdd3} satisfies
the following differential equation:
\be
\label{delsointG3}
\begin{split}
\left[ \Delta_{SO(d,d)} + \tfrac32 d(d-4)\right]\ \RN \int_{\cF_3} \de\mu_3\, \Gamma_{d,d,3}=&
\zeta(3)\, \delta_{d,4}
+ \tfrac{5\pi}{6}  \RN \int_{\cF_1} \de\mu_1\, \Gamma_{5,5,1}\, \delta_{d,5}
\\& 
+3\,  \RN \int_{\cF_2} \de\mu_2\, \Gamma_{6,6,2}\,\delta_{d,6}\ ,
\end{split}
\ee
where $\Delta_{SO(d,d)}$ denotes the Laplace-Beltrami operator on the Grassmannian $G_{d,d}$
parametrizing even selfdual lattices of signature $(d,d)$. As explained in  \cite{Pioline:2015yea},
the renormalized integral $\RN\int_{\cF_3} \de\mu_3 \Gamma_{d,d,3}$ arises in the low-energy expansion of the four-graviton scattering amplitude in type II strings compactified on a torus $T^d$ at order $D^6\cR^4$, and Eq. \eqref{delsointG3} can be understood as a consequence of supersymmetry. 
Here, we would like to establish \eqref{delsointG3} based on  properties of the integrand near the boundaries of the fundamental domain $\cF_3$. Our analysis will extend the study in \cite{Pioline:2015nfa} of similar one-loop and two-loop modular integrals which appear at order $\cR^4$ and 
$D^4\cR^4$ in the low-energy expansion of the four-graviton scattering amplitude, and satisfy \cite{Pioline:2015nfa}\footnote{In comparing \eqref{delsointG3} and \eqref{delsointG} with
Eq. (2.23-25) in  \cite{Pioline:2015yea} and Eq. (1.5), (3.2) in \cite{Pioline:2015nfa}, one should take into account the difference of normalization of the
integration measure, $\de\mu_h^{\rm here} = 2^{-h(h+1)/2} \de\mu_h^{\rm there}$.}
\be
\label{delsointG}
\begin{split}
\left[ \Delta_{SO(d,d)} + \tfrac12 d(d-2)\right]\ \RN \int_{\cF_1} \de\mu_1\, \Gamma_{d,d,1}=&
2\, \delta_{d,2} \,,\\
\left[ \Delta_{SO(d,d)} + d(d-3)\right]\ 
\RN \int_{\cF_2} \de\mu_2\, \Gamma_{d,d,2}=&
\pi\, \delta_{d,3} + 2 \,\RN \int_{\cF_1} \de\mu_1\, \Gamma_{4,4,1}\, \delta_{d,1} \ .
\end{split}
\ee
In fact, \eqref{delsointG} will be needed for the proof of \eqref{delsointG3}. We note that the physical origin of the source terms in Eq. \eqref{delsointG3} was independently discussed in \cite{Basu:2015dqa}.

We start from the definition \eqref{defRNintlat3} of the renormalized integral.
Then, using \cite{Obers:1999um}
\be
\left[ \Delta_{SO(d,d)} - 2 \Delta_{Sp(6)} + \tfrac{3}{2}d(d-4) \right]\, \Gamma_{d,d,3}=0\,,
\ee
along with the identities \eqref{delsointG} for $h=1$ and $h=2$, we find
\be
\begin{split}
&\left[ \Delta_{SO(d,d)} + \tfrac{3}{2}d(d-4) \right]  \RN \int_{\cF_3} \de\mu_3\, \Gamma_{d,d,3}= 
\lim_{\Lambda\to\infty} \Biggr[ 2
\int_{\cF_3^\Lambda} \de\mu_3\, \Delta_{Sp(6)}  \Gamma_{d,d,3} \\
&-\frac12 \left( \frac{\Lambda^{\tfrac{d}{2}-3}}{\tfrac{d}{2}-3}\Theta(d-6)  +  \log\Lambda\, \delta_{d,6}\right)
\left[ \frac12 d(d-6) \RN \int_{\cF_2} \de\mu_2\, \Gamma_{d,d,2} +\pi \delta_{d,3} + 2 
\int_{\cF_1} \de\mu_1\, \Gamma_{4,4,1} \,\delta_{d,4} \right]\,
\\
&-\left ( \frac{c(\tfrac{d-4}{2}) \Lambda^{d-5}}{(d-4)(d-5)} \Theta(d-5) + \frac{\pi}{6} \log\Lambda
\, \delta_{d,5} \right)\, \left[ d(d-5) \RN \int_{\cF_1} \de\mu_1\, \Gamma_{d,d,1} + 2 \,\delta_{d,2} \right]
\\
&-
\frac{3}{2}d(d-4) 
\left(\frac{6\, c(6-\tfrac{3d}{2}) \Lambda^{\tfrac32(d-4)}}{\frac{3}{2}(d-4)} \Theta(d-4) +
\frac14\zeta(3)\, \log\Lambda\, \delta_{d,4} \right)
\Biggr]\,.
\end{split} 
\ee
The terms proportional to $\log\Lambda$ all drop as they have vanishing coefficient. The remaining anomalous terms also disappear in the limit $\Lambda\to\infty$. Thus, the above simplifies to
\be
\label{Delta99}
\begin{split}
\left[ \Delta_{SO(d,d)} + \tfrac{3}{2}d(d-4) \right]  \RN \int_{\cF_3} \de\mu_3\, \Gamma_{d,d,3}= 
\lim_{\Lambda\to\infty} \left[ 2
\int_{\cF_3^\Lambda} \de\mu_3\, \Delta_{Sp(6)}  \Gamma_{d,d,3} \right.\\
-\frac{d}{2} \Lambda^{\tfrac{d}{2}-3} \, \Theta(d-6)  \, \RN \int_{\cF_2} \de\mu_2\, \Gamma_{d,d,2}
\\
- \frac{d\, c(\tfrac{d-4}{2}) \Lambda^{d-5}}{d-4} \Theta(d-5)  \, \RN \int_{\cF_1} \de\mu_1\, \Gamma_{d,d,1} 
\\
 -6\,d\, c_3(6-\tfrac{3d}{2}) \Lambda^{\tfrac32(d-4)} \, \Theta(d-4) 
\Biggr]\,.
\end{split} 
\ee
The first term can be  integrated by parts and gives boundary terms from regions I, II and III,
which are easily computed using the asymptotic forms \eqref{mu3asI}, \eqref{mu3asII}, \eqref{mu3asIII}, \eqref{DeltaSp6asI},\eqref{DeltaSp6asII},\eqref{DeltaSp6asIII} of the measure 
$\de\mu_3$ 
and of the Laplacian $\Delta_{Sp(6)}$:
\be
\label{DelSp6Gparts}
\begin{split}
2\int_{\cF_3^\Lambda} \de\mu_3\, \Delta_{Sp(6)}  \Gamma_{d,d,3}=&
\int_{\cF_2} \de\mu_2(\tilde\Omega)\, \Gamma_{d,d,2}(\tilde\Omega)\,
 \left[t^{-2} \partial_t\, t^{d/2} \right]_{t=\Lambda}
 \\
&- \frac12 \int_{\cF_1} \de\mu_1(\tau)\, \int_{\cF_1} \de\mu_1(\rho)\, \Gamma_{d,d,1}(\rho)\, 
\left[ V^6 \partial_V\, V^{-d} \right]_{V=\tau_2/\Lambda} 
\\
&- 4 \int_{\hat \cF_{PGL(3,\IZ)}} \de\hat\mu_3\, 
\left[ \cV^7\,\partial_{\cV} \cV^{-3d/2} \right]_{\cV=L\tau_2/\Lambda}
\\
=& \frac{d}{2} \Lambda^{\tfrac{d}{2}-3} \, \RN \int_{\cF_2} \de\mu_2\, \Gamma_{d,d,2}
+ \frac{d\, c(\tfrac{d-4}{2}) \Lambda^{d-5}}{d-4}  \, \RN \int_{\cF_1} \de\mu_1\, \Gamma_{d,d,1}
\\&
+ 6\,d\, c_3(6-\tfrac{3d}{2}) \Lambda^{\tfrac32(d-4)}\,.
\end{split} 
\ee
These boundary terms precisely cancel the divergent terms in \eqref{Delta99}, leaving a finite
reminder in $d=4$, $d=5$ and $d=6$ as $\Lambda\to\infty$. Using $c(\tfrac12)=\tfrac{\pi}{6}$, $c_3(0)=\tfrac{1}{24}\zeta(3)$ we find precise agreement with \eqref{delsointG3}.

\providecommand{\href}[2]{#2}\begingroup\raggedright\endgroup


\end{document}